\documentclass[%
 preprint,
superscriptaddress,
 amsmath,amssymb,
 aps,
]{revtex4-2}

\usepackage{graphicx}
\usepackage{dcolumn}
\usepackage{bm}
\usepackage{mhchem}

\begin{document}


\title{Conditions for electronic hybridization between transition-metal dichalcogenide monolayers and physisorbed carbon-conjugated molecules}

\author{Jannis Krumland}
\affiliation{%
 Physics Department and IRIS Adlershof, Humboldt-Universit\"at zu Berlin, 12489 Berlin, Germany}%
\author{Caterina Cocchi}
 \affiliation{Physics Department and IRIS Adlershof, Humboldt-Universit\"at zu Berlin, 12489 Berlin, Germany}
\affiliation{Institute of Physics, Carl von Ossietzky Universit\"at Oldenburg, 26129 Oldenburg, Germany
}%
\email{caterina.cocchi@uni-oldenburg.de}

\date{\today}
             
\newpage

\begin{abstract}
Hybridization effects play a crucial role in determining the electronic properties of hybrid inorganic/organic interfaces. 
To gain insight into these important interactions, we perform a first-principles study based on hybrid density-functional theory including spin-orbit coupling, focusing on eight representative systems formed by two carbon-conjugated molecules -- pyrene and perylene -- physisorbed on the transition-metal dichalcogenide monolayers (TMDCs) \ce{MoS2}, \ce{MoSe2}, \ce{WS2}, and \ce{WSe2}.
By means of band unfolding techniques, we analyze the band structures of the considered materials, identifying the contributions of the individual constituents as well as the signatures of their hybridization.
Based on symmetry and energetic arguments, we derive general conditions for electronic hybridization between conjugated molecules and underlying TMDCs even when the former do not lie planar on the latter, thus providing the key to predict how their mutual arrangement affects their electronic interactions. 
\end{abstract}

\maketitle

\newpage
\section{Introduction}

Hybrid interfaces formed by organic semiconductors deposited on inorganic substrates have received considerable attention in the last couple of decades: Combining light-absorbing molecules with inorganic crystals characterized by enhanced charge-carrier mobility is particularly promising in view of novel opto-electronic applications~\cite{agra+11cr,wrig12sesmc,koch12pssrrl,liu14,hewl-mcla16am,stae-rink17cp}.
Since the first attempts at the beginning of this century to decorate oxide surfaces with light-harvesting molecules~\cite{grat03jppc,meng+03lang,lee+04jjap}, countless hybrid materials and interfaces have been explored in order to identify optimal combinations for the desired targets~\cite{rao+08jpcm,hsu+12nano,jnaw+15nano,tsai+15nano,paro+16afm,mcke+17am,boot+17cm,xu+18am}.
In parallel, the rise of low-dimensional materials~\cite{novo+04sci,geim-grig13nature,mak+14sci} has opened new horizons in this field.
Among others, the unique electronic and optical properties discovered for transition metal dichalcogenide (TMDC) monolayers~\cite{mak+10prl,mak+14sci,koza+14natcom,mak+18natph} have unveiled new perspectives to realize novel hybrid materials with unprecedented opto-electronic performance~\cite{mour+13nl,he+15apl,bett+16nl,cai+16cm,choi+16nano,jari+16nl,peto+16nano,zhen+16nano,kafl+17nano,liu+17nl,zhon+18jpcl,zhu+18sa,gu+18acsp,wang+18afm,zhan+18am,gobb+18am,amst+19nano,mutz+20jpcc,liao+20lang,dreh+20cm,park+21am,qiao+212DM}. 

The electronic coupling between their constituents represents one of the crucial aspects ruling the characteristics of inorganic/organic interfaces.
When molecules are physisorbed onto inorganic semiconducting substrates, chemical interactions are typically weak, especially if they do not induce charge transfer in the ground state~\cite{wang+19aem,jaco+20apx}.
However, even in this scenario, the band structure of the hybrid system is rarely the mere superposition of the features of its building blocks~\cite{cai+16cm,fu+17pccp,shen-tao17ami,habi+20ats}.
Very often, hybridization effects are large enough to generate new electronic states that are unique of the new structure, and that are responsible for the peculiar types of excitations emerging at hybrid inorganic/organic interfaces~\cite{drax+14acr,schl+15natcom,mowb-miga16jctc,ljun+17njp,turk+19ats,sula+20ees}.

\textit{Ab initio} studies have played a decisive role in understanding electronic interactions~\cite{dell+11prl,xu+13prl,schu+14afm,matt+14aem,grue+15jpcc,zhen+16nano,amst+19nano} and spectroscopic signatures~\cite{liu+17nl,wei+19jpcc,jono+20jpcc} of hybrid materials. 
In particular, many-body perturbation theory methods applied on top of density-functional theory (DFT) are capable of providing a quantitative description of the electronic and optical proprieties of hybrid materials in excellent agreement with experiments~\cite{stae-rink17cp,wei+19jpcc}. 
Unfortunately, these calculations are extremely expensive when performed on systems approaching 100 atoms in their unit cells.  
Employing DFT with range-separated hybrid functionals represents a reliable and yet numerically sustainable strategy to obtain an accurate description of the electronic structure of hybrid materials~\cite{xu+13prl,hofm+13jcp,amst+19nano}.
However, even with the reduced computational costs provided by this approach, the use of large supercells that is needed to simulate such complex systems represents a serious limitation for the interpretation of the results, as the computed electronic bands are folded with respect to those obtained in the unit cell of the inorganic substrate~\cite{trem-hoff87jacs,yang+18apx}.
This inhibits an immediate, visual identification of the electronic states of the constituents as well as of hybridization signatures.

A way to overcome these limitations is offered by band unfolding techniques, which enable mapping band structures computed within supercells in the Brillouin zone of the reference system simulated in its primitive cell~\cite{darg+97prb,wang+98prl,boyk-gerh05prb,boyk+07jpcm,ku+10prl,mayo+20jpcm,popescuZunger2012prb}.
These schemes have been successfully employed on top of first-principles calculations to decipher the electronic structure of several complex materials~\cite{made+14prb,liu+16prb,tan+16apl,chen+17prb,iwat+17prb}. 
Adopting them in the context of hybrid inorganic/organic interfaces is not only useful to obtain a more accessible representation of the band structures of these systems.
Most importantly, band unfolding methods can be used to identify and rationalize the signatures of electronic interactions between organic and inorganic constituents. 

\begin{figure*}[h!]
    \centering
    \includegraphics[width=\textwidth]{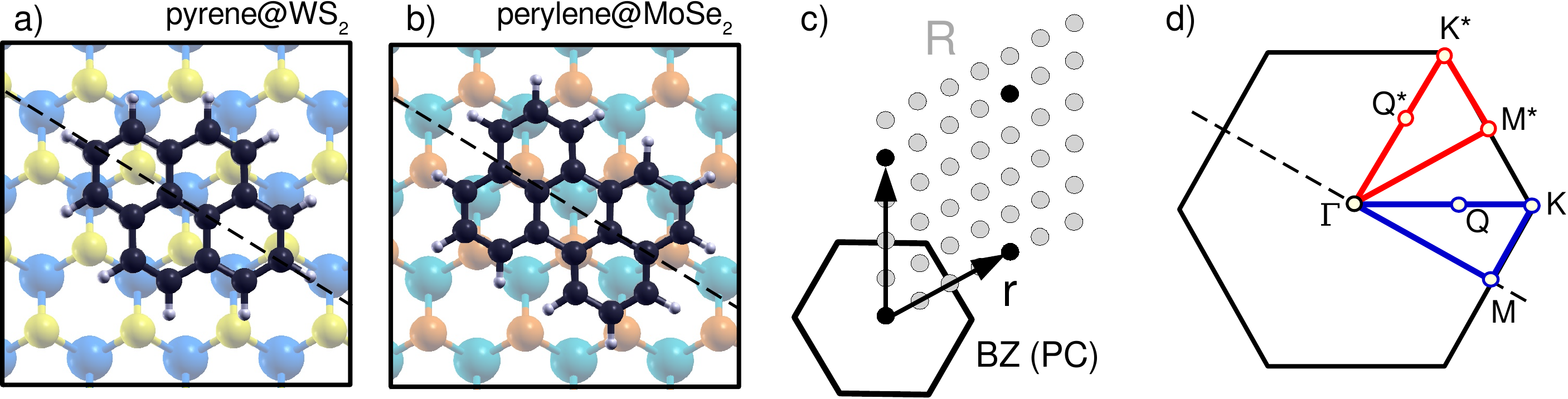}
    \caption{
    Top view of a) pyrene adsorbed on WS$_2$ and b) perylene adsorbed on MoSe$_2$. W atoms are depicted in blue, Mo atoms in turquoise, S atoms in yellow, Se ones in orange, C atoms in grey, and H in white. c) Relationship between the reciprocal lattices of a 2D hexagonal Bravais lattice described in the primitive cell (r, black points) and in a 4$\times$4 supercell (R, grey points). The arrows denote the primitive reciprocal lattice vectors of the primitive cell. d) Brillouin zone of the considered hexagonal systems with two inequivalent paths highlighted in color. Note that $Q$ and $Q^*$ are not high-symmetry points but are included in panel d) to mark the position of the local conduction band minimum in the band structures of the TMDCs. The dashed lines in panels a), b), and d) indicate the direction of the long molecular axis.}
    \label{fig.structuresAndBZ}
\end{figure*}

In this paper, we present a detailed first-principles study, based on hybrid DFT, of the electronic properties of hybrid inorganic/organic interfaces formed by two representative polycyclic aromatic hydrocarbons (PAHs) physisorbed on the four monolayer TMDCs, \ce{MoS2}, \ce{MoSe2}, \ce{WS2}, and \ce{WSe2} (see figures~\ref{fig.structuresAndBZ}a-b).
The considered molecules, pyrene and perylene, are of particular interest in the context of hybrid materials.
The former is frequently adopted as an luminescent probe on several inorganic substrates~\cite{bhow+16jpcc,ritt+20acsaem}; the latter and its derivatives have been extensively used in hybrid interfaces~\cite{azum+02jap,neub+11jpcc,abde+19acsaem} as they absorb visible radiation.
In order to achieve a rigorous description of the band structures of the considered systems, spin-orbit coupling (SOC) effects, which are known to be crucial in the above-mentioned TMDCs, are explicitly accounted for.
By applying appropriate band unfolding techniques, developed in an in-house implemented post-processing script, we identify the electronic contributions of the individual constituents as well as the signatures of their hybridization.
In this analysis, we discuss the role of the metal and of the chalcogen atoms in the electronic structure of the TMDCs, and highlight how the planar anisotropy of the physisorbed molecules affects their interactions with the substrate. 
We finally derive conditions for electronic hybridization between conjugated molecules and underlying TMDCs even when the former do not lie planar on the latter. 

\section{Methodology}

\subsection{Theoretical Background}
\label{sec.theory}
The electronic properties of the systems considered in this work are calculated using DFT \cite{hohenbergKohn1964pr} in the Kohn-Sham (KS) scheme \cite{kohnSham1965pr}. This implies self-consistently solving the KS equation,
\begin{align}\label{eq.ks}
\left(\frac{\textbf{p}^2}{2m}+V_\text{KS}\right)|n\textbf{k}\rangle = E_n(\textbf{k})|n\textbf{k}\rangle,
\end{align}
for Bloch states $|n\textbf{k}\rangle$ with band index $n$, wave vector $\textbf{k}$, and energy $E_n(\textbf{k})$. In equation~\eqref{eq.ks},
 $m$ is the electron mass, $\textbf{p}$ is the single-particle momentum operator, and $V_\text{KS} = V_\text{ext}+V_\text{H}+V_\text{xc}$ is the KS potential, which contains the nuclear (pseudo)potential ($V_\text{ext}$), the Hartree potential ($V_\text{H}$), and the exchange-correlation term ($V_\text{xc}$), the exact form of which is not known and thus requires approximations.
The solution of equation~\eqref{eq.ks} also gives us access to the total energy of the system, $E_\text{tot}$, which can be used to determine molecular adsorption energies, $E_\text{ads}$.
In the context of the investigated hybrid interfaces constituted by a TMDC monolayer and a PAH, this is achieved by computing
\begin{align}
    E_\text{ads} = E_\text{tot}^\text{(PAH+TMDC)}-E_\text{tot}^\text{(PAH)}-E_\text{tot}^\text{(TMDC)},
\end{align}
where $E^\text{(PAH+TMDC)}$ is evaluated for the hybrid system in its relaxed geometry, and $E^\text{(PAH)}$ and $E^\text{(TMDC)}$ are calculated in the same simulation cell as the hybrid system after the removal of the TMDC and of the PAH atoms, respectively, without performing any additional structural optimization.

The analysis of the electronic structure computed from DFT is conducted using unfolding techniques, \textit{i.e.}, by mapping Bloch-vector-dependent quantities defined in the supercell (SC) calculations into the primitive cell (PC). This transformation can be performed even if the PC translational invariance is broken in the SC, \textit{e.g.}, by a lattice impurity or an adsorbed molecule. Using the band energies $E=E_n(\textbf{K})$, with band index $n$ and SC wave vector $\textbf{K}$, we define the spectral function as \cite{popescuZunger2012prb}
\begin{align}
    W(\textbf{k},E) =\sum_nW_n(\textbf{k})\delta(E-E_{n}(\textbf{K}))
    \label{eq.W}
\end{align}
with 
\begin{align}
    W_n(\textbf{k}) = W_n(\textbf{K}+\textbf{G}_0) = \sum_{\sigma=\uparrow,\downarrow}\sum_{\textbf{g}\in{\text r}}|c_{n\sigma,\textbf{K}}(\textbf{G}_0+\textbf{g})|^2,
    \label{eq.Wn}
\end{align}
where $\textbf{k}$ is the wave vector in the Brillouin zone (BZ) of the PC, which is folded into $\textbf{K}$ by means the SC reciprocal lattice vector $\textbf{G}_0$, such that $\textbf{k} = \textbf{K} + \textbf{G}_0$. The sum over $\textbf{g}$ encompasses the PC reciprocal lattice (r), which is a subset of the SC one (R) (Fig.~\ref{fig.structuresAndBZ}c). The coefficients being thus selectively summed originate from the plane-wave representation of the spinor $|n\textbf{K}\rangle$ associated with the band index $n$ and the SC \textbf{k}-point $\textbf{K}$:
\begin{equation}
    |n\textbf{K}\rangle = \sum_{\sigma=\uparrow,\downarrow}\left(\sum_{\textbf{G}\in{\text R}}c_{n\sigma,\textbf{K}}(\textbf{G})|\textbf{K}+\textbf{G}\rangle\right)\otimes|\sigma\rangle,
\end{equation}
where $\langle\textbf{r}|\textbf{G}\rangle=\Omega^{-3/2}\text e^{i\textbf{G}\cdot\textbf{r}}$ is the plane wave with wave vector $\textbf{G}$ normalized to the volume $\Omega$ of the SC, and $|\sigma\rangle$ = $|\uparrow\rangle$ or $|\downarrow\rangle$ represent spin-up and spin-down channels, which are mixed through SOC.

For SCs without perturbations breaking the PC translation symmetry, $W_n(\textbf{k})\in\{0,1\}$, and $W(\textbf{k},E)$ reduces to the band structure of the PC. In the presence of impurities breaking the PC lattice-translational invariance, $W_n(\textbf{K})$ can assume values between 0 and 1, giving an indication about the degree of Bloch character of the band. Values close to 1 show that the corresponding wave function has plane-wave character (\textit{modulo} PC-periodic functions), whereas low values, thinly spread throughout the BZ, indicate that the state must be represented by a superposition of many plane waves.
This could be for example the case of localized defect states, but not necessarily of molecular orbitals. Indeed, some of them can come quite close to plane waves, as we will see in the following. 
Although not relevant for the purpose of the present analysis, we note in passing the analogy between mapping the molecular orbitals in the Brillouin zone of the substrate and analyzing exciton contributions in reciprocal-space~\cite{fu+17pccp,cocc+18pccp}.

The ionization potential of the building blocks of the hybrid system is of chief importance in the context of heterostructures, as it is a key factor determining the mutual level alignment of their components. While this is generally not an easily accessible quantity after imposing periodic boundary conditions, it can be estimated here as 
\begin{align}\label{eq.ip}
 \text{IP} = \lim_{z\rightarrow\infty}[V_\text{ext}(\textbf{r})+V_\text{H}(\textbf{r})]-E_\text{VBM}(\textbf{k}_\text{VBM}),
\end{align}
since in all considered cases there is at least one dimension, marked as $z$ in equation~\eqref{eq.ip}, along which the system is non periodic.
Inspection of the electrostatic potential ($V_\text{ext}+V_\text{H}$) along $z$ shows that it converges rapidly in the vacuum layer separating replicas, such that the the vacuum level, which is the asymptotic value in equation~\eqref{eq.ip}, can be well approximated by the electrostatic potential assumed halfway between periodic replicas along $z$.

\subsection{Computational Details}
DFT calculations are performed with the Quantum Espresso suite~\cite{qe2009, qe2020}, version 6.7. The wave-function cutoff is set to 30~Ry and 40~Ry for Mo- and W-containing systems, respectively, while the cutoff for the density is four times as high. The $c$ parameter of the hexagonal lattice is set to 20~\AA, providing a sufficient amount of vacuum to decouple periodic replicas along the $z$ direction. 
Geometries are optimized using the Perdew-Burke-Ernzerhof (PBE) approximation \cite{pbe} for $V_\mathrm{xc}$, together with norm-conserving, scalar-relativistic SG15 pseudopotentials~\cite{sg15_2015} for $V_\text{ext}$; the pairwise Tkatschenko-Scheffler (TS) scheme~\cite{ts} is applied to account for van der Waals interactions. Band structures are computed with the Heyd-Scuseria-Ernzerhof (HSE06) functional~\cite{hse06} in conjunction with fully-relativistic versions of the SG15 pseudopotentials~\cite{sg15_2016} including SOC. Band structures computed in the PC are interpolated with the Wannier90 code~\cite{wannier90}; the character of the electronic states is determined by projection of the corresponding wave functions onto the Wannier functions~\cite{marzari+2012rmp} localized at the chalcogen atoms. In the SC, we instead add the \textbf{k}-points along the desired path with zero weight to the mesh of the self-consistent calculation, thus gaining direct access to the plane-wave representations of the wave functions needed for unfolding. In the PC calculations, we use a 8$\times$8$\times$1 \textbf{k}-mesh to sample the BZ and a 4$\times$4$\times$1 \textbf{q}-mesh for the evaluation of the Fock exchange; in the SC ones, corresponding 2$\times$2$\times$1 and 1$\times$1$\times$1 grids are employed, respectively. For molecule-only calculations in the SC of the hybrid system, we use PBE employing a 8$\times$8$\times$1 \textbf{k}-mesh in the non-self-consistent calculation, which allows us to map the orbitals to the whole BZ with satisfactory resolution. We deem PBE to be sufficient in this context, as we are mainly interested in the character of the orbitals, which is insensitive to the choice of the exchange-correlation potential.


\section{Results and Discussion}

\subsection{Structural properties of the hybrid systems}
\label{sec.struct}
The initial geometry of the hybrid systems is constructed by expanding the optimized PC of the considered TMDC monolayers into a 4$\times$4 SC and placing the molecule above the transition-metal layer at a distance of 1.4 times the in-plane lattice constant of the TMDC.
The long molecular axis is aligned with the long diagonal of the hexagonal SC, maximizing the distance between molecular replicas in neighboring cells (see figure~\ref{fig.structuresAndBZ}a-b). This symmetric alignment is maintained throughout the structural optimization. Monolayer TMDCs have $D_{3h}$ symmetry; neighboring high-symmetry points K and K' as well as M and M' along the boundary of the hexagonal Brillouin zone are inequivalent. However, as a consequence of the 3-fold rotation axis and time-reversal symmetry, the energy dispersion along the $\Gamma$-M-K-$\Gamma$ and $\Gamma$-M'-K'-$\Gamma$ paths is identical, although there are differences in the spin structure \cite{dresselhaus2007group, xiao2012prl}. For our purposes, we can neglect these details and represent the band structures of the TMDCs as if the systems had hexagonal $D_{6h}$ symmetry in reciprocal space. 
However, belonging to the $D_{2h}$ point group, the adsorbed molecule lowers this symmetry, giving rise to different energy dispersions along the directions of the long and short molecular axis. We refer to corresponding \textbf{k}-paths as $\Gamma$-M-K-$\Gamma$ (along long axis) and $\Gamma$-M$^*$-K$^*$-$\Gamma$ (along short axis, see figure \ref{fig.structuresAndBZ}d). We emphasize that K$^*$ and M$^*$ do not coincide with the previously mentioned K' and M'. To gain a full understanding of the band structure of the hybrid system, both paths should be considered. 
In this work, we mainly focus on the dispersion along the $\Gamma$-M-K-$\Gamma$ path, but dedicate Section~\ref{sec.paths} to exploring and rationalizing the differences emerging for the two paths in two selected systems.

\begin{table*}[h]
\caption{Primitive-cell lattice constants $a$, transition metal-chalcogen bond length $l$, distance between the chalcogen layers ($2z$), TMDC-molecule separation $d$, and molecular adsorption energies $E_\text{ads}$, calculated at the PBE+TS theory level.}
\vspace{0.3cm}
    \centering
\begin{tabular}{p{3.2cm}p{2cm}p{2cm}p{2cm}p{2cm}}
\hline
      &  \centering MoS$_2$  & \centering MoSe$_2$ & \centering WS$_2$ & \centering WSe$_2$ \arraybackslash\\
      \hline
      \centering $a$ (\AA) & \centering 3.18 & \centering 3.32 &
      \centering 3.19 & \centering 3.32\arraybackslash\\
      \centering $l$ (\AA) & \centering 2.42 & \centering 2.55 &
      \centering 2.41 & \centering 2.54\arraybackslash\\
      \centering $2z$ (\AA) & \centering 3.13 & \centering 3.33 &
      \centering 3.12 & \centering 3.33\arraybackslash\\
      \centering $d^\text{(Py)}$ (\AA) & \centering 3.30 & \centering 3.42 & \centering 3.34 & \centering 3.40
       \arraybackslash\\
         \centering $d^\text{(Pe)}$ (\AA) & \centering 3.36 & \centering 3.48 & \centering 3.33 & \centering 3.52
       \arraybackslash\\
      \centering $E_\text{ads}^\text{(Py)}$ (eV) & \centering-1.48 & \centering -1.39 & \centering -1.42 & \centering -1.65 \arraybackslash\\
      \centering $E_\text{ads}^\text{(Pe)}$ (eV) & \centering -1.71 & \centering -1.62 & \centering -1.37 & \centering -1.58 \arraybackslash\\
     \hline 
\end{tabular}
\label{tab.alatAndAdsorption}
\end{table*}
%

Among the four TMDCs, the diselenides have larger lattice constants than the disulfides, as known from previous experimental and \textit{ab initio} studies~\cite{boek+01prb,pisa+21prb}, while the transition-metal species barely affect these values (see table \ref{tab.alatAndAdsorption}). The same is true for the bond lengths between the transition-metal and the chalcogen atoms, as well as for the distances between the chalcogen layers within the TMDC. The presence of adsorbed PAHs affects these values by less than 0.01~\AA{}. 
The C-C bond lengths in the PAHs (1.40~$\pm$~0.03~\AA{}) are incompatible with those in the TMDC, making the conjugated network incommensurate with respect to the underlying TMDC lattice. For this reason, the adsorption configuration of the PAH is arbitrary.
The distances between the absorbed molecules and the TMDCs range from 3.3~\AA{} to 3.5~\AA{}.
These values are close to the interlayer separation of corresponding TMDC-graphene heterostructures~\cite{ma+2011ns, sun+2019md}. In general, such distances are larger for the Se-containing hybrid systems than for the S-based ones, due to the larger size of the Se atoms.

Inspecting the adsorption energies (table \ref{tab.alatAndAdsorption}), we find values between -1.4~eV and -1.7~eV, but no clear correlation with the atomic masses of the TMDC-constituting elements nor with the size of the PAHs. This is the first hint at the presence of interface-specific and non-trivial interactions between the organic and inorganic components, going beyond the common picture of a system held together purely by van der Waals forces.

\subsection{Electronic properties}
In the next step of our analysis, we investigate the electronic properties of the hybrid systems described in Section~\ref{sec.struct}.
To set the stage, we first inspect the band structures of the isolated monolayer TMDCs computed in their unit cells (Section~\ref{sec.tmdc}) and then move on considering the effects of pyrene (Section~\ref{sec.pyrene}) and perylene adsorption (Section~\ref{sec.peryl})

\subsubsection{Monolayer TMDCs}
\label{sec.tmdc}

\begin{figure*}[h]
    \centering
    \includegraphics[width=\textwidth]{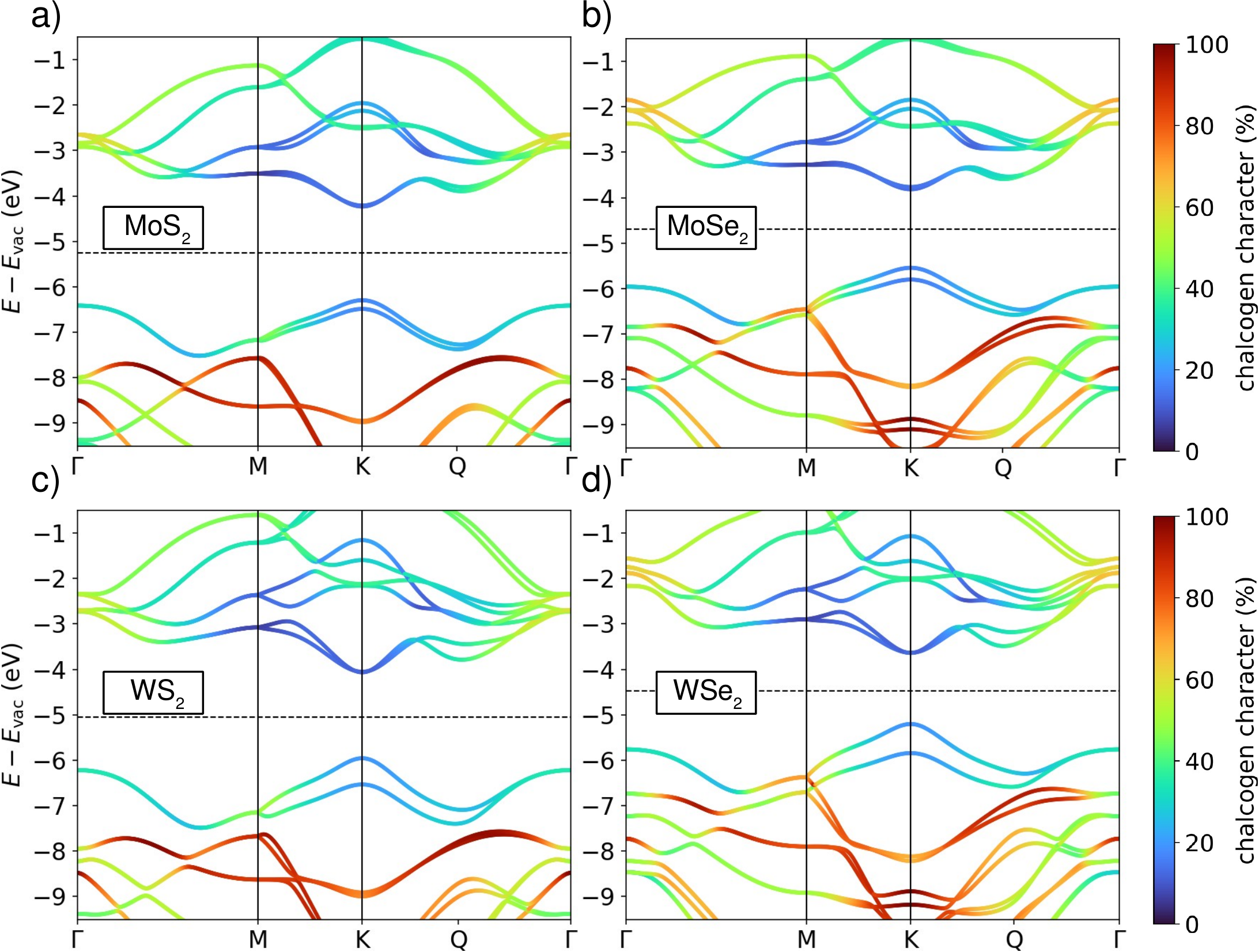}
    \caption{Band structure of the the four TDMCs in their unit cells, calculated with HSE06 and plotted with Wannier interpolation. The color code indicates the degree of chalcogen character of the bands, which is calculated by projection onto the Wannier functions centered on the corresponding atoms. Energies are expressed with respect to the vacuum level ($E_\text{vac}$), calculated from the electrostatic potential.}
    \label{fig.unitCell}
\end{figure*}

We first review the electronic band structure of the four considered TMDC monolayers, focusing in particular on the character of the electronic states (see figure~\ref{fig.unitCell}).
Computed band gaps and band dispersions are in good agreement with previous results obtained at analogous level of theory~\cite{kang+2013apl,gusakova+2017pa,ramasubramaniam2012prb}.
In all systems, the fundamental gap is direct and located at the high-symmetry point K, with the corresponding values ranging from 1.6~eV to 2.1~eV, inversely related to the atomic masses of the involved species (see table~\ref{tab.IP-gap}). The ionization potential is strongly influenced by the chalcogen species, with diselenides having values 0.75~eV smaller than their disulfide relatives. Thus, the bands of the former are shifted up in energy, such that heterojunctions between disulfides and diselenides tend to exhibit a staggered (type-II) level alignment~\cite{kang+2013apl}. This shift is somewhat smaller for the conduction bands than for the valence bands, resulting in smaller bandgaps for the diselenides. 

Apart from these overall shifts, the chalcogen species affects also the energetic dispersion. This is evident upon comparison of the valence bands, but to lesser extent also in the conduction bands, which appear vertically compressed in the diselenide monolayers. This results in a smaller conduction bandwidth, which we define as the difference of the highest and the lowest energies within the first manifold of conduction bands. The opposite is true when replacing Mo with W, which instead increases this value. Furthermore, the transition-metal species affects the size of the indirect gap between K and Q, as the local conduction band minimum at Q is subject to a stronger spin-orbit split in the W-containing monolayers compared to the Mo-based ones. We note that in \ce{WSe2}, the local conduction band minimum (CBm) at Q is almost degenerate with the global CBm at K. In fact, experimental results~\cite{zhang+2015nl, hsu+2017natcomm} as well as many-body perturbation theory studies~\cite{ramasubramaniam2012prb} suggest that \ce{WSe2} actually has an indirect bandgap. Generally, however, the effects of SOC in the W monolayers are most evident around K, where corresponding fingerprints are easily identified in higher conduction bands upon comparison to the Mo-containing counterparts, while the CBm is always unaffected. The highest valence bands in the W-based monolayers are parted by up to 0.6~eV due to SOC and thus significantly more than in the Mo-containing TMDCs, which is the root of the smaller bandgaps of the W-based siblings.

\begin{table*}[h!]
\caption{Ionization potentials (IP), direct $(\mathrm{K}\rightarrow\mathrm{K})$ and indirect $(\mathrm{K}\rightarrow\mathrm{Q})$ energy gaps $E_g$, and the conduction bandwidth $\Delta E_c$, calculated at the HSE06+SOC theory level. All values are in eV.}
\vspace{0.3cm}
    \centering
\begin{tabular}{p{3.2cm}p{2cm}p{2cm}p{2cm}p{2cm}}
\hline
      &  \centering MoS$_2$  & \centering MoSe$_2$ & \centering WS$_2$ & \centering WSe$_2$ \arraybackslash\\
      \hline
      \centering
      \centering IP & \centering 6.29 & \centering 5.54 & \centering 5.95 & \centering 5.20 \arraybackslash\\
      \centering $E_g^{(\mathrm{K}\rightarrow\mathrm{K})}$ & \centering 2.06 & \centering 1.73 & \centering 1.90 & \centering 1.57\arraybackslash\\
      \centering $E_g^{(\mathrm{K}\rightarrow\mathrm{Q})}$ & \centering 2.39 & \centering 1.96 & \centering 2.17 & \centering 1.71 \arraybackslash\\
      \centering $\Delta E_c$ & \centering 3.81 & \centering 3.40 & \centering 4.43 & \centering 3.96 \arraybackslash\\
     \hline 
\end{tabular}
\label{tab.IP-gap}
\end{table*}

The character of the electronic bands is expected to play a key role in the interactions with adsorbed molecules, as it is correlated to the localization of the wave functions within the TMDC. Due to the structural characteristics of the monolayers, states dominated by atomic orbitals of the chalcogen species are situated on the two outer atomic layers of the TMDC, while those with transition-metal character lie inside and moreover are strongly confined to the nuclei, since they correspond to $d$ electrons.
Hence, wave functions localized on the chalcogen atoms will more likely interact with the orbitals of the adsorbed molecule due to the enhanced overlap.
This also implies that the interaction between the organic and inorganic parts are unlikely to be ruled by strong correlations typically associated with $d$ or $f$ electrons. 
We note in passing that, likewise, phenomena like surface reconstruction, which lead to reduced coordination at the surface, can be excluded from this discussion: in TMDCs, they appear at very high temperatures~\cite{tiwa+08ss} when, however, molecules would desorb from the substrate.
Regarding the molecular part, only $\pi$ orbitals can be expected to have significant overlap with TMDC wave functions.
As shown in figure~\ref{fig.unitCell}, the two highest valence bands of the TMDCs bear a predominant transition-metal $d$ character; only around the high-symmetry point M, some admixture of chalcogen $p$ states are visible. The valence states directly underneath, on the other hand, are mainly constituted by chalcogen $p$ states. The lowest conduction bands are dominated by transition-metal states around K;
all eight lowest conduction bands are proportionally mixed at $\Gamma$.

%

\subsubsection{Pyrene.}
\label{sec.pyrene}

\begin{figure*}[h]
    \centering
    \includegraphics[width=\textwidth]{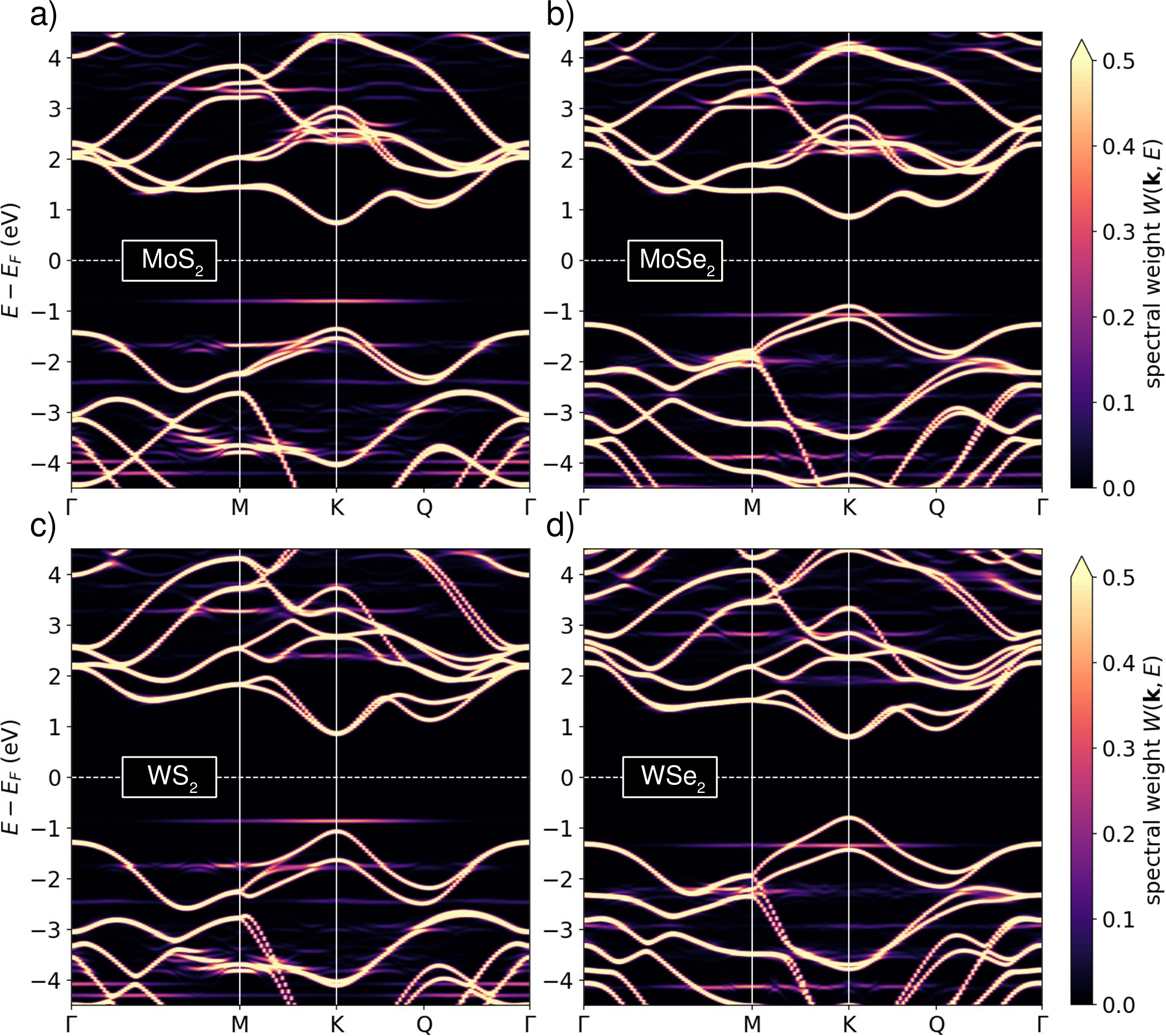}
    \caption{Unfolded band structures of pyrene adsorbed on a) MoS$_2$, b) MoSe$_2$, c) WS$_2$, and d) WSe$_2$, computed with HSE06 and spin-orbit coupling. The energy scales are shifted with respect to the calculated Fermi energy ($E_F$).}
    \label{fig.pyreneUnfolded}
\end{figure*}

We start the analysis of the electronic structure of the hybrid systems considering pyrene (Py) as an adsorbant.
Physisorbed on \ce{MoS2}, this molecule gives rise to a heterostructure with type-II level alignment, with the highest-occupied molecular orbital (HOMO) being $\sim$0.5~eV above the valence band maximum (VBM) of the TMDC, and localized around the high-symmetry point K (see figure~\ref{fig.pyreneUnfolded}a). Conversely, the HOMO-1, found at approximately -1.7~eV, is evidently hybridized with the \ce{MoS2} bands halfway between M and K.
Considering even deeper molecular states, we find strong interactions with TMDC bands close to -3.8~eV in the vicinity of M, while the lowest, $\Gamma$-centered orbitals do not exhibit signatures of hybridization with the substrate. 
In the conduction region, we see that both the lowest unoccupied molecular orbital (LUMO) around +2.5~eV as well as the LUMO+1 at +3.4~eV couple with the TMDC bands at K and M, respectively.

When Py is adsorbed onto \ce{MoSe2} (see figure \ref{fig.pyreneUnfolded}b), the resulting hybrid system exhibits a type-I level alignment, which is a consequence of the decreased ionization potential in the \ce{MoSe2} monolayer compared to its disulfide counterpart (table~\ref{tab.IP-gap}).   
Although in this electronic configuration the HOMO of Py is found inside the valence region of the TMDC, there is no interaction with these states.
This result can be understood considering the prevailing transition-metal character of the TMDC bands around K (see figure~\ref{fig.unitCell}), as well as to the comparatively low spectral weight of the HOMO (figure \ref{fig.pyreneUnfolded}b).
The increased chalcogen contribution at M enables instead the hybridization between the HOMO-1 and the bands of \ce{MoSe2} at -2~eV.
Contrary to these differences regarding the valence region, the conduction bands of Py@MoSe$_2$ look remarkably similar to those of Py@MoS$_2$: In both cases, the LUMO interacts with TMDC conduction bands around K, and the LUMO+1 shows signs of hybridization in the vicinity of M.
As mentioned before, the lower ionization potential in the diselenide corresponds to an upshift of the bands. However, this effect is compensated in the conduction band by the smaller bandgap and bandwidth, resulting in almost identical level alignments and hybridization effects in the conduction region of Py@\ce{MoS2} and Py@\ce{MoSe2}.

Considering now Py adsorbed on the W-based TMDCs, differences arise mainly as a consequence of the stronger SOC in the substrate, as well as of the increased conduction bandwidth (see figure~\ref{fig.pyreneUnfolded}c-d).
In Py@WS$_2$, the energetic distance between the HOMO and the VBM is reduced when compared to Py@MoS$_2$, which is a result of the larger VBM split in \ce{WS2}. 
Similar to Py@\ce{MoS2}, there is hybridization between the HOMO-1 of Py and the highest valence bands of \ce{WS2} in the region between M and K. However, while in the former system both spin-split bands partake in these interactions, mainly the upper one in \ce{WS2} is involved, since the lower one overlaps less with the orbital. The strong interaction with the upper band results in a smooth transition from the HOMO-1 of Py at M into the upper valence band towards K, possibly constituting an efficient pathway for charge carrier exchange between the TMDC and the PAH. Deeper in the valence region, the picture is fairly similar to Py@MoS$_2$, as spin-orbit splits are less dramatic. The conduction bands, on the other hand, look quite different. Especially in the region between 2~eV and 3.5~eV, hybridization effects with the unoccupied orbitals of Py are no longer as salient as in the Mo-containing hybrid systems. The alignment of bands and orbitals is different due to the vertical stretch of the conduction bands in the W-featuring monolayer, reflected in its higher bandwidth (table~\ref{tab.IP-gap}). Py@WSe$_2$ follows similar trends: The valence region is qualitatively similar to that of Py@MoSe$_2$, as the larger SOC in the former does not influence the level alignment; the conduction bands resemble those of Py@WS$_2$, as the lower ionization potential of WSe$_2$ is compensated by bandgap and bandwidth reductions.

\begin{figure*}[h]
    \centering
    \includegraphics[width=\textwidth]{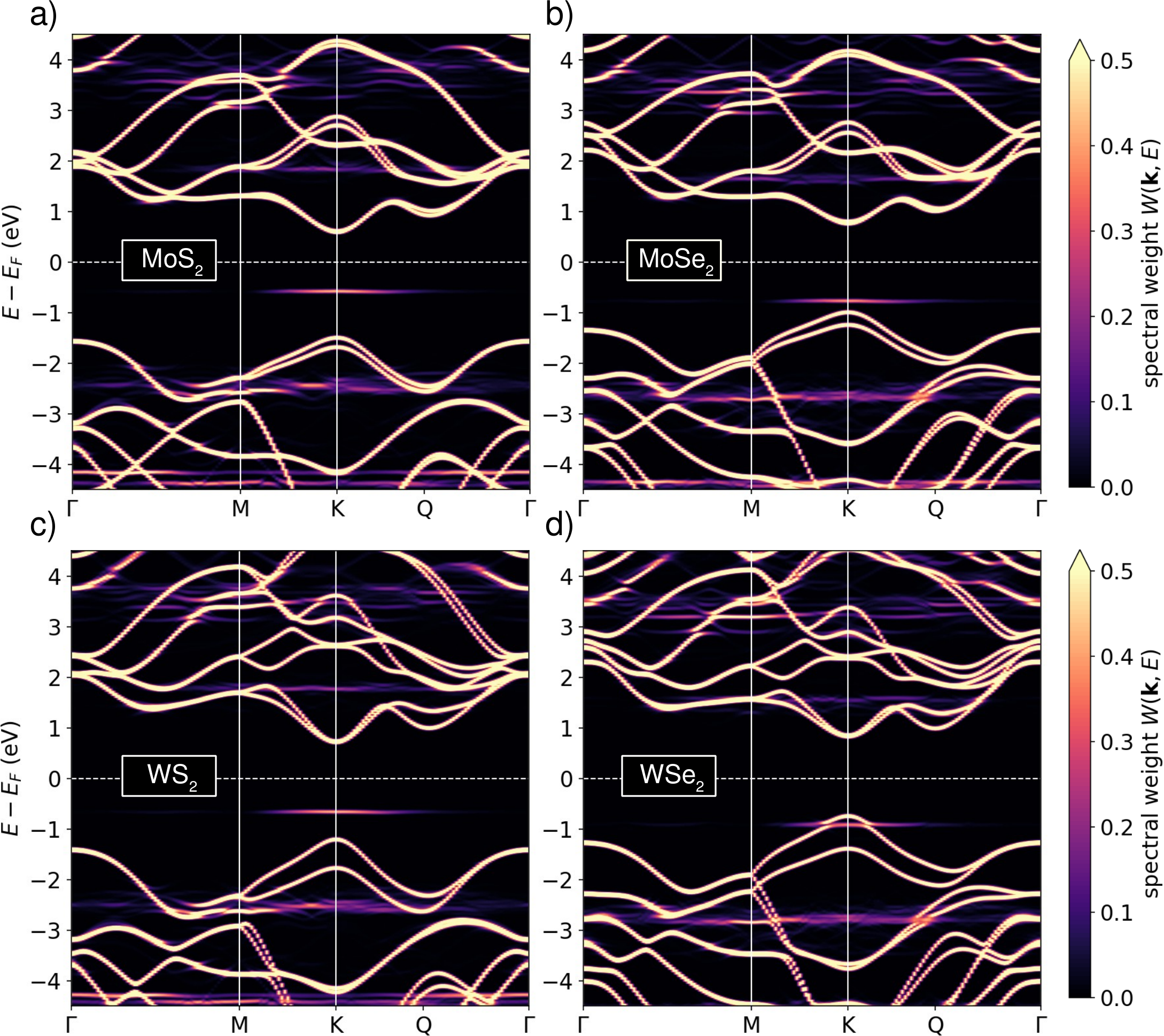}
    \caption{Unfolded band structure of perylene adsorbed on a) MoS$_2$, b) MoSe$_2$, c) WS$_2$, and d) WSe$_2$, computed with HSE06 and spin-orbit coupling. The energy scales are shifted with respect to the Fermi energy ($E_F$), placed mid-gap.}
    \label{fig.peryleneUnfolded}
\end{figure*}
%

\subsubsection{Perylene.} 
\label{sec.peryl}

The energy levels of the second considered adsorbant molecule, perylene (Pe), align quite differently with the TMDC ones when compared to those of Py. We find the HOMO of Pe much higher in the energy gap of MoS$_2$ (figure~\ref{fig.peryleneUnfolded}a), resulting in a clear type-II level alignment. No molecular levels unfold close to the VBM, but a mixture thereof appears around -2.5~eV. At this energy, signs of hybridization are evident in the vicinity of M, where the bands of MoS$_2$ carry notable chalcogen character (see figure~\ref{fig.unitCell}).
In the conduction region, only bands well above the CBm are affected. Moreover, no virtual molecular orbitals unfold close to TMDC states at the high-symmetry point K. An overall very similar picture is seen also in Pe@WS$_2$ (figure~\ref{fig.peryleneUnfolded}c).

Moving on to the hybrid systems with diselenide substrates, we find for Pe@MoSe$_2$ the only type-II alignment among the Se-featuring hybrid systems (figure~\ref{fig.peryleneUnfolded}b).
Interactions between lower-lying states are present at -2.7~eV close to M, but they are fairly weak despite the high Se contribution to the TMDC band at that energy. This is likely a consequence of the significant slope of the band, such that the electronic states overlap only within a very limited \textbf{k}-region. In Pe@WSe$_2$, there are actually some interactions close to the VBM, in spite of the comparatively low chalcogen character of the TMDC band, giving rise to an avoided crossing. We explain the presence of these interactions in this system and their absence in the similarly aligned Py@MoSe$_2$ (see figure~\ref{fig.pyreneUnfolded}b) by the slightly higher chalcogen character at the VBM in WSe$_2$ compared to MoSe$_2$, as well as by the more concentrated spectral weight of the HOMO of Pe with respect to the one of Py.

\subsection{Patterns.}
From the previous analysis, we can derive three conditions that have to be met for hybridization between molecular orbitals and TMDC bands to occur:

    \begin{enumerate}
        \item the energies of the PAH orbital and the TMDC wave function have to be similar;
        \item the nodal structure of the PAH orbital has to be compatible with the plane-wave part of the TMDC wave function: this corresponds to their \textbf{k}-points matching in the unfolded band structure;
        \item the TMDC wave function must have a significant chalcogen  contribution, and the PAH orbital must have $\pi$ character         with a locally high spectral weight.
    \end{enumerate}
    Point (i) can be rationalized in terms of perturbation theory, which shows that the degree of mixture of states caused by their interaction is inversely proportional to their energy difference. Regarding points (ii) and (iii), we note that chemical interactions occur if there is wave function overlap \cite{szabo1082},
    \begin{align}
        \int\text d^3r\,\psi^*_\text{PAH}(\textbf{r})\psi_\text{TMDC}(\textbf{r}),
    \end{align}
    which, in turn, requires significant probability density overlap,
    \begin{align}\label{eq.probDensOverlap}
        \int\text d^3r\,|\psi_\text{PAH}(\textbf{r})|^2|\psi_\text{TMDC}(\textbf{r})|^2,
    \end{align}
    as well as non-orthogonality between $\psi_\text{PAH}$ and $\psi_\text{TMDC}$. This implies an at least partial match in the phases of these wave functions, thus enabling constructive interference.    
    These two requirements are met when points (iii) and (ii) are fulfilled, respectively.
    
    As a consequence of the proposed conditions, no strong hybridization can be expected to occur at the VBM and at the CBm of the TMDC, as the corresponding bands bear mainly  transition-metal character (see figure~\ref{fig.unitCell}). 
    Yet, the remaining chalcogen fraction might lead to weak mixing when the weight of the molecular orbital is strongly concentrated at K, as is the case for Pe@\ce{WSe2} (see figure~\ref{fig.peryleneUnfolded}d). As far as we can extrapolate from the examined hybrid systems including Py and Pe, most of the interactions between the PAH orbitals and the highest valence bands of the TMDCs occur close to M, where the latter have pronounced chalcogen character and where many 
    orbitals tend to unfold to.
    The chalcogen species in the TMDC controls the level alignment and thus the 
    interactions with the physisorbed molecule in the valence region by setting the ionization potential of the inorganic monolayer; the conduction bands are less affected due to counteracting bandgap and bandwidth reductions. The opposite is true for the transition-metal type, which determines the conduction bandwidth and thus corresponding level alignments, but barely affects the valence bands. Overall, the varying size of SOC-related splits in both valence and conduction regions can in some cases give rise to slightly different interaction patterns in Mo- and W-based systems (e.g. hybridization in the highest valence band of Py@\ce{WS2} compared to Py@\ce{MoS2}, see figure~\ref{fig.pyreneUnfolded}a and c).

\subsection{The role of the Brillouin zone path}\label{sec.paths}
\subsection{The role of the Brillouin zone path}\label{sec.paths}
\begin{figure*}[h!]
    \centering
    \includegraphics[width=\textwidth]{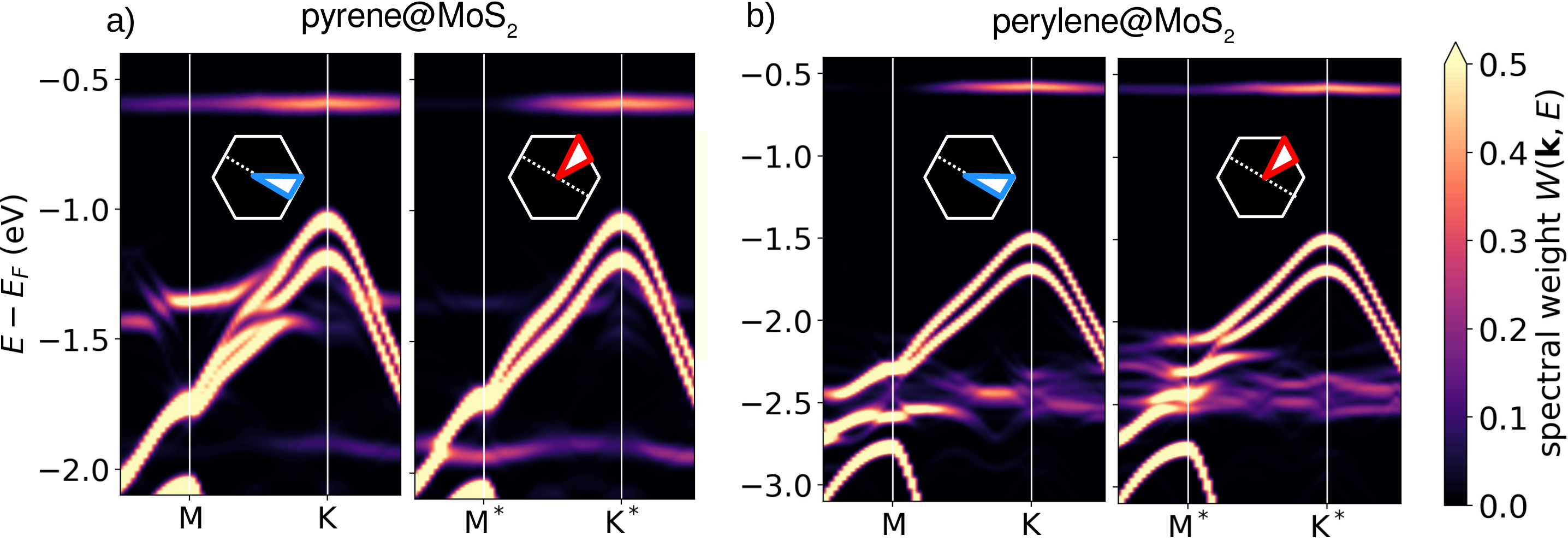}
    \caption{Cuts of the unfolded band structures of a) pyrene@MoS$_2$ and b) perylene@MoS$_2$ along the $\Gamma$-M-K-$\Gamma$ (blue inset) and $\Gamma$-M$^*$-K$^*$-$\Gamma$ (red inset) paths. The dotted lines in the insets indicate the direction of the respective long molecular axis.}
    \label{fig.anisotropy}
\end{figure*}

As mentioned above in Section~\ref{sec.struct}, the considered hybrid systems are characterized by two \textbf{k}-paths with different energy dispersions. This anisotropy strongly depends on the nature of the orbital in question. A large difference can be observed in the case of the HOMO-1 of Py. As extensively discussed in Section~\ref{sec.pyrene}, 
this orbital hybridizes with MoS$_2$ bands between M and K (see left panel figure~\ref{fig.anisotropy}a). However, along the $\Gamma$-M$^*$-K$^*$-$\Gamma$ path, the HOMO-1 is only faintly visible and the MoS$_2$ bands are unaffected (right panel). 
Also  the HOMO and HOMO-2 appear quite different along the two paths. 

For Pe, such anisotropies seem less pronounced, as there is strong mixing along both paths (figure~\ref{fig.anisotropy}b). However, it is clear that there are substantial differences in the details: Along the $\Gamma$-M-K-$\Gamma$ path, a molecular orbital induces a clean split of the highest valence band (left panel), whereas along the $\Gamma$-M$^*$-K$^*$-$\Gamma$ path, the band structure is strongly distorted slightly higher in energy. In any case, the perturbing influence of the orbitals on the TMDC bands is remarkable. This is partly a consequence of the non-negligible chalcogen character of the highest valence band close to M. However, it also comes down to the specific electronic structure of the molecular orbitals at play, as analyzed in Section~\ref{sec.MOs} below.

\subsection{Molecular orbitals}
\label{sec.MOs}

In order to rationalize how the character of the molecular orbitals affects their hybridization with TMDC bands, we map the orbitals of the isolated molecules into the TMDC unit cells. To do so, the molecule is placed in a hexagonal unit cell with a lattice constant that is a multiple of the lattice constant of the primitive cell of a TMDC. In the following, we consider Py and Pe in a 4$\times$4 MoS$_2$ SC. The orbital energies do not change across the BZ as there are negligible interactions between the replicas. Thus, we can consider the state-specific $W_n(\textbf{k})$ instead of the whole $W(\textbf{k},E)$ (compare equations~\ref{eq.W} and \ref{eq.Wn}), with $n$ being the orbital index, and plot $W_n(\textbf{k})$ as a function of $k_x$ and $k_y$ across the whole BZ. 

\begin{figure*}[h]
    \centering
    \includegraphics[width=\textwidth]{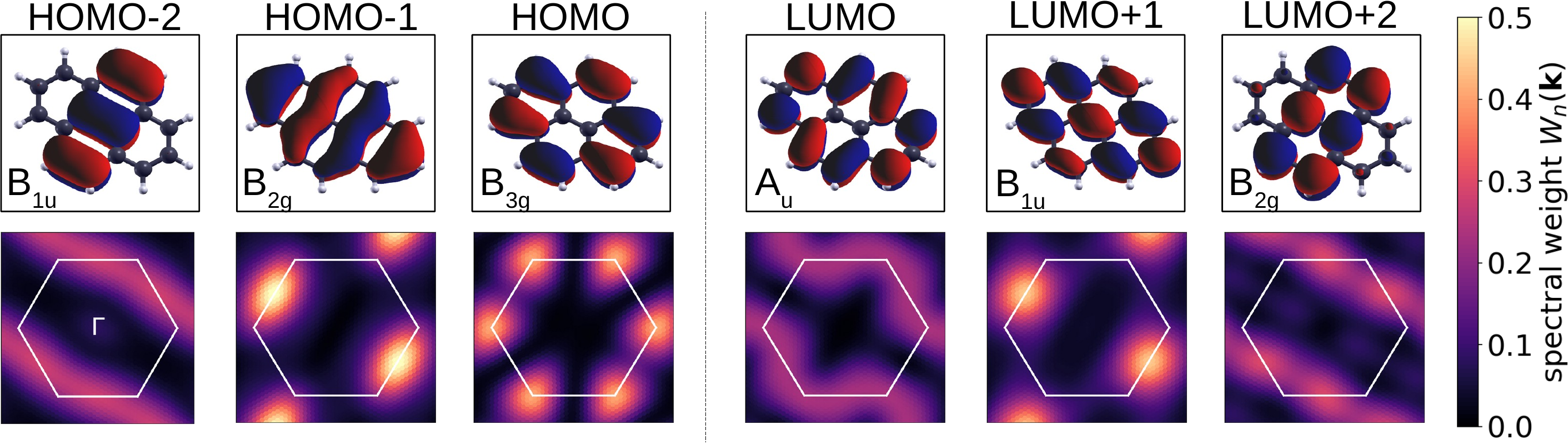}
    \caption{Molecular orbitals of pyrene represented in real space (upper panels), and unfolded to the MoS$_2$ unit cell (lower panels). The spectral weights, indicated according to the color bar on the right, are multiplied by a factor of 2 to make them comparable to the unfolded band structures in Fig.s~\ref{fig.pyreneUnfolded} to \ref{fig.anisotropy}, where the presence of degenerate spin-up and spin-down orbitals doubles the values.}
    \label{fig.orbitalsPyrene}
\end{figure*}

The spectral weight of all considered orbitals for Py, ranging from the HOMO-2 to the LUMO+2 (see figure~\ref{fig.orbitalsPyrene}), is localized at the zone edges. The orbitals can be grouped 
into two main classes. The first one, encompassing the HOMO-2, the LUMO, and the LUMO+2, is characterized by states that are fairly delocalized over the edge of the BZ, such that spectral weights are rather low (maximum $\sim$20\%).
This implies that either the orbitals lack any plane-wave character, or the orientation of the molecule prohibits a stronger \textbf{k}-localization. The latter possibility will be addressed towards the end of this section. The other three orbitals, \textit{i.e.}, the HOMO-1, the HOMO, and the LUMO+1, show instead distinct areas of weight accumulation. Among them, we highlight the HOMO-1 and LUMO+1, whose weights are concentrated at one single point in the BZ. These orbitals are characterized by the absence of nodal planes along the short molecular axis. As such, they are most similar to actual Bloch states. Due to the high local values of $W_n$ ($\sim$50\%), these orbitals are prone to interact with the TMDCs. Indeed, both the HOMO-1 and the LUMO+1 of Py show signs of hybridization with all four considered substrates 
(figure~\ref{fig.pyreneUnfolded}). However, interactions occur only along one direction, explaining the strong path dependence of the unfolded band structure (figure~\ref{fig.anisotropy}a).

\begin{figure*}[h]
    \centering
    \includegraphics[width=\textwidth]{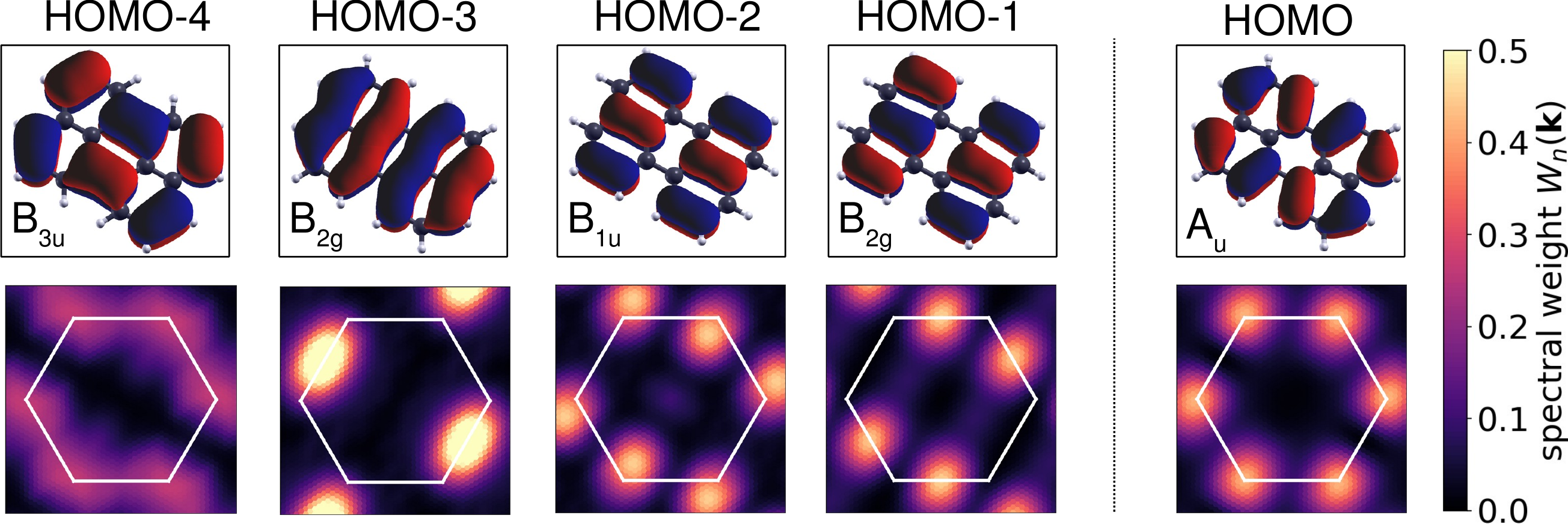}
    \caption{Highest occupied molecular orbitals of perylene: real-space representation (top) and Bloch transformation (bottom), according to the color scale for the spectral weights reported on the right. The line between the HOMO and the HOMO-1 emphasizes the energetic proximity of the four lower levels.}
    \label{fig.orbitalsPerylene}
\end{figure*}

In order to rationalize the coupling signatures seen in figure~\ref{fig.anisotropy}b), we focus on the five highst occupied molecular orbitals of Pe (figure~\ref{fig.orbitalsPerylene}). The HOMO is situated either right below the VBM (for \ce{WSe2}), or within the energy gap (for the other three TMDCs), and has spectral weight at K similar to the one of the HOMO of Py. 
Interactions with the valence bands of the monolayers occur mainly via the next four occupied non-frontier orbitals in the vicinity of the high-symmetry points M and M$^*$.
The energies of these orbitals lie within a range of 200~meV. With the help of the BZ-mapped orbitals, we conclude that the split of the highest occupied TMDC band along the $\Gamma$-M-K-$\Gamma$ path and the distortion of the bands in the hybrid system along $\Gamma$-M$^*$-K$^*$-$\Gamma$ can be traced back to the HOMO-3 and HOMO-1, respectively, as they have strong weight accumulations at M (HOMO-3) and M$^*$ (HOMO-1).

\begin{figure*}[h]
    \centering
    \includegraphics[width=\textwidth]{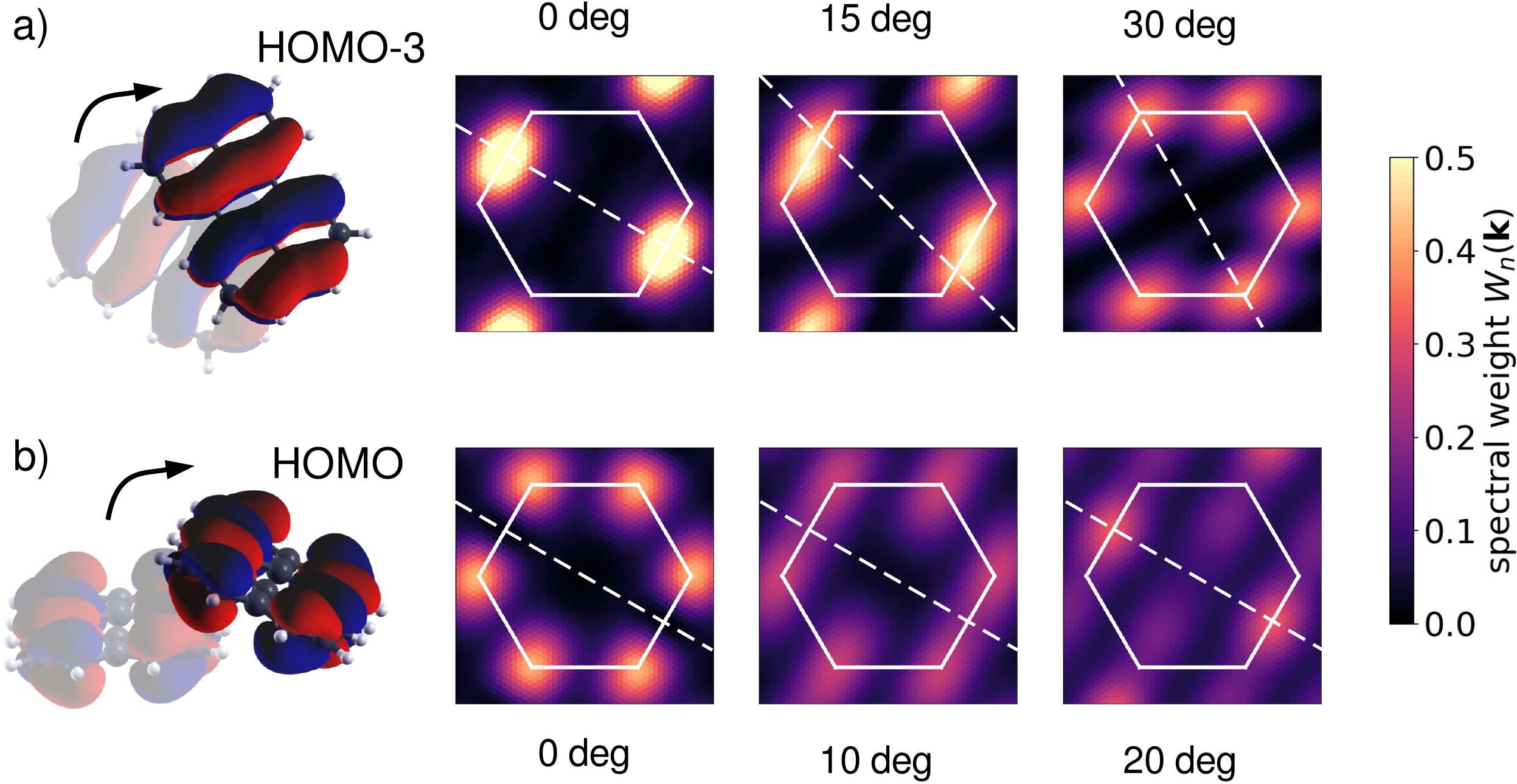}
    \caption{Spectral weight distribution a) of the HOMO-3 of perylene rotating around the out-of-plane axis, and b) of its HOMO upon rotation around the long molecular axis, indicated by the white dashed line.}
    \label{fig.peryRotation}
\end{figure*}

We note that the distribution of spectral weight in the BZ is not an intrinsic property of the orbital, but also depends on the underlying substrate through its lattice constant as well as via the orientation of the molecule with respect to it. Contrary to a normal, continuous Fourier transform, $W_n(\mathbf{k})$ is periodic in reciprocal space; the former yields the momentum representation of the orbital, whereas the latter can be considered a crystal momentum distribution in a scenario in which the environment (the TMDC, in this case) dictates a certain periodicity. Orbital components having the unit-cell periodicity are effectively projected out as large-wave-vector features of the Fourier transform are folded back into the BZ. Since the BZ is not spherical in shape, a rotation of the molecule around the $z$ axis does not entail a corresponding rotation of the spectral weight distribution, which would be the case for a normal Fourier transform, and thus can result in a spectral redistribution. Thus, a change of orientation can transform \textbf{k}-delocalized orbitals into \textbf{k}-localized ones, and \textit{vice versa}. To exemplify, we consider the HOMO-3 of Pe, which was shown to interact strongly with \ce{MoS2} (figure~\ref{fig.anisotropy}b) and to have a concentration of spectral weight at M (figure~\ref{fig.orbitalsPerylene}). Upon rotation around the $z$ axis, the narrow feature first elongates, then splits (figure~\ref{fig.peryRotation}a); the spectral weight is shifted from M to \textit{all} K, thus spreading the spectral weights more thinly and decreasing the orbital's ability to hybridize. Conversely, it seems likely that the presence or absence of these interactions for different angles are a determining factor in the azimuthal orientation of the molecule.

While cofacial arrangements of planar molecules and TMDCs cover the most common and intuitive cases,
some molecules and particularly organic crystalline thin films have been observed to be adsorbed on-edge or in a herringbone manner \cite{breu+16pssrrl,mrky+19apl, padg+19jpcc,kach+21cs,amst+21jpcl}. 
In order to get an idea to which extent the spectral weights are influenced by a variation of the polar angle, we consider the HOMO of perylene. When the molecule is rotated around its long axis, the orbital spectral weight is redistributed from K to M (figure~\ref{fig.peryRotation}b). A qualitatively similar observation has been made for the HOMO of acenes~\cite{puschnig+2009sci, saet+21jpcc}, which have a similar nodal structure, suggesting that this is a general feature of PAHs with $D_{2h}$ symmetry. Although this particular orbital does not interact with the TMDC in either orientation, this weight localization shows that also the polar angle can determine which PAH and TMDC states interact with each other. In turn, this means that the distribution of spectral weights in \textbf{k}-space can be viewed as a determining factor for the polar orientation of the molecule with respect to the substrate.
Other aspects to consider in this context include the interfacial area, which is responsible for the wave function overlap and decreases when departing from the cofacial arrangement, as well as competing intermolecular interactions in the cases of dense coverage or crystalline thin films.

\section{Summary and Conclusions}
In summary, we have investigated the electronic properties of eight hybrid inorganic/organic interfaces formed by two representative PAHs (pyrene and perylene) physisorbed on four TMDC monolayers (\ce{MoS2}, \ce{MoSe2}, \ce{WS2}, and \ce{WSe2}), focusing on the  conditions that induce electronic hybridization between the constituents. 
To this end, we have employed a state-of-the-art first-principles methodology based on hybrid DFT explicitly accounting for spin-orbit coupling. 
The latter effects play a role especially in the presence of W-based TMDCs, where the induced splittings are as large as a few hundreds meV. 
To gain quantitative insight into the computed band structures, we developed and applied a band unfolding technique to map the results obtained for the hybrid interfaces into the unit cell of the TMDC monolayers. 
In this way, we could estimate the level alignment between the organic and inorganic components and identify interaction patterns between the electronic states within the hybrid systems.
A staggered lineup is found in all the considered interfaces including the disulfide monolayers, while \ce{WSe2} forms a straddling heterostructure with both pyrene and perylene; in the borderline case of \ce{MoSe2}, the valence band maximum is found slightly above (below) the HOMO of pyrene (perylene), giving rise to a type-I (type-II) level alignment.
Although the relative energies of the electronic levels of the constituents naturally play a role in the electronic properties of the heterostructure, the hybridization conditions identified in the examined systems are general.
Specifically, we found that hybridization occurs primarily close to M, involving the highest valence band of the former and non-frontier occupied orbitals of the latter. The derived conditions are general to this class of materials: they are based not only on symmetry arguments and on the energetic proximity between the involved states, but also on the spatial distribution of the TMDC bands and of the molecular orbitals in the unit cell of the substrate.
By inspecting the distribution of the molecular orbitals across the Brillouin zone, we could explain the interaction mechanisms with the underlying TMDCs. 
We extended this analysis also to molecules which are rotated with respect to the equilibrium adsorption geometry, finding a sensitive dependence of the weights on both azimuthal and polar rotation angles.

By unraveling the microscopic mechanisms leading to hybridization in inorganic/organic interfaces, our study provides the keys to understand the electronic structure of these materials and to predict their optical response in the linear regime.
In particular, the analysis of the \textbf{k}-projected molecular orbitals represents a computationally efficient and yet reliable tool to predict whether electronic hybridization is likely to occur with the sole knowledge of the molecular orbital wave function and the crystal structure of the substrate, thereby avoiding, at least for a first screening, costly simulations of the whole interface. 
As such, this approach is complementary to a recent development of an effective treatment of substrate-induced electrostatic interactions~\cite{krum+21jcp}.
Possible case-studies for further investigations include functionalized counterparts of the considered PAHs, larger carbon-conjugated molecules absorbing visible light, as well as organic thin films.

\section*{Acknowledgement}
The authors are grateful to M. Sufyan Ramzan for his valuable feedback on the unpublished manuscript. 
This work was funded by the German Research Foundation (DFG), project number 182087777 -- CRC 951. Additional financial support is acknowledged by C.C. to the German Federal Ministry of Education and Research (Professorinnenprogramm III), and by the State of Lower Saxony (Professorinnen für Niedersachsen).
Computational resources were provided by the North-German Supercomputing Alliance (HLRN), project bep00076.

\section*{Data Availability Statement}
The data that support the findings of this study are openly available at the following DOI: 10.5281/zenodo.5362454 (record number 5362454).

\newpage
\bibliographystyle{apsrev4-1}

\begin{thebibliography}{116}%
\makeatletter
\providecommand \@ifxundefined [1]{%
 \@ifx{#1\undefined}
}%
\providecommand \@ifnum [1]{%
 \ifnum #1\expandafter \@firstoftwo
 \else \expandafter \@secondoftwo
 \fi
}%
\providecommand \@ifx [1]{%
 \ifx #1\expandafter \@firstoftwo
 \else \expandafter \@secondoftwo
 \fi
}%
\providecommand \natexlab [1]{#1}%
\providecommand \enquote  [1]{``#1''}%
\providecommand \bibnamefont  [1]{#1}%
\providecommand \bibfnamefont [1]{#1}%
\providecommand \citenamefont [1]{#1}%
\providecommand \href@noop [0]{\@secondoftwo}%
\providecommand \href [0]{\begingroup \@sanitize@url \@href}%
\providecommand \@href[1]{\@@startlink{#1}\@@href}%
\providecommand \@@href[1]{\endgroup#1\@@endlink}%
\providecommand \@sanitize@url [0]{\catcode `\\12\catcode `\$12\catcode
  `\&12\catcode `\#12\catcode `\^12\catcode `\_12\catcode `\%12\relax}%
\providecommand \@@startlink[1]{}%
\providecommand \@@endlink[0]{}%
\providecommand \url  [0]{\begingroup\@sanitize@url \@url }%
\providecommand \@url [1]{\endgroup\@href {#1}{\urlprefix }}%
\providecommand \urlprefix  [0]{URL }%
\providecommand \Eprint [0]{\href }%
\providecommand \doibase [0]{http://dx.doi.org/}%
\providecommand \selectlanguage [0]{\@gobble}%
\providecommand \bibinfo  [0]{\@secondoftwo}%
\providecommand \bibfield  [0]{\@secondoftwo}%
\providecommand \translation [1]{[#1]}%
\providecommand \BibitemOpen [0]{}%
\providecommand \bibitemStop [0]{}%
\providecommand \bibitemNoStop [0]{.\EOS\space}%
\providecommand \EOS [0]{\spacefactor3000\relax}%
\providecommand \BibitemShut  [1]{\csname bibitem#1\endcsname}%
\let\auto@bib@innerbib\@empty
\bibitem [{\citenamefont {Agranovich}\ \emph {et~al.}(2011)\citenamefont
  {Agranovich}, \citenamefont {Gartstein},\ and\ \citenamefont
  {Litinskaya}}]{agra+11cr}%
  \BibitemOpen
  \bibfield  {author} {\bibinfo {author} {\bibfnamefont {V.}~\bibnamefont
  {Agranovich}}, \bibinfo {author} {\bibfnamefont {Y.~N.}\ \bibnamefont
  {Gartstein}}, \ and\ \bibinfo {author} {\bibfnamefont {M.}~\bibnamefont
  {Litinskaya}},\ }\href@noop {} {\bibfield  {journal} {\bibinfo  {journal}
  {Chem.~Rev.~}\ }\textbf {\bibinfo {volume} {111}},\ \bibinfo {pages} {5179}
  (\bibinfo {year} {2011})}\BibitemShut {NoStop}%
\bibitem [{\citenamefont {Wright}\ and\ \citenamefont
  {Uddin}(2012)}]{wrig12sesmc}%
  \BibitemOpen
  \bibfield  {author} {\bibinfo {author} {\bibfnamefont {M.}~\bibnamefont
  {Wright}}\ and\ \bibinfo {author} {\bibfnamefont {A.}~\bibnamefont {Uddin}},\
  }\href@noop {} {\bibfield  {journal} {\bibinfo  {journal}
  {Sol.~Energy~Mater.~Sol.~Cells}\ }\textbf {\bibinfo {volume} {107}},\
  \bibinfo {pages} {87} (\bibinfo {year} {2012})}\BibitemShut {NoStop}%
\bibitem [{\citenamefont {Koch}(2012)}]{koch12pssrrl}%
  \BibitemOpen
  \bibfield  {author} {\bibinfo {author} {\bibfnamefont {N.}~\bibnamefont
  {Koch}},\ }\href@noop {} {\bibfield  {journal} {\bibinfo  {journal}
  {Phys.~Status~Solidi~--RRL}\ }\textbf {\bibinfo {volume} {6}},\ \bibinfo
  {pages} {277} (\bibinfo {year} {2012})}\BibitemShut {NoStop}%
\bibitem [{\citenamefont {Liu}(2014)}]{liu14}%
  \BibitemOpen
  \bibfield  {author} {\bibinfo {author} {\bibfnamefont {R.}~\bibnamefont
  {Liu}},\ }\href@noop {} {\bibfield  {journal} {\bibinfo  {journal}
  {Materials}\ }\textbf {\bibinfo {volume} {7}},\ \bibinfo {pages} {2747}
  (\bibinfo {year} {2014})}\BibitemShut {NoStop}%
\bibitem [{\citenamefont {Hewlett}\ and\ \citenamefont
  {McLachlan}(2016)}]{hewl-mcla16am}%
  \BibitemOpen
  \bibfield  {author} {\bibinfo {author} {\bibfnamefont {R.~M.}\ \bibnamefont
  {Hewlett}}\ and\ \bibinfo {author} {\bibfnamefont {M.~A.}\ \bibnamefont
  {McLachlan}},\ }\href@noop {} {\bibfield  {journal} {\bibinfo  {journal}
  {Adv.~Mater.~}\ }\textbf {\bibinfo {volume} {28}},\ \bibinfo {pages} {3893}
  (\bibinfo {year} {2016})}\BibitemShut {NoStop}%
\bibitem [{\citenamefont {St{\"a}hler}\ and\ \citenamefont
  {Rinke}(2017)}]{stae-rink17cp}%
  \BibitemOpen
  \bibfield  {author} {\bibinfo {author} {\bibfnamefont {J.}~\bibnamefont
  {St{\"a}hler}}\ and\ \bibinfo {author} {\bibfnamefont {P.}~\bibnamefont
  {Rinke}},\ }\href@noop {} {\bibfield  {journal} {\bibinfo  {journal}
  {Chem.~Phys.~}\ }\textbf {\bibinfo {volume} {485}},\ \bibinfo {pages} {149}
  (\bibinfo {year} {2017})}\BibitemShut {NoStop}%
\bibitem [{\citenamefont {Gr{\"a}tzel}(2003)}]{grat03jppc}%
  \BibitemOpen
  \bibfield  {author} {\bibinfo {author} {\bibfnamefont {M.}~\bibnamefont
  {Gr{\"a}tzel}},\ }\href@noop {} {\bibfield  {journal} {\bibinfo  {journal}
  {J.~Photoch.~Photobiol.~C}\ }\textbf {\bibinfo {volume} {4}},\ \bibinfo
  {pages} {145} (\bibinfo {year} {2003})}\BibitemShut {NoStop}%
\bibitem [{\citenamefont {Meng}\ \emph {et~al.}(2003)\citenamefont {Meng},
  \citenamefont {Takahashi}, \citenamefont {Zhang}, \citenamefont {Sutanto},
  \citenamefont {Rao}, \citenamefont {Sato}, \citenamefont {Fujishima},
  \citenamefont {Watanabe}, \citenamefont {Nakamori},\ and\ \citenamefont
  {Uragami}}]{meng+03lang}%
  \BibitemOpen
  \bibfield  {author} {\bibinfo {author} {\bibfnamefont {Q.-B.}\ \bibnamefont
  {Meng}}, \bibinfo {author} {\bibfnamefont {K.}~\bibnamefont {Takahashi}},
  \bibinfo {author} {\bibfnamefont {X.-T.}\ \bibnamefont {Zhang}}, \bibinfo
  {author} {\bibfnamefont {I.}~\bibnamefont {Sutanto}}, \bibinfo {author}
  {\bibfnamefont {T.}~\bibnamefont {Rao}}, \bibinfo {author} {\bibfnamefont
  {O.}~\bibnamefont {Sato}}, \bibinfo {author} {\bibfnamefont {A.}~\bibnamefont
  {Fujishima}}, \bibinfo {author} {\bibfnamefont {H.}~\bibnamefont {Watanabe}},
  \bibinfo {author} {\bibfnamefont {T.}~\bibnamefont {Nakamori}}, \ and\
  \bibinfo {author} {\bibfnamefont {M.}~\bibnamefont {Uragami}},\ }\href@noop
  {} {\bibfield  {journal} {\bibinfo  {journal} {Langmuir}\ }\textbf {\bibinfo
  {volume} {19}},\ \bibinfo {pages} {3572} (\bibinfo {year}
  {2003})}\BibitemShut {NoStop}%
\bibitem [{\citenamefont {Lee}\ \emph {et~al.}(2004)\citenamefont {Lee},
  \citenamefont {Suzuki}, \citenamefont {Imaeda}, \citenamefont {Okada},
  \citenamefont {Wakahara},\ and\ \citenamefont {Yoshida}}]{lee+04jjap}%
  \BibitemOpen
  \bibfield  {author} {\bibinfo {author} {\bibfnamefont {W.~J.}\ \bibnamefont
  {Lee}}, \bibinfo {author} {\bibfnamefont {A.}~\bibnamefont {Suzuki}},
  \bibinfo {author} {\bibfnamefont {K.}~\bibnamefont {Imaeda}}, \bibinfo
  {author} {\bibfnamefont {H.}~\bibnamefont {Okada}}, \bibinfo {author}
  {\bibfnamefont {A.}~\bibnamefont {Wakahara}}, \ and\ \bibinfo {author}
  {\bibfnamefont {A.}~\bibnamefont {Yoshida}},\ }\href@noop {} {\bibfield
  {journal} {\bibinfo  {journal} {Jpn.~J.~Appl.~Phys.~}\ }\textbf {\bibinfo
  {volume} {43}},\ \bibinfo {pages} {152} (\bibinfo {year} {2004})}\BibitemShut
  {NoStop}%
\bibitem [{\citenamefont {Rao}\ \emph {et~al.}(2008)\citenamefont {Rao},
  \citenamefont {Cheetham},\ and\ \citenamefont {Thirumurugan}}]{rao+08jpcm}%
  \BibitemOpen
  \bibfield  {author} {\bibinfo {author} {\bibfnamefont {C.}~\bibnamefont
  {Rao}}, \bibinfo {author} {\bibfnamefont {A.}~\bibnamefont {Cheetham}}, \
  and\ \bibinfo {author} {\bibfnamefont {A.}~\bibnamefont {Thirumurugan}},\
  }\href@noop {} {\bibfield  {journal} {\bibinfo  {journal}
  {J.~Phys.~Condens.~Matter.~}\ }\textbf {\bibinfo {volume} {20}},\ \bibinfo
  {pages} {083202} (\bibinfo {year} {2008})}\BibitemShut {NoStop}%
\bibitem [{\citenamefont {Hsu}\ \emph {et~al.}(2012)\citenamefont {Hsu},
  \citenamefont {Lin}, \citenamefont {Huang}, \citenamefont {Chu},
  \citenamefont {Wei},\ and\ \citenamefont {Li}}]{hsu+12nano}%
  \BibitemOpen
  \bibfield  {author} {\bibinfo {author} {\bibfnamefont {C.-L.}\ \bibnamefont
  {Hsu}}, \bibinfo {author} {\bibfnamefont {C.-T.}\ \bibnamefont {Lin}},
  \bibinfo {author} {\bibfnamefont {J.-H.}\ \bibnamefont {Huang}}, \bibinfo
  {author} {\bibfnamefont {C.-W.}\ \bibnamefont {Chu}}, \bibinfo {author}
  {\bibfnamefont {K.-H.}\ \bibnamefont {Wei}}, \ and\ \bibinfo {author}
  {\bibfnamefont {L.-J.}\ \bibnamefont {Li}},\ }\href@noop {} {\bibfield
  {journal} {\bibinfo  {journal} {ACS~Nano}\ }\textbf {\bibinfo {volume} {6}},\
  \bibinfo {pages} {5031} (\bibinfo {year} {2012})}\BibitemShut {NoStop}%
\bibitem [{\citenamefont {Jnawali}\ \emph {et~al.}(2015)\citenamefont
  {Jnawali}, \citenamefont {Rao}, \citenamefont {Beck}, \citenamefont
  {Petrone}, \citenamefont {Kymissis}, \citenamefont {Hone},\ and\
  \citenamefont {Heinz}}]{jnaw+15nano}%
  \BibitemOpen
  \bibfield  {author} {\bibinfo {author} {\bibfnamefont {G.}~\bibnamefont
  {Jnawali}}, \bibinfo {author} {\bibfnamefont {Y.}~\bibnamefont {Rao}},
  \bibinfo {author} {\bibfnamefont {J.~H.}\ \bibnamefont {Beck}}, \bibinfo
  {author} {\bibfnamefont {N.}~\bibnamefont {Petrone}}, \bibinfo {author}
  {\bibfnamefont {I.}~\bibnamefont {Kymissis}}, \bibinfo {author}
  {\bibfnamefont {J.}~\bibnamefont {Hone}}, \ and\ \bibinfo {author}
  {\bibfnamefont {T.~F.}\ \bibnamefont {Heinz}},\ }\href@noop {} {\bibfield
  {journal} {\bibinfo  {journal} {ACS~Nano}\ }\textbf {\bibinfo {volume} {9}},\
  \bibinfo {pages} {7175} (\bibinfo {year} {2015})}\BibitemShut {NoStop}%
\bibitem [{\citenamefont {Tsai}\ \emph {et~al.}(2015)\citenamefont {Tsai},
  \citenamefont {Omrani}, \citenamefont {Coh}, \citenamefont {Oh},
  \citenamefont {Wickenburg}, \citenamefont {Son}, \citenamefont {Wong},
  \citenamefont {Riss}, \citenamefont {Jung}, \citenamefont {Nguyen},
  \citenamefont {Rodgers}, \citenamefont {Aikawa}, \citenamefont {Taniguchi},
  \citenamefont {Watanabe}, \citenamefont {Zettl}, \citenamefont {Louie},
  \citenamefont {Lu}, \citenamefont {Cohen},\ and\ \citenamefont
  {Crommie}}]{tsai+15nano}%
  \BibitemOpen
  \bibfield  {author} {\bibinfo {author} {\bibfnamefont {H.-Z.}\ \bibnamefont
  {Tsai}}, \bibinfo {author} {\bibfnamefont {A.~A.}\ \bibnamefont {Omrani}},
  \bibinfo {author} {\bibfnamefont {S.}~\bibnamefont {Coh}}, \bibinfo {author}
  {\bibfnamefont {H.}~\bibnamefont {Oh}}, \bibinfo {author} {\bibfnamefont
  {S.}~\bibnamefont {Wickenburg}}, \bibinfo {author} {\bibfnamefont {Y.-W.}\
  \bibnamefont {Son}}, \bibinfo {author} {\bibfnamefont {D.}~\bibnamefont
  {Wong}}, \bibinfo {author} {\bibfnamefont {A.}~\bibnamefont {Riss}}, \bibinfo
  {author} {\bibfnamefont {H.~S.}\ \bibnamefont {Jung}}, \bibinfo {author}
  {\bibfnamefont {G.~D.}\ \bibnamefont {Nguyen}}, \bibinfo {author}
  {\bibfnamefont {G.~F.}\ \bibnamefont {Rodgers}}, \bibinfo {author}
  {\bibfnamefont {A.~S.}\ \bibnamefont {Aikawa}}, \bibinfo {author}
  {\bibfnamefont {T.}~\bibnamefont {Taniguchi}}, \bibinfo {author}
  {\bibfnamefont {K.}~\bibnamefont {Watanabe}}, \bibinfo {author}
  {\bibfnamefont {A.}~\bibnamefont {Zettl}}, \bibinfo {author} {\bibfnamefont
  {S.~G.}\ \bibnamefont {Louie}}, \bibinfo {author} {\bibfnamefont
  {J.}~\bibnamefont {Lu}}, \bibinfo {author} {\bibfnamefont {M.~L.}\
  \bibnamefont {Cohen}}, \ and\ \bibinfo {author} {\bibfnamefont {M.~F.}\
  \bibnamefont {Crommie}},\ }\href {\doibase 10.1021/acsnano.5b05322}
  {\bibfield  {journal} {\bibinfo  {journal} {ACS~Nano}\ }\textbf {\bibinfo
  {volume} {9}},\ \bibinfo {pages} {12168} (\bibinfo {year} {2015})},\ \bibinfo
  {note} {pMID: 26482218},\ \Eprint
  {http://arxiv.org/abs/https://doi.org/10.1021/acsnano.5b05322}
  {https://doi.org/10.1021/acsnano.5b05322} \BibitemShut {NoStop}%
\bibitem [{\citenamefont {Parola}\ \emph {et~al.}(2016)\citenamefont {Parola},
  \citenamefont {Juli{\'a}n-L{\'o}pez}, \citenamefont {Carlos},\ and\
  \citenamefont {Sanchez}}]{paro+16afm}%
  \BibitemOpen
  \bibfield  {author} {\bibinfo {author} {\bibfnamefont {S.}~\bibnamefont
  {Parola}}, \bibinfo {author} {\bibfnamefont {B.}~\bibnamefont
  {Juli{\'a}n-L{\'o}pez}}, \bibinfo {author} {\bibfnamefont {L.~D.}\
  \bibnamefont {Carlos}}, \ and\ \bibinfo {author} {\bibfnamefont
  {C.}~\bibnamefont {Sanchez}},\ }\href@noop {} {\bibfield  {journal} {\bibinfo
   {journal} {Advanced Functional Materials}\ }\textbf {\bibinfo {volume}
  {26}},\ \bibinfo {pages} {6506} (\bibinfo {year} {2016})}\BibitemShut
  {NoStop}%
\bibitem [{\citenamefont {McKenna}\ and\ \citenamefont
  {Evans}(2017)}]{mcke+17am}%
  \BibitemOpen
  \bibfield  {author} {\bibinfo {author} {\bibfnamefont {B.}~\bibnamefont
  {McKenna}}\ and\ \bibinfo {author} {\bibfnamefont {R.~C.}\ \bibnamefont
  {Evans}},\ }\href@noop {} {\bibfield  {journal} {\bibinfo  {journal}
  {Adv.~Mater.~}\ }\textbf {\bibinfo {volume} {29}},\ \bibinfo {pages}
  {1606491} (\bibinfo {year} {2017})}\BibitemShut {NoStop}%
\bibitem [{\citenamefont {Boota}\ \emph {et~al.}(2017)\citenamefont {Boota},
  \citenamefont {Pasini}, \citenamefont {Galeotti}, \citenamefont {Porzio},
  \citenamefont {Zhao}, \citenamefont {Halim},\ and\ \citenamefont
  {Gogotsi}}]{boot+17cm}%
  \BibitemOpen
  \bibfield  {author} {\bibinfo {author} {\bibfnamefont {M.}~\bibnamefont
  {Boota}}, \bibinfo {author} {\bibfnamefont {M.}~\bibnamefont {Pasini}},
  \bibinfo {author} {\bibfnamefont {F.}~\bibnamefont {Galeotti}}, \bibinfo
  {author} {\bibfnamefont {W.}~\bibnamefont {Porzio}}, \bibinfo {author}
  {\bibfnamefont {M.-Q.}\ \bibnamefont {Zhao}}, \bibinfo {author}
  {\bibfnamefont {J.}~\bibnamefont {Halim}}, \ and\ \bibinfo {author}
  {\bibfnamefont {Y.}~\bibnamefont {Gogotsi}},\ }\href@noop {} {\bibfield
  {journal} {\bibinfo  {journal} {Chem.~Mater.~}\ }\textbf {\bibinfo {volume}
  {29}},\ \bibinfo {pages} {2731} (\bibinfo {year} {2017})}\BibitemShut
  {NoStop}%
\bibitem [{\citenamefont {Xu}\ \emph {et~al.}(2018)\citenamefont {Xu},
  \citenamefont {Yin}, \citenamefont {Liu}, \citenamefont {Sheng},\ and\
  \citenamefont {Zhao}}]{xu+18am}%
  \BibitemOpen
  \bibfield  {author} {\bibinfo {author} {\bibfnamefont {H.}~\bibnamefont
  {Xu}}, \bibinfo {author} {\bibfnamefont {L.}~\bibnamefont {Yin}}, \bibinfo
  {author} {\bibfnamefont {C.}~\bibnamefont {Liu}}, \bibinfo {author}
  {\bibfnamefont {X.}~\bibnamefont {Sheng}}, \ and\ \bibinfo {author}
  {\bibfnamefont {N.}~\bibnamefont {Zhao}},\ }\href@noop {} {\bibfield
  {journal} {\bibinfo  {journal} {Adv.~Mater.~}\ }\textbf {\bibinfo {volume}
  {30}},\ \bibinfo {pages} {1800156} (\bibinfo {year} {2018})}\BibitemShut
  {NoStop}%
\bibitem [{\citenamefont {Novoselov}\ \emph {et~al.}(2004)\citenamefont
  {Novoselov}, \citenamefont {Geim}, \citenamefont {Morozov}, \citenamefont
  {Jiang}, \citenamefont {Zhang}, \citenamefont {Dubonos}, \citenamefont
  {Grigorieva},\ and\ \citenamefont {Firsov}}]{novo+04sci}%
  \BibitemOpen
  \bibfield  {author} {\bibinfo {author} {\bibfnamefont {K.~S.}\ \bibnamefont
  {Novoselov}}, \bibinfo {author} {\bibfnamefont {A.~K.}\ \bibnamefont {Geim}},
  \bibinfo {author} {\bibfnamefont {S.~V.}\ \bibnamefont {Morozov}}, \bibinfo
  {author} {\bibfnamefont {D.-e.}\ \bibnamefont {Jiang}}, \bibinfo {author}
  {\bibfnamefont {Y.}~\bibnamefont {Zhang}}, \bibinfo {author} {\bibfnamefont
  {S.~V.}\ \bibnamefont {Dubonos}}, \bibinfo {author} {\bibfnamefont {I.~V.}\
  \bibnamefont {Grigorieva}}, \ and\ \bibinfo {author} {\bibfnamefont {A.~A.}\
  \bibnamefont {Firsov}},\ }\href@noop {} {\bibfield  {journal} {\bibinfo
  {journal} {Science}\ }\textbf {\bibinfo {volume} {306}},\ \bibinfo {pages}
  {666} (\bibinfo {year} {2004})}\BibitemShut {NoStop}%
\bibitem [{\citenamefont {Geim}\ and\ \citenamefont
  {Grigorieva}(2013)}]{geim-grig13nature}%
  \BibitemOpen
  \bibfield  {author} {\bibinfo {author} {\bibfnamefont {A.~K.}\ \bibnamefont
  {Geim}}\ and\ \bibinfo {author} {\bibfnamefont {I.~V.}\ \bibnamefont
  {Grigorieva}},\ }\href@noop {} {\bibfield  {journal} {\bibinfo  {journal}
  {Nature}\ }\textbf {\bibinfo {volume} {499}},\ \bibinfo {pages} {419}
  (\bibinfo {year} {2013})}\BibitemShut {NoStop}%
\bibitem [{\citenamefont {Mak}\ \emph {et~al.}(2014)\citenamefont {Mak},
  \citenamefont {McGill}, \citenamefont {Park},\ and\ \citenamefont
  {McEuen}}]{mak+14sci}%
  \BibitemOpen
  \bibfield  {author} {\bibinfo {author} {\bibfnamefont {K.~F.}\ \bibnamefont
  {Mak}}, \bibinfo {author} {\bibfnamefont {K.~L.}\ \bibnamefont {McGill}},
  \bibinfo {author} {\bibfnamefont {J.}~\bibnamefont {Park}}, \ and\ \bibinfo
  {author} {\bibfnamefont {P.~L.}\ \bibnamefont {McEuen}},\ }\href@noop {}
  {\bibfield  {journal} {\bibinfo  {journal} {Science}\ }\textbf {\bibinfo
  {volume} {344}},\ \bibinfo {pages} {1489} (\bibinfo {year}
  {2014})}\BibitemShut {NoStop}%
\bibitem [{\citenamefont {Mak}\ \emph {et~al.}(2010)\citenamefont {Mak},
  \citenamefont {Lee}, \citenamefont {Hone}, \citenamefont {Shan},\ and\
  \citenamefont {Heinz}}]{mak+10prl}%
  \BibitemOpen
  \bibfield  {author} {\bibinfo {author} {\bibfnamefont {K.~F.}\ \bibnamefont
  {Mak}}, \bibinfo {author} {\bibfnamefont {C.}~\bibnamefont {Lee}}, \bibinfo
  {author} {\bibfnamefont {J.}~\bibnamefont {Hone}}, \bibinfo {author}
  {\bibfnamefont {J.}~\bibnamefont {Shan}}, \ and\ \bibinfo {author}
  {\bibfnamefont {T.~F.}\ \bibnamefont {Heinz}},\ }\href@noop {} {\bibfield
  {journal} {\bibinfo  {journal} {Phys.~Rev.~Lett.~}\ }\textbf {\bibinfo
  {volume} {105}},\ \bibinfo {pages} {136805} (\bibinfo {year}
  {2010})}\BibitemShut {NoStop}%
\bibitem [{\citenamefont {Kozawa}\ \emph {et~al.}(2014)\citenamefont {Kozawa},
  \citenamefont {Kumar}, \citenamefont {Carvalho}, \citenamefont {Amara},
  \citenamefont {Zhao}, \citenamefont {Wang}, \citenamefont {Toh},
  \citenamefont {Ribeiro}, \citenamefont {Neto}, \citenamefont {Matsuda},\ and\
  \citenamefont {Eda}}]{koza+14natcom}%
  \BibitemOpen
  \bibfield  {author} {\bibinfo {author} {\bibfnamefont {D.}~\bibnamefont
  {Kozawa}}, \bibinfo {author} {\bibfnamefont {R.}~\bibnamefont {Kumar}},
  \bibinfo {author} {\bibfnamefont {A.}~\bibnamefont {Carvalho}}, \bibinfo
  {author} {\bibfnamefont {K.~K.}\ \bibnamefont {Amara}}, \bibinfo {author}
  {\bibfnamefont {W.}~\bibnamefont {Zhao}}, \bibinfo {author} {\bibfnamefont
  {S.}~\bibnamefont {Wang}}, \bibinfo {author} {\bibfnamefont {M.}~\bibnamefont
  {Toh}}, \bibinfo {author} {\bibfnamefont {R.~M.}\ \bibnamefont {Ribeiro}},
  \bibinfo {author} {\bibfnamefont {A.~C.}\ \bibnamefont {Neto}}, \bibinfo
  {author} {\bibfnamefont {K.}~\bibnamefont {Matsuda}}, \ and\ \bibinfo
  {author} {\bibfnamefont {G.}~\bibnamefont {Eda}},\ }\href@noop {} {\bibfield
  {journal} {\bibinfo  {journal} {Nature~Comm.}\ }\textbf {\bibinfo {volume}
  {5}},\ \bibinfo {pages} {1} (\bibinfo {year} {2014})}\BibitemShut {NoStop}%
\bibitem [{\citenamefont {Mak}\ \emph {et~al.}(2018)\citenamefont {Mak},
  \citenamefont {Xiao},\ and\ \citenamefont {Shan}}]{mak+18natph}%
  \BibitemOpen
  \bibfield  {author} {\bibinfo {author} {\bibfnamefont {K.~F.}\ \bibnamefont
  {Mak}}, \bibinfo {author} {\bibfnamefont {D.}~\bibnamefont {Xiao}}, \ and\
  \bibinfo {author} {\bibfnamefont {J.}~\bibnamefont {Shan}},\ }\href@noop {}
  {\bibfield  {journal} {\bibinfo  {journal} {Nature~Photon.}\ }\textbf
  {\bibinfo {volume} {12}},\ \bibinfo {pages} {451} (\bibinfo {year}
  {2018})}\BibitemShut {NoStop}%
\bibitem [{\citenamefont {Mouri}\ \emph {et~al.}(2013)\citenamefont {Mouri},
  \citenamefont {Miyauchi},\ and\ \citenamefont {Matsuda}}]{mour+13nl}%
  \BibitemOpen
  \bibfield  {author} {\bibinfo {author} {\bibfnamefont {S.}~\bibnamefont
  {Mouri}}, \bibinfo {author} {\bibfnamefont {Y.}~\bibnamefont {Miyauchi}}, \
  and\ \bibinfo {author} {\bibfnamefont {K.}~\bibnamefont {Matsuda}},\
  }\href@noop {} {\bibfield  {journal} {\bibinfo  {journal} {Nano~Lett.~}\
  }\textbf {\bibinfo {volume} {13}},\ \bibinfo {pages} {5944} (\bibinfo {year}
  {2013})}\BibitemShut {NoStop}%
\bibitem [{\citenamefont {He}\ \emph {et~al.}(2015)\citenamefont {He},
  \citenamefont {Pan}, \citenamefont {Nan}, \citenamefont {Gu}, \citenamefont
  {Yang}, \citenamefont {Wu}, \citenamefont {Luo}, \citenamefont {Xu},
  \citenamefont {Zhang}, \citenamefont {Li}, \citenamefont {Ni}, \citenamefont
  {Wang}, \citenamefont {Zhu}, \citenamefont {Chai}, \citenamefont {Shi},\ and\
  \citenamefont {Wang}}]{he+15apl}%
  \BibitemOpen
  \bibfield  {author} {\bibinfo {author} {\bibfnamefont {D.}~\bibnamefont
  {He}}, \bibinfo {author} {\bibfnamefont {Y.}~\bibnamefont {Pan}}, \bibinfo
  {author} {\bibfnamefont {H.}~\bibnamefont {Nan}}, \bibinfo {author}
  {\bibfnamefont {S.}~\bibnamefont {Gu}}, \bibinfo {author} {\bibfnamefont
  {Z.}~\bibnamefont {Yang}}, \bibinfo {author} {\bibfnamefont {B.}~\bibnamefont
  {Wu}}, \bibinfo {author} {\bibfnamefont {X.}~\bibnamefont {Luo}}, \bibinfo
  {author} {\bibfnamefont {B.}~\bibnamefont {Xu}}, \bibinfo {author}
  {\bibfnamefont {Y.}~\bibnamefont {Zhang}}, \bibinfo {author} {\bibfnamefont
  {Y.}~\bibnamefont {Li}}, \bibinfo {author} {\bibfnamefont {Z.}~\bibnamefont
  {Ni}}, \bibinfo {author} {\bibfnamefont {B.}~\bibnamefont {Wang}}, \bibinfo
  {author} {\bibfnamefont {J.}~\bibnamefont {Zhu}}, \bibinfo {author}
  {\bibfnamefont {Y.}~\bibnamefont {Chai}}, \bibinfo {author} {\bibfnamefont
  {Y.}~\bibnamefont {Shi}}, \ and\ \bibinfo {author} {\bibfnamefont
  {X.}~\bibnamefont {Wang}},\ }\href {\doibase 10.1063/1.4935028} {\bibfield
  {journal} {\bibinfo  {journal} {Applied Physics Letters}\ }\textbf {\bibinfo
  {volume} {107}},\ \bibinfo {pages} {183103} (\bibinfo {year} {2015})},\
  \Eprint {http://arxiv.org/abs/https://doi.org/10.1063/1.4935028}
  {https://doi.org/10.1063/1.4935028} \BibitemShut {NoStop}%
\bibitem [{\citenamefont {Bettis~Homan}\ \emph {et~al.}(2016)\citenamefont
  {Bettis~Homan}, \citenamefont {Sangwan}, \citenamefont {Balla}, \citenamefont
  {Bergeron}, \citenamefont {Weiss},\ and\ \citenamefont {Hersam}}]{bett+16nl}%
  \BibitemOpen
  \bibfield  {author} {\bibinfo {author} {\bibfnamefont {S.}~\bibnamefont
  {Bettis~Homan}}, \bibinfo {author} {\bibfnamefont {V.~K.}\ \bibnamefont
  {Sangwan}}, \bibinfo {author} {\bibfnamefont {I.}~\bibnamefont {Balla}},
  \bibinfo {author} {\bibfnamefont {H.}~\bibnamefont {Bergeron}}, \bibinfo
  {author} {\bibfnamefont {E.~A.}\ \bibnamefont {Weiss}}, \ and\ \bibinfo
  {author} {\bibfnamefont {M.~C.}\ \bibnamefont {Hersam}},\ }\href@noop {}
  {\bibfield  {journal} {\bibinfo  {journal} {Nano~Lett.~}\ }\textbf {\bibinfo
  {volume} {17}},\ \bibinfo {pages} {164} (\bibinfo {year} {2016})}\BibitemShut
  {NoStop}%
\bibitem [{\citenamefont {Cai}\ \emph {et~al.}(2016)\citenamefont {Cai},
  \citenamefont {Zhou}, \citenamefont {Zhang},\ and\ \citenamefont
  {Zhang}}]{cai+16cm}%
  \BibitemOpen
  \bibfield  {author} {\bibinfo {author} {\bibfnamefont {Y.}~\bibnamefont
  {Cai}}, \bibinfo {author} {\bibfnamefont {H.}~\bibnamefont {Zhou}}, \bibinfo
  {author} {\bibfnamefont {G.}~\bibnamefont {Zhang}}, \ and\ \bibinfo {author}
  {\bibfnamefont {Y.-W.}\ \bibnamefont {Zhang}},\ }\href@noop {} {\bibfield
  {journal} {\bibinfo  {journal} {Chem.~Mater.~}\ }\textbf {\bibinfo {volume}
  {28}},\ \bibinfo {pages} {8611} (\bibinfo {year} {2016})}\BibitemShut
  {NoStop}%
\bibitem [{\citenamefont {Choi}\ \emph {et~al.}(2016)\citenamefont {Choi},
  \citenamefont {Zhang},\ and\ \citenamefont {Choi}}]{choi+16nano}%
  \BibitemOpen
  \bibfield  {author} {\bibinfo {author} {\bibfnamefont {J.}~\bibnamefont
  {Choi}}, \bibinfo {author} {\bibfnamefont {H.}~\bibnamefont {Zhang}}, \ and\
  \bibinfo {author} {\bibfnamefont {J.~H.}\ \bibnamefont {Choi}},\ }\href@noop
  {} {\bibfield  {journal} {\bibinfo  {journal} {ACS~Nano}\ }\textbf {\bibinfo
  {volume} {10}},\ \bibinfo {pages} {1671} (\bibinfo {year}
  {2016})}\BibitemShut {NoStop}%
\bibitem [{\citenamefont {Jariwala}\ \emph {et~al.}(2016)\citenamefont
  {Jariwala}, \citenamefont {Howell}, \citenamefont {Chen}, \citenamefont
  {Kang}, \citenamefont {Sangwan}, \citenamefont {Filippone}, \citenamefont
  {Turrisi}, \citenamefont {Marks}, \citenamefont {Lauhon},\ and\ \citenamefont
  {Hersam}}]{jari+16nl}%
  \BibitemOpen
  \bibfield  {author} {\bibinfo {author} {\bibfnamefont {D.}~\bibnamefont
  {Jariwala}}, \bibinfo {author} {\bibfnamefont {S.~L.}\ \bibnamefont
  {Howell}}, \bibinfo {author} {\bibfnamefont {K.-S.}\ \bibnamefont {Chen}},
  \bibinfo {author} {\bibfnamefont {J.}~\bibnamefont {Kang}}, \bibinfo {author}
  {\bibfnamefont {V.~K.}\ \bibnamefont {Sangwan}}, \bibinfo {author}
  {\bibfnamefont {S.~A.}\ \bibnamefont {Filippone}}, \bibinfo {author}
  {\bibfnamefont {R.}~\bibnamefont {Turrisi}}, \bibinfo {author} {\bibfnamefont
  {T.~J.}\ \bibnamefont {Marks}}, \bibinfo {author} {\bibfnamefont {L.~J.}\
  \bibnamefont {Lauhon}}, \ and\ \bibinfo {author} {\bibfnamefont {M.~C.}\
  \bibnamefont {Hersam}},\ }\href@noop {} {\bibfield  {journal} {\bibinfo
  {journal} {Nano~Lett.~}\ }\textbf {\bibinfo {volume} {16}},\ \bibinfo {pages}
  {497} (\bibinfo {year} {2016})}\BibitemShut {NoStop}%
\bibitem [{\citenamefont {Petoukhoff}\ \emph {et~al.}(2016)\citenamefont
  {Petoukhoff}, \citenamefont {Krishna}, \citenamefont {Voiry}, \citenamefont
  {Bozkurt}, \citenamefont {Deckoff-Jones}, \citenamefont {Chhowalla},
  \citenamefont {O’Carroll},\ and\ \citenamefont {Dani}}]{peto+16nano}%
  \BibitemOpen
  \bibfield  {author} {\bibinfo {author} {\bibfnamefont {C.~E.}\ \bibnamefont
  {Petoukhoff}}, \bibinfo {author} {\bibfnamefont {M.~B.~M.}\ \bibnamefont
  {Krishna}}, \bibinfo {author} {\bibfnamefont {D.}~\bibnamefont {Voiry}},
  \bibinfo {author} {\bibfnamefont {I.}~\bibnamefont {Bozkurt}}, \bibinfo
  {author} {\bibfnamefont {S.}~\bibnamefont {Deckoff-Jones}}, \bibinfo {author}
  {\bibfnamefont {M.}~\bibnamefont {Chhowalla}}, \bibinfo {author}
  {\bibfnamefont {D.~M.}\ \bibnamefont {O’Carroll}}, \ and\ \bibinfo {author}
  {\bibfnamefont {K.~M.}\ \bibnamefont {Dani}},\ }\href@noop {} {\bibfield
  {journal} {\bibinfo  {journal} {ACS~Nano}\ }\textbf {\bibinfo {volume}
  {10}},\ \bibinfo {pages} {9899} (\bibinfo {year} {2016})}\BibitemShut
  {NoStop}%
\bibitem [{\citenamefont {Zheng}\ \emph {et~al.}(2016)\citenamefont {Zheng},
  \citenamefont {Huang}, \citenamefont {Chen}, \citenamefont {Zhao},
  \citenamefont {Eda}, \citenamefont {Spataru}, \citenamefont {Zhang},
  \citenamefont {Chang}, \citenamefont {Li}, \citenamefont {Chi}, \citenamefont
  {Quek},\ and\ \citenamefont {Wee}}]{zhen+16nano}%
  \BibitemOpen
  \bibfield  {author} {\bibinfo {author} {\bibfnamefont {Y.~J.}\ \bibnamefont
  {Zheng}}, \bibinfo {author} {\bibfnamefont {Y.~L.}\ \bibnamefont {Huang}},
  \bibinfo {author} {\bibfnamefont {Y.}~\bibnamefont {Chen}}, \bibinfo {author}
  {\bibfnamefont {W.}~\bibnamefont {Zhao}}, \bibinfo {author} {\bibfnamefont
  {G.}~\bibnamefont {Eda}}, \bibinfo {author} {\bibfnamefont {C.~D.}\
  \bibnamefont {Spataru}}, \bibinfo {author} {\bibfnamefont {W.}~\bibnamefont
  {Zhang}}, \bibinfo {author} {\bibfnamefont {Y.-H.}\ \bibnamefont {Chang}},
  \bibinfo {author} {\bibfnamefont {L.-J.}\ \bibnamefont {Li}}, \bibinfo
  {author} {\bibfnamefont {D.}~\bibnamefont {Chi}}, \bibinfo {author}
  {\bibfnamefont {S.~Y.}\ \bibnamefont {Quek}}, \ and\ \bibinfo {author}
  {\bibfnamefont {A.~T.~S.}\ \bibnamefont {Wee}},\ }\href@noop {} {\bibfield
  {journal} {\bibinfo  {journal} {ACS~Nano}\ }\textbf {\bibinfo {volume}
  {10}},\ \bibinfo {pages} {2476} (\bibinfo {year} {2016})}\BibitemShut
  {NoStop}%
\bibitem [{\citenamefont {Kafle}\ \emph {et~al.}(2017)\citenamefont {Kafle},
  \citenamefont {Kattel}, \citenamefont {Lane}, \citenamefont {Wang},
  \citenamefont {Zhao},\ and\ \citenamefont {Chan}}]{kafl+17nano}%
  \BibitemOpen
  \bibfield  {author} {\bibinfo {author} {\bibfnamefont {T.~R.}\ \bibnamefont
  {Kafle}}, \bibinfo {author} {\bibfnamefont {B.}~\bibnamefont {Kattel}},
  \bibinfo {author} {\bibfnamefont {S.~D.}\ \bibnamefont {Lane}}, \bibinfo
  {author} {\bibfnamefont {T.}~\bibnamefont {Wang}}, \bibinfo {author}
  {\bibfnamefont {H.}~\bibnamefont {Zhao}}, \ and\ \bibinfo {author}
  {\bibfnamefont {W.-L.}\ \bibnamefont {Chan}},\ }\href@noop {} {\bibfield
  {journal} {\bibinfo  {journal} {ACS~Nano}\ }\textbf {\bibinfo {volume}
  {11}},\ \bibinfo {pages} {10184} (\bibinfo {year} {2017})}\BibitemShut
  {NoStop}%
\bibitem [{\citenamefont {Liu}\ \emph {et~al.}(2017)\citenamefont {Liu},
  \citenamefont {Gu}, \citenamefont {Ding}, \citenamefont {Fan}, \citenamefont
  {Hu}, \citenamefont {Tseng}, \citenamefont {Lee}, \citenamefont {Menon},\
  and\ \citenamefont {Forrest}}]{liu+17nl}%
  \BibitemOpen
  \bibfield  {author} {\bibinfo {author} {\bibfnamefont {X.}~\bibnamefont
  {Liu}}, \bibinfo {author} {\bibfnamefont {J.}~\bibnamefont {Gu}}, \bibinfo
  {author} {\bibfnamefont {K.}~\bibnamefont {Ding}}, \bibinfo {author}
  {\bibfnamefont {D.}~\bibnamefont {Fan}}, \bibinfo {author} {\bibfnamefont
  {X.}~\bibnamefont {Hu}}, \bibinfo {author} {\bibfnamefont {Y.-W.}\
  \bibnamefont {Tseng}}, \bibinfo {author} {\bibfnamefont {Y.-H.}\ \bibnamefont
  {Lee}}, \bibinfo {author} {\bibfnamefont {V.}~\bibnamefont {Menon}}, \ and\
  \bibinfo {author} {\bibfnamefont {S.~R.}\ \bibnamefont {Forrest}},\
  }\href@noop {} {\bibfield  {journal} {\bibinfo  {journal} {Nano~Lett.~}\
  }\textbf {\bibinfo {volume} {17}},\ \bibinfo {pages} {3176} (\bibinfo {year}
  {2017})}\BibitemShut {NoStop}%
\bibitem [{\citenamefont {Zhong}\ \emph {et~al.}(2018)\citenamefont {Zhong},
  \citenamefont {Sangwan}, \citenamefont {Wang}, \citenamefont {Bergeron},
  \citenamefont {Hersam},\ and\ \citenamefont {Weiss}}]{zhon+18jpcl}%
  \BibitemOpen
  \bibfield  {author} {\bibinfo {author} {\bibfnamefont {C.}~\bibnamefont
  {Zhong}}, \bibinfo {author} {\bibfnamefont {V.~K.}\ \bibnamefont {Sangwan}},
  \bibinfo {author} {\bibfnamefont {C.}~\bibnamefont {Wang}}, \bibinfo {author}
  {\bibfnamefont {H.}~\bibnamefont {Bergeron}}, \bibinfo {author}
  {\bibfnamefont {M.~C.}\ \bibnamefont {Hersam}}, \ and\ \bibinfo {author}
  {\bibfnamefont {E.~A.}\ \bibnamefont {Weiss}},\ }\href@noop {} {\bibfield
  {journal} {\bibinfo  {journal} {J.~Phys.~Chem.~Lett.}\ }\textbf {\bibinfo
  {volume} {9}},\ \bibinfo {pages} {2484} (\bibinfo {year} {2018})}\BibitemShut
  {NoStop}%
\bibitem [{\citenamefont {Zhu}\ \emph {et~al.}(2018)\citenamefont {Zhu},
  \citenamefont {Yuan}, \citenamefont {Zhao}, \citenamefont {Zhou},
  \citenamefont {Wan}, \citenamefont {Mei},\ and\ \citenamefont
  {Huang}}]{zhu+18sa}%
  \BibitemOpen
  \bibfield  {author} {\bibinfo {author} {\bibfnamefont {T.}~\bibnamefont
  {Zhu}}, \bibinfo {author} {\bibfnamefont {L.}~\bibnamefont {Yuan}}, \bibinfo
  {author} {\bibfnamefont {Y.}~\bibnamefont {Zhao}}, \bibinfo {author}
  {\bibfnamefont {M.}~\bibnamefont {Zhou}}, \bibinfo {author} {\bibfnamefont
  {Y.}~\bibnamefont {Wan}}, \bibinfo {author} {\bibfnamefont {J.}~\bibnamefont
  {Mei}}, \ and\ \bibinfo {author} {\bibfnamefont {L.}~\bibnamefont {Huang}},\
  }\href@noop {} {\bibfield  {journal} {\bibinfo  {journal} {Sci.~Adv.}\
  }\textbf {\bibinfo {volume} {4}},\ \bibinfo {pages} {eaao3104} (\bibinfo
  {year} {2018})}\BibitemShut {NoStop}%
\bibitem [{\citenamefont {Gu}\ \emph {et~al.}(2018)\citenamefont {Gu},
  \citenamefont {Liu}, \citenamefont {Lin}, \citenamefont {Lee}, \citenamefont
  {Forrest},\ and\ \citenamefont {Menon}}]{gu+18acsp}%
  \BibitemOpen
  \bibfield  {author} {\bibinfo {author} {\bibfnamefont {J.}~\bibnamefont
  {Gu}}, \bibinfo {author} {\bibfnamefont {X.}~\bibnamefont {Liu}}, \bibinfo
  {author} {\bibfnamefont {E.-c.}\ \bibnamefont {Lin}}, \bibinfo {author}
  {\bibfnamefont {Y.-H.}\ \bibnamefont {Lee}}, \bibinfo {author} {\bibfnamefont
  {S.~R.}\ \bibnamefont {Forrest}}, \ and\ \bibinfo {author} {\bibfnamefont
  {V.~M.}\ \bibnamefont {Menon}},\ }\href@noop {} {\bibfield  {journal}
  {\bibinfo  {journal} {ACS~Photon.}\ }\textbf {\bibinfo {volume} {5}},\
  \bibinfo {pages} {100} (\bibinfo {year} {2018})}\BibitemShut {NoStop}%
\bibitem [{\citenamefont {Wang}\ \emph {et~al.}(2018)\citenamefont {Wang},
  \citenamefont {Ji}, \citenamefont {Yang}, \citenamefont {Chuai},
  \citenamefont {Liu}, \citenamefont {Zhou}, \citenamefont {Lu}, \citenamefont
  {Wei}, \citenamefont {Shi}, \citenamefont {Niu}, \citenamefont {Wang},
  \citenamefont {Wang}, \citenamefont {Chen}, \citenamefont {Lu}, \citenamefont
  {Jiang}, \citenamefont {Li},\ and\ \citenamefont {Liu}}]{wang+18afm}%
  \BibitemOpen
  \bibfield  {author} {\bibinfo {author} {\bibfnamefont {J.}~\bibnamefont
  {Wang}}, \bibinfo {author} {\bibfnamefont {Z.}~\bibnamefont {Ji}}, \bibinfo
  {author} {\bibfnamefont {G.}~\bibnamefont {Yang}}, \bibinfo {author}
  {\bibfnamefont {X.}~\bibnamefont {Chuai}}, \bibinfo {author} {\bibfnamefont
  {F.}~\bibnamefont {Liu}}, \bibinfo {author} {\bibfnamefont {Z.}~\bibnamefont
  {Zhou}}, \bibinfo {author} {\bibfnamefont {C.}~\bibnamefont {Lu}}, \bibinfo
  {author} {\bibfnamefont {W.}~\bibnamefont {Wei}}, \bibinfo {author}
  {\bibfnamefont {X.}~\bibnamefont {Shi}}, \bibinfo {author} {\bibfnamefont
  {J.}~\bibnamefont {Niu}}, \bibinfo {author} {\bibfnamefont {L.}~\bibnamefont
  {Wang}}, \bibinfo {author} {\bibfnamefont {H.}~\bibnamefont {Wang}}, \bibinfo
  {author} {\bibfnamefont {J.}~\bibnamefont {Chen}}, \bibinfo {author}
  {\bibfnamefont {N.}~\bibnamefont {Lu}}, \bibinfo {author} {\bibfnamefont
  {C.}~\bibnamefont {Jiang}}, \bibinfo {author} {\bibfnamefont
  {L.}~\bibnamefont {Li}}, \ and\ \bibinfo {author} {\bibfnamefont
  {M.}~\bibnamefont {Liu}},\ }\href@noop {} {\bibfield  {journal} {\bibinfo
  {journal} {Adv.~Funct.~Mater.~}\ }\textbf {\bibinfo {volume} {28}},\ \bibinfo
  {pages} {1806244} (\bibinfo {year} {2018})}\BibitemShut {NoStop}%
\bibitem [{\citenamefont {Zhang}\ \emph {et~al.}(2018)\citenamefont {Zhang},
  \citenamefont {Sharma}, \citenamefont {Zhu}, \citenamefont {Zhang},
  \citenamefont {Wang}, \citenamefont {Dong}, \citenamefont {Nguyen},
  \citenamefont {Wang}, \citenamefont {Wen}, \citenamefont {Cao}, \citenamefont
  {Liu}, \citenamefont {Sun}, \citenamefont {Yang}, \citenamefont {Li},
  \citenamefont {Kar}, \citenamefont {Shi}, \citenamefont {Macdonald},
  \citenamefont {Yu}, \citenamefont {Wang},\ and\ \citenamefont
  {Lu}}]{zhan+18am}%
  \BibitemOpen
  \bibfield  {author} {\bibinfo {author} {\bibfnamefont {L.}~\bibnamefont
  {Zhang}}, \bibinfo {author} {\bibfnamefont {A.}~\bibnamefont {Sharma}},
  \bibinfo {author} {\bibfnamefont {Y.}~\bibnamefont {Zhu}}, \bibinfo {author}
  {\bibfnamefont {Y.}~\bibnamefont {Zhang}}, \bibinfo {author} {\bibfnamefont
  {B.}~\bibnamefont {Wang}}, \bibinfo {author} {\bibfnamefont {M.}~\bibnamefont
  {Dong}}, \bibinfo {author} {\bibfnamefont {H.~T.}\ \bibnamefont {Nguyen}},
  \bibinfo {author} {\bibfnamefont {Z.}~\bibnamefont {Wang}}, \bibinfo {author}
  {\bibfnamefont {B.}~\bibnamefont {Wen}}, \bibinfo {author} {\bibfnamefont
  {Y.}~\bibnamefont {Cao}}, \bibinfo {author} {\bibfnamefont {B.}~\bibnamefont
  {Liu}}, \bibinfo {author} {\bibfnamefont {X.}~\bibnamefont {Sun}}, \bibinfo
  {author} {\bibfnamefont {J.}~\bibnamefont {Yang}}, \bibinfo {author}
  {\bibfnamefont {Z.}~\bibnamefont {Li}}, \bibinfo {author} {\bibfnamefont
  {A.}~\bibnamefont {Kar}}, \bibinfo {author} {\bibfnamefont {Y.}~\bibnamefont
  {Shi}}, \bibinfo {author} {\bibfnamefont {D.}~\bibnamefont {Macdonald}},
  \bibinfo {author} {\bibfnamefont {Z.}~\bibnamefont {Yu}}, \bibinfo {author}
  {\bibfnamefont {X.}~\bibnamefont {Wang}}, \ and\ \bibinfo {author}
  {\bibfnamefont {Y.}~\bibnamefont {Lu}},\ }\href {\doibase
  https://doi.org/10.1002/adma.201803986} {\bibfield  {journal} {\bibinfo
  {journal} {Adv.~Mater.~}\ }\textbf {\bibinfo {volume} {30}},\ \bibinfo
  {pages} {1803986} (\bibinfo {year} {2018})},\ \Eprint
  {http://arxiv.org/abs/https://onlinelibrary.wiley.com/doi/pdf/10.1002/adma.201803986}
  {https://onlinelibrary.wiley.com/doi/pdf/10.1002/adma.201803986} \BibitemShut
  {NoStop}%
\bibitem [{\citenamefont {Gobbi}\ \emph {et~al.}(2018)\citenamefont {Gobbi},
  \citenamefont {Orgiu},\ and\ \citenamefont {Samor{\`\i}}}]{gobb+18am}%
  \BibitemOpen
  \bibfield  {author} {\bibinfo {author} {\bibfnamefont {M.}~\bibnamefont
  {Gobbi}}, \bibinfo {author} {\bibfnamefont {E.}~\bibnamefont {Orgiu}}, \ and\
  \bibinfo {author} {\bibfnamefont {P.}~\bibnamefont {Samor{\`\i}}},\
  }\href@noop {} {\bibfield  {journal} {\bibinfo  {journal} {Adv.~Mater.~}\
  }\textbf {\bibinfo {volume} {30}},\ \bibinfo {pages} {1706103} (\bibinfo
  {year} {2018})}\BibitemShut {NoStop}%
\bibitem [{\citenamefont {Amsterdam}\ \emph {et~al.}(2019)\citenamefont
  {Amsterdam}, \citenamefont {Stanev}, \citenamefont {Zhou}, \citenamefont
  {Lou}, \citenamefont {Bergeron}, \citenamefont {Darancet}, \citenamefont
  {Hersam}, \citenamefont {Stern},\ and\ \citenamefont {Marks}}]{amst+19nano}%
  \BibitemOpen
  \bibfield  {author} {\bibinfo {author} {\bibfnamefont {S.~H.}\ \bibnamefont
  {Amsterdam}}, \bibinfo {author} {\bibfnamefont {T.~K.}\ \bibnamefont
  {Stanev}}, \bibinfo {author} {\bibfnamefont {Q.}~\bibnamefont {Zhou}},
  \bibinfo {author} {\bibfnamefont {A.~J.-T.}\ \bibnamefont {Lou}}, \bibinfo
  {author} {\bibfnamefont {H.}~\bibnamefont {Bergeron}}, \bibinfo {author}
  {\bibfnamefont {P.}~\bibnamefont {Darancet}}, \bibinfo {author}
  {\bibfnamefont {M.~C.}\ \bibnamefont {Hersam}}, \bibinfo {author}
  {\bibfnamefont {N.~P.}\ \bibnamefont {Stern}}, \ and\ \bibinfo {author}
  {\bibfnamefont {T.~J.}\ \bibnamefont {Marks}},\ }\href@noop {} {\bibfield
  {journal} {\bibinfo  {journal} {ACS~Nano}\ }\textbf {\bibinfo {volume}
  {13}},\ \bibinfo {pages} {4183} (\bibinfo {year} {2019})}\BibitemShut
  {NoStop}%
\bibitem [{\citenamefont {Mutz}\ \emph {et~al.}(2020)\citenamefont {Mutz},
  \citenamefont {Park}, \citenamefont {Schultz}, \citenamefont {Sadofev},
  \citenamefont {Dalgleish}, \citenamefont {Reissig}, \citenamefont {Koch},
  \citenamefont {List-Kratochvil},\ and\ \citenamefont
  {Blumstengel}}]{mutz+20jpcc}%
  \BibitemOpen
  \bibfield  {author} {\bibinfo {author} {\bibfnamefont {N.}~\bibnamefont
  {Mutz}}, \bibinfo {author} {\bibfnamefont {S.}~\bibnamefont {Park}}, \bibinfo
  {author} {\bibfnamefont {T.}~\bibnamefont {Schultz}}, \bibinfo {author}
  {\bibfnamefont {S.}~\bibnamefont {Sadofev}}, \bibinfo {author} {\bibfnamefont
  {S.}~\bibnamefont {Dalgleish}}, \bibinfo {author} {\bibfnamefont
  {L.}~\bibnamefont {Reissig}}, \bibinfo {author} {\bibfnamefont
  {N.}~\bibnamefont {Koch}}, \bibinfo {author} {\bibfnamefont {E.~J.}\
  \bibnamefont {List-Kratochvil}}, \ and\ \bibinfo {author} {\bibfnamefont
  {S.}~\bibnamefont {Blumstengel}},\ }\href@noop {} {\bibfield  {journal}
  {\bibinfo  {journal} {J.~Phys.~Chem.~C}\ }\textbf {\bibinfo {volume} {124}},\
  \bibinfo {pages} {2837} (\bibinfo {year} {2020})}\BibitemShut {NoStop}%
\bibitem [{\citenamefont {Liao}\ \emph {et~al.}(2020)\citenamefont {Liao},
  \citenamefont {Phan}, \citenamefont {Martinez-Barron},\ and\ \citenamefont
  {Mahmoud}}]{liao+20lang}%
  \BibitemOpen
  \bibfield  {author} {\bibinfo {author} {\bibfnamefont {C.-K.}\ \bibnamefont
  {Liao}}, \bibinfo {author} {\bibfnamefont {J.}~\bibnamefont {Phan}}, \bibinfo
  {author} {\bibfnamefont {H.}~\bibnamefont {Martinez-Barron}}, \ and\ \bibinfo
  {author} {\bibfnamefont {M.~A.}\ \bibnamefont {Mahmoud}},\ }\href@noop {}
  {\bibfield  {journal} {\bibinfo  {journal} {Langmuir}\ }\textbf {\bibinfo
  {volume} {36}},\ \bibinfo {pages} {2574} (\bibinfo {year}
  {2020})}\BibitemShut {NoStop}%
\bibitem [{\citenamefont {Dreher}\ \emph {et~al.}(2020)\citenamefont {Dreher},
  \citenamefont {G\"under}, \citenamefont {Z\"orb},\ and\ \citenamefont
  {Witte}}]{dreh+20cm}%
  \BibitemOpen
  \bibfield  {author} {\bibinfo {author} {\bibfnamefont {M.}~\bibnamefont
  {Dreher}}, \bibinfo {author} {\bibfnamefont {D.}~\bibnamefont {G\"under}},
  \bibinfo {author} {\bibfnamefont {S.}~\bibnamefont {Z\"orb}}, \ and\ \bibinfo
  {author} {\bibfnamefont {G.}~\bibnamefont {Witte}},\ }\href@noop {}
  {\bibfield  {journal} {\bibinfo  {journal} {Chem.~Mater.~}\ }\textbf
  {\bibinfo {volume} {32}},\ \bibinfo {pages} {9034} (\bibinfo {year}
  {2020})}\BibitemShut {NoStop}%
\bibitem [{\citenamefont {Park}\ \emph {et~al.}(2021)\citenamefont {Park},
  \citenamefont {Wang}, \citenamefont {Schultz}, \citenamefont {Shin},
  \citenamefont {Ovsyannikov}, \citenamefont {Zacharias}, \citenamefont
  {Maksimov}, \citenamefont {Meissner}, \citenamefont {Hasegawa}, \citenamefont
  {Yamaguchi}, \citenamefont {Kera}, \citenamefont {Aljarb}, \citenamefont
  {Hakami}, \citenamefont {Li}, \citenamefont {Tung}, \citenamefont {Amsalem},
  \citenamefont {Rossi},\ and\ \citenamefont {Koch}}]{park+21am}%
  \BibitemOpen
  \bibfield  {author} {\bibinfo {author} {\bibfnamefont {S.}~\bibnamefont
  {Park}}, \bibinfo {author} {\bibfnamefont {H.}~\bibnamefont {Wang}}, \bibinfo
  {author} {\bibfnamefont {T.}~\bibnamefont {Schultz}}, \bibinfo {author}
  {\bibfnamefont {D.}~\bibnamefont {Shin}}, \bibinfo {author} {\bibfnamefont
  {R.}~\bibnamefont {Ovsyannikov}}, \bibinfo {author} {\bibfnamefont
  {M.}~\bibnamefont {Zacharias}}, \bibinfo {author} {\bibfnamefont
  {D.}~\bibnamefont {Maksimov}}, \bibinfo {author} {\bibfnamefont
  {M.}~\bibnamefont {Meissner}}, \bibinfo {author} {\bibfnamefont
  {Y.}~\bibnamefont {Hasegawa}}, \bibinfo {author} {\bibfnamefont
  {T.}~\bibnamefont {Yamaguchi}}, \bibinfo {author} {\bibfnamefont
  {S.}~\bibnamefont {Kera}}, \bibinfo {author} {\bibfnamefont {A.}~\bibnamefont
  {Aljarb}}, \bibinfo {author} {\bibfnamefont {M.}~\bibnamefont {Hakami}},
  \bibinfo {author} {\bibfnamefont {L.-J.}\ \bibnamefont {Li}}, \bibinfo
  {author} {\bibfnamefont {V.}~\bibnamefont {Tung}}, \bibinfo {author}
  {\bibfnamefont {P.}~\bibnamefont {Amsalem}}, \bibinfo {author} {\bibfnamefont
  {M.}~\bibnamefont {Rossi}}, \ and\ \bibinfo {author} {\bibfnamefont
  {N.}~\bibnamefont {Koch}},\ }\href {\doibase
  https://doi.org/10.1002/adma.202008677} {\bibfield  {journal} {\bibinfo
  {journal} {Adv.~Mater.~}\ }\textbf {\bibinfo {volume} {33}},\ \bibinfo
  {pages} {2008677} (\bibinfo {year} {2021})},\ \Eprint
  {http://arxiv.org/abs/https://onlinelibrary.wiley.com/doi/pdf/10.1002/adma.202008677}
  {https://onlinelibrary.wiley.com/doi/pdf/10.1002/adma.202008677} \BibitemShut
  {NoStop}%
\bibitem [{\citenamefont {Qiao}\ \emph {et~al.}(2021)\citenamefont {Qiao},
  \citenamefont {Niu}, \citenamefont {Wen}, \citenamefont {Yang}, \citenamefont
  {Chen}, \citenamefont {Wang}, \citenamefont {Feng}, \citenamefont {Qin},\
  and\ \citenamefont {Hao}}]{qiao+212DM}%
  \BibitemOpen
  \bibfield  {author} {\bibinfo {author} {\bibfnamefont {J.-W.}\ \bibnamefont
  {Qiao}}, \bibinfo {author} {\bibfnamefont {M.-S.}\ \bibnamefont {Niu}},
  \bibinfo {author} {\bibfnamefont {Z.-C.}\ \bibnamefont {Wen}}, \bibinfo
  {author} {\bibfnamefont {X.-K.}\ \bibnamefont {Yang}}, \bibinfo {author}
  {\bibfnamefont {Z.-H.}\ \bibnamefont {Chen}}, \bibinfo {author}
  {\bibfnamefont {Y.-X.}\ \bibnamefont {Wang}}, \bibinfo {author}
  {\bibfnamefont {L.}~\bibnamefont {Feng}}, \bibinfo {author} {\bibfnamefont
  {W.}~\bibnamefont {Qin}}, \ and\ \bibinfo {author} {\bibfnamefont {X.-T.}\
  \bibnamefont {Hao}},\ }\href@noop {} {\bibfield  {journal} {\bibinfo
  {journal} {2D Mater.}\ }\textbf {\bibinfo {volume} {8}},\ \bibinfo {pages}
  {025026} (\bibinfo {year} {2021})}\BibitemShut {NoStop}%
\bibitem [{\citenamefont {Wang}\ \emph {et~al.}(2019)\citenamefont {Wang},
  \citenamefont {Levchenko}, \citenamefont {Schultz}, \citenamefont {Koch},
  \citenamefont {Scheffler},\ and\ \citenamefont {Rossi}}]{wang+19aem}%
  \BibitemOpen
  \bibfield  {author} {\bibinfo {author} {\bibfnamefont {H.}~\bibnamefont
  {Wang}}, \bibinfo {author} {\bibfnamefont {S.~V.}\ \bibnamefont {Levchenko}},
  \bibinfo {author} {\bibfnamefont {T.}~\bibnamefont {Schultz}}, \bibinfo
  {author} {\bibfnamefont {N.}~\bibnamefont {Koch}}, \bibinfo {author}
  {\bibfnamefont {M.}~\bibnamefont {Scheffler}}, \ and\ \bibinfo {author}
  {\bibfnamefont {M.}~\bibnamefont {Rossi}},\ }\href {\doibase
  https://doi.org/10.1002/aelm.201800891} {\bibfield  {journal} {\bibinfo
  {journal} {Adv.~Energy~Mater.~}\ }\textbf {\bibinfo {volume} {5}},\ \bibinfo
  {pages} {1800891} (\bibinfo {year} {2019})},\ \Eprint
  {http://arxiv.org/abs/https://onlinelibrary.wiley.com/doi/pdf/10.1002/aelm.201800891}
  {https://onlinelibrary.wiley.com/doi/pdf/10.1002/aelm.201800891} \BibitemShut
  {NoStop}%
\bibitem [{\citenamefont {Jacobs}\ \emph {et~al.}(2020)\citenamefont {Jacobs},
  \citenamefont {Krumland}, \citenamefont {Valencia}, \citenamefont {Wang},
  \citenamefont {Rossi},\ and\ \citenamefont {Cocchi}}]{jaco+20apx}%
  \BibitemOpen
  \bibfield  {author} {\bibinfo {author} {\bibfnamefont {M.}~\bibnamefont
  {Jacobs}}, \bibinfo {author} {\bibfnamefont {J.}~\bibnamefont {Krumland}},
  \bibinfo {author} {\bibfnamefont {A.~M.}\ \bibnamefont {Valencia}}, \bibinfo
  {author} {\bibfnamefont {H.}~\bibnamefont {Wang}}, \bibinfo {author}
  {\bibfnamefont {M.}~\bibnamefont {Rossi}}, \ and\ \bibinfo {author}
  {\bibfnamefont {C.}~\bibnamefont {Cocchi}},\ }\href@noop {} {\bibfield
  {journal} {\bibinfo  {journal} {Adv.~Phys.~X}\ }\textbf {\bibinfo {volume}
  {5}},\ \bibinfo {pages} {1749883} (\bibinfo {year} {2020})}\BibitemShut
  {NoStop}%
\bibitem [{\citenamefont {Fu}\ \emph {et~al.}(2017)\citenamefont {Fu},
  \citenamefont {Cocchi}, \citenamefont {Nabok}, \citenamefont {Gulans},\ and\
  \citenamefont {Draxl}}]{fu+17pccp}%
  \BibitemOpen
  \bibfield  {author} {\bibinfo {author} {\bibfnamefont {Q.}~\bibnamefont
  {Fu}}, \bibinfo {author} {\bibfnamefont {C.}~\bibnamefont {Cocchi}}, \bibinfo
  {author} {\bibfnamefont {D.}~\bibnamefont {Nabok}}, \bibinfo {author}
  {\bibfnamefont {A.}~\bibnamefont {Gulans}}, \ and\ \bibinfo {author}
  {\bibfnamefont {C.}~\bibnamefont {Draxl}},\ }\href@noop {} {\bibfield
  {journal} {\bibinfo  {journal} {Phys.~Chem.~Chem.~Phys.~}\ }\textbf {\bibinfo
  {volume} {19}},\ \bibinfo {pages} {6196} (\bibinfo {year}
  {2017})}\BibitemShut {NoStop}%
\bibitem [{\citenamefont {Shen}\ and\ \citenamefont
  {Tao}(2017)}]{shen-tao17ami}%
  \BibitemOpen
  \bibfield  {author} {\bibinfo {author} {\bibfnamefont {N.}~\bibnamefont
  {Shen}}\ and\ \bibinfo {author} {\bibfnamefont {G.}~\bibnamefont {Tao}},\
  }\href@noop {} {\bibfield  {journal} {\bibinfo  {journal}
  {Adv.~Mater.~Interfaces}\ }\textbf {\bibinfo {volume} {4}},\ \bibinfo {pages}
  {1601083} (\bibinfo {year} {2017})}\BibitemShut {NoStop}%
\bibitem [{\citenamefont {Habib}\ \emph {et~al.}(2020)\citenamefont {Habib},
  \citenamefont {Wang}, \citenamefont {Khan}, \citenamefont {Khan},
  \citenamefont {Obaidulla}, \citenamefont {Pi},\ and\ \citenamefont
  {Xu}}]{habi+20ats}%
  \BibitemOpen
  \bibfield  {author} {\bibinfo {author} {\bibfnamefont {M.~R.}\ \bibnamefont
  {Habib}}, \bibinfo {author} {\bibfnamefont {W.}~\bibnamefont {Wang}},
  \bibinfo {author} {\bibfnamefont {A.}~\bibnamefont {Khan}}, \bibinfo {author}
  {\bibfnamefont {Y.}~\bibnamefont {Khan}}, \bibinfo {author} {\bibfnamefont
  {S.~M.}\ \bibnamefont {Obaidulla}}, \bibinfo {author} {\bibfnamefont
  {X.}~\bibnamefont {Pi}}, \ and\ \bibinfo {author} {\bibfnamefont
  {M.}~\bibnamefont {Xu}},\ }\href@noop {} {\bibfield  {journal} {\bibinfo
  {journal} {Adv.~Theory~Simul.}\ }\textbf {\bibinfo {volume} {3}},\ \bibinfo
  {pages} {2000045} (\bibinfo {year} {2020})}\BibitemShut {NoStop}%
\bibitem [{\citenamefont {Draxl}\ \emph {et~al.}(2014)\citenamefont {Draxl},
  \citenamefont {Nabok},\ and\ \citenamefont {Hannewald}}]{drax+14acr}%
  \BibitemOpen
  \bibfield  {author} {\bibinfo {author} {\bibfnamefont {C.}~\bibnamefont
  {Draxl}}, \bibinfo {author} {\bibfnamefont {D.}~\bibnamefont {Nabok}}, \ and\
  \bibinfo {author} {\bibfnamefont {K.}~\bibnamefont {Hannewald}},\ }\href@noop
  {} {\bibfield  {journal} {\bibinfo  {journal} {Acc.~Chem.~Res.~}\ }\textbf
  {\bibinfo {volume} {47}},\ \bibinfo {pages} {3225} (\bibinfo {year}
  {2014})}\BibitemShut {NoStop}%
\bibitem [{\citenamefont {Schlesinger}\ \emph {et~al.}(2015)\citenamefont
  {Schlesinger}, \citenamefont {Bianchi}, \citenamefont {Blumstengel},
  \citenamefont {Christodoulou}, \citenamefont {Ovsyannikov}, \citenamefont
  {Kobin}, \citenamefont {Moudgil}, \citenamefont {Barlow}, \citenamefont
  {Hecht}, \citenamefont {Marder}, \citenamefont {Henneberger},\ and\
  \citenamefont {Koch}}]{schl+15natcom}%
  \BibitemOpen
  \bibfield  {author} {\bibinfo {author} {\bibfnamefont {R.}~\bibnamefont
  {Schlesinger}}, \bibinfo {author} {\bibfnamefont {F.}~\bibnamefont
  {Bianchi}}, \bibinfo {author} {\bibfnamefont {S.}~\bibnamefont
  {Blumstengel}}, \bibinfo {author} {\bibfnamefont {C.}~\bibnamefont
  {Christodoulou}}, \bibinfo {author} {\bibfnamefont {R.}~\bibnamefont
  {Ovsyannikov}}, \bibinfo {author} {\bibfnamefont {B.}~\bibnamefont {Kobin}},
  \bibinfo {author} {\bibfnamefont {K.}~\bibnamefont {Moudgil}}, \bibinfo
  {author} {\bibfnamefont {S.}~\bibnamefont {Barlow}}, \bibinfo {author}
  {\bibfnamefont {S.}~\bibnamefont {Hecht}}, \bibinfo {author} {\bibfnamefont
  {S.}~\bibnamefont {Marder}}, \bibinfo {author} {\bibfnamefont
  {F.}~\bibnamefont {Henneberger}}, \ and\ \bibinfo {author} {\bibfnamefont
  {N.}~\bibnamefont {Koch}},\ }\href@noop {} {\bibfield  {journal} {\bibinfo
  {journal} {Nature~Comm.}\ }\textbf {\bibinfo {volume} {6}},\ \bibinfo {pages}
  {1} (\bibinfo {year} {2015})}\BibitemShut {NoStop}%
\bibitem [{\citenamefont {Mowbray}\ and\ \citenamefont
  {Migani}(2016)}]{mowb-miga16jctc}%
  \BibitemOpen
  \bibfield  {author} {\bibinfo {author} {\bibfnamefont {D.~J.}\ \bibnamefont
  {Mowbray}}\ and\ \bibinfo {author} {\bibfnamefont {A.}~\bibnamefont
  {Migani}},\ }\href@noop {} {\bibfield  {journal} {\bibinfo  {journal}
  {J.~Chem.~Theory.~Comput.~}\ }\textbf {\bibinfo {volume} {12}},\ \bibinfo
  {pages} {2843} (\bibinfo {year} {2016})}\BibitemShut {NoStop}%
\bibitem [{\citenamefont {Ljungberg}\ \emph {et~al.}(2017)\citenamefont
  {Ljungberg}, \citenamefont {V{\"a}nsk{\"a}}, \citenamefont {Koval},
  \citenamefont {Koch}, \citenamefont {Kira},\ and\ \citenamefont
  {S{\'a}nchez-Portal}}]{ljun+17njp}%
  \BibitemOpen
  \bibfield  {author} {\bibinfo {author} {\bibfnamefont {M.~P.}\ \bibnamefont
  {Ljungberg}}, \bibinfo {author} {\bibfnamefont {O.}~\bibnamefont
  {V{\"a}nsk{\"a}}}, \bibinfo {author} {\bibfnamefont {P.}~\bibnamefont
  {Koval}}, \bibinfo {author} {\bibfnamefont {S.~W.}\ \bibnamefont {Koch}},
  \bibinfo {author} {\bibfnamefont {M.}~\bibnamefont {Kira}}, \ and\ \bibinfo
  {author} {\bibfnamefont {D.}~\bibnamefont {S{\'a}nchez-Portal}},\ }\href@noop
  {} {\bibfield  {journal} {\bibinfo  {journal} {New.~J.~Phys.~}\ }\textbf
  {\bibinfo {volume} {19}},\ \bibinfo {pages} {033019} (\bibinfo {year}
  {2017})}\BibitemShut {NoStop}%
\bibitem [{\citenamefont {Turkina}\ \emph {et~al.}(2019)\citenamefont
  {Turkina}, \citenamefont {Nabok}, \citenamefont {Gulans}, \citenamefont
  {Cocchi},\ and\ \citenamefont {Draxl}}]{turk+19ats}%
  \BibitemOpen
  \bibfield  {author} {\bibinfo {author} {\bibfnamefont {O.}~\bibnamefont
  {Turkina}}, \bibinfo {author} {\bibfnamefont {D.}~\bibnamefont {Nabok}},
  \bibinfo {author} {\bibfnamefont {A.}~\bibnamefont {Gulans}}, \bibinfo
  {author} {\bibfnamefont {C.}~\bibnamefont {Cocchi}}, \ and\ \bibinfo {author}
  {\bibfnamefont {C.}~\bibnamefont {Draxl}},\ }\href@noop {} {\bibfield
  {journal} {\bibinfo  {journal} {Adv.~Theory~Simul.}\ }\textbf {\bibinfo
  {volume} {2}},\ \bibinfo {pages} {1800108} (\bibinfo {year}
  {2019})}\BibitemShut {NoStop}%
\bibitem [{\citenamefont {Sulas-Kern}\ \emph {et~al.}(2020)\citenamefont
  {Sulas-Kern}, \citenamefont {Miller},\ and\ \citenamefont
  {Blackburn}}]{sula+20ees}%
  \BibitemOpen
  \bibfield  {author} {\bibinfo {author} {\bibfnamefont {D.~B.}\ \bibnamefont
  {Sulas-Kern}}, \bibinfo {author} {\bibfnamefont {E.~M.}\ \bibnamefont
  {Miller}}, \ and\ \bibinfo {author} {\bibfnamefont {J.~L.}\ \bibnamefont
  {Blackburn}},\ }\href@noop {} {\bibfield  {journal} {\bibinfo  {journal}
  {Energy~Environ.~Sci.}\ }\textbf {\bibinfo {volume} {13}},\ \bibinfo {pages}
  {2684} (\bibinfo {year} {2020})}\BibitemShut {NoStop}%
\bibitem [{\citenamefont {Della~Sala}\ \emph {et~al.}(2011)\citenamefont
  {Della~Sala}, \citenamefont {Blumstengel},\ and\ \citenamefont
  {Henneberger}}]{dell+11prl}%
  \BibitemOpen
  \bibfield  {author} {\bibinfo {author} {\bibfnamefont {F.}~\bibnamefont
  {Della~Sala}}, \bibinfo {author} {\bibfnamefont {S.}~\bibnamefont
  {Blumstengel}}, \ and\ \bibinfo {author} {\bibfnamefont {F.}~\bibnamefont
  {Henneberger}},\ }\href@noop {} {\bibfield  {journal} {\bibinfo  {journal}
  {Phys.~Rev.~Lett.~}\ }\textbf {\bibinfo {volume} {107}},\ \bibinfo {pages}
  {146401} (\bibinfo {year} {2011})}\BibitemShut {NoStop}%
\bibitem [{\citenamefont {Xu}\ \emph {et~al.}(2013)\citenamefont {Xu},
  \citenamefont {Hofmann}, \citenamefont {Schlesinger}, \citenamefont
  {Winkler}, \citenamefont {Frisch}, \citenamefont {Niederhausen},
  \citenamefont {Vollmer}, \citenamefont {Blumstengel}, \citenamefont
  {Henneberger}, \citenamefont {Koch}, \citenamefont {Rinke},\ and\
  \citenamefont {Scheffler}}]{xu+13prl}%
  \BibitemOpen
  \bibfield  {author} {\bibinfo {author} {\bibfnamefont {Y.}~\bibnamefont
  {Xu}}, \bibinfo {author} {\bibfnamefont {O.~T.}\ \bibnamefont {Hofmann}},
  \bibinfo {author} {\bibfnamefont {R.}~\bibnamefont {Schlesinger}}, \bibinfo
  {author} {\bibfnamefont {S.}~\bibnamefont {Winkler}}, \bibinfo {author}
  {\bibfnamefont {J.}~\bibnamefont {Frisch}}, \bibinfo {author} {\bibfnamefont
  {J.}~\bibnamefont {Niederhausen}}, \bibinfo {author} {\bibfnamefont
  {A.}~\bibnamefont {Vollmer}}, \bibinfo {author} {\bibfnamefont
  {S.}~\bibnamefont {Blumstengel}}, \bibinfo {author} {\bibfnamefont
  {F.}~\bibnamefont {Henneberger}}, \bibinfo {author} {\bibfnamefont
  {N.}~\bibnamefont {Koch}}, \bibinfo {author} {\bibfnamefont {P.}~\bibnamefont
  {Rinke}}, \ and\ \bibinfo {author} {\bibfnamefont {M.}~\bibnamefont
  {Scheffler}},\ }\href@noop {} {\bibfield  {journal} {\bibinfo  {journal}
  {Phys.~Rev.~Lett.~}\ }\textbf {\bibinfo {volume} {111}},\ \bibinfo {pages}
  {226802} (\bibinfo {year} {2013})}\BibitemShut {NoStop}%
\bibitem [{\citenamefont {Schulz}\ \emph {et~al.}(2014)\citenamefont {Schulz},
  \citenamefont {Kelly}, \citenamefont {Winget}, \citenamefont {Li},
  \citenamefont {Kim}, \citenamefont {Ndione}, \citenamefont {Sigdel},
  \citenamefont {Berry}, \citenamefont {Graham}, \citenamefont {Br{\'e}das},
  \citenamefont {Kahn},\ and\ \citenamefont {Monti}}]{schu+14afm}%
  \BibitemOpen
  \bibfield  {author} {\bibinfo {author} {\bibfnamefont {P.}~\bibnamefont
  {Schulz}}, \bibinfo {author} {\bibfnamefont {L.~L.}\ \bibnamefont {Kelly}},
  \bibinfo {author} {\bibfnamefont {P.}~\bibnamefont {Winget}}, \bibinfo
  {author} {\bibfnamefont {H.}~\bibnamefont {Li}}, \bibinfo {author}
  {\bibfnamefont {H.}~\bibnamefont {Kim}}, \bibinfo {author} {\bibfnamefont
  {P.~F.}\ \bibnamefont {Ndione}}, \bibinfo {author} {\bibfnamefont {A.~K.}\
  \bibnamefont {Sigdel}}, \bibinfo {author} {\bibfnamefont {J.~J.}\
  \bibnamefont {Berry}}, \bibinfo {author} {\bibfnamefont {S.}~\bibnamefont
  {Graham}}, \bibinfo {author} {\bibfnamefont {J.-L.}\ \bibnamefont
  {Br{\'e}das}}, \bibinfo {author} {\bibfnamefont {A.}~\bibnamefont {Kahn}}, \
  and\ \bibinfo {author} {\bibfnamefont {O.}~\bibnamefont {Monti}},\
  }\href@noop {} {\bibfield  {journal} {\bibinfo  {journal}
  {Adv.~Funct.~Mater.~}\ }\textbf {\bibinfo {volume} {24}},\ \bibinfo {pages}
  {7381} (\bibinfo {year} {2014})}\BibitemShut {NoStop}%
\bibitem [{\citenamefont {Mattioli}\ \emph {et~al.}(2014)\citenamefont
  {Mattioli}, \citenamefont {Dkhil}, \citenamefont {Saba}, \citenamefont
  {Malloci}, \citenamefont {Melis}, \citenamefont {Alippi}, \citenamefont
  {Filippone}, \citenamefont {Giannozzi}, \citenamefont {Thakur}, \citenamefont
  {Gaceur}, \citenamefont {Margeat}, \citenamefont {Diallo}, \citenamefont
  {Videlot-Ackermann}, \citenamefont {Ackermann}, \citenamefont
  {Amore~Bonapasta},\ and\ \citenamefont {Mattoni}}]{matt+14aem}%
  \BibitemOpen
  \bibfield  {author} {\bibinfo {author} {\bibfnamefont {G.}~\bibnamefont
  {Mattioli}}, \bibinfo {author} {\bibfnamefont {S.~B.}\ \bibnamefont {Dkhil}},
  \bibinfo {author} {\bibfnamefont {M.~I.}\ \bibnamefont {Saba}}, \bibinfo
  {author} {\bibfnamefont {G.}~\bibnamefont {Malloci}}, \bibinfo {author}
  {\bibfnamefont {C.}~\bibnamefont {Melis}}, \bibinfo {author} {\bibfnamefont
  {P.}~\bibnamefont {Alippi}}, \bibinfo {author} {\bibfnamefont
  {F.}~\bibnamefont {Filippone}}, \bibinfo {author} {\bibfnamefont
  {P.}~\bibnamefont {Giannozzi}}, \bibinfo {author} {\bibfnamefont {A.~K.}\
  \bibnamefont {Thakur}}, \bibinfo {author} {\bibfnamefont {M.}~\bibnamefont
  {Gaceur}}, \bibinfo {author} {\bibfnamefont {O.}~\bibnamefont {Margeat}},
  \bibinfo {author} {\bibfnamefont {A.~K.}\ \bibnamefont {Diallo}}, \bibinfo
  {author} {\bibfnamefont {C.}~\bibnamefont {Videlot-Ackermann}}, \bibinfo
  {author} {\bibfnamefont {J.}~\bibnamefont {Ackermann}}, \bibinfo {author}
  {\bibfnamefont {A.}~\bibnamefont {Amore~Bonapasta}}, \ and\ \bibinfo {author}
  {\bibfnamefont {A.}~\bibnamefont {Mattoni}},\ }\href@noop {} {\bibfield
  {journal} {\bibinfo  {journal} {Adv.~Energy~Mater.~}\ }\textbf {\bibinfo
  {volume} {4}},\ \bibinfo {pages} {1301694} (\bibinfo {year}
  {2014})}\BibitemShut {NoStop}%
\bibitem [{\citenamefont {Gruenewald}\ \emph {et~al.}(2015)\citenamefont
  {Gruenewald}, \citenamefont {Schirra}, \citenamefont {Winget}, \citenamefont
  {Kozlik}, \citenamefont {Ndione}, \citenamefont {Sigdel}, \citenamefont
  {Berry}, \citenamefont {Forker}, \citenamefont {Bredas}, \citenamefont
  {Fritz},\ and\ \citenamefont {Monti}}]{grue+15jpcc}%
  \BibitemOpen
  \bibfield  {author} {\bibinfo {author} {\bibfnamefont {M.}~\bibnamefont
  {Gruenewald}}, \bibinfo {author} {\bibfnamefont {L.~K.}\ \bibnamefont
  {Schirra}}, \bibinfo {author} {\bibfnamefont {P.}~\bibnamefont {Winget}},
  \bibinfo {author} {\bibfnamefont {M.}~\bibnamefont {Kozlik}}, \bibinfo
  {author} {\bibfnamefont {P.~F.}\ \bibnamefont {Ndione}}, \bibinfo {author}
  {\bibfnamefont {A.~K.}\ \bibnamefont {Sigdel}}, \bibinfo {author}
  {\bibfnamefont {J.~J.}\ \bibnamefont {Berry}}, \bibinfo {author}
  {\bibfnamefont {R.}~\bibnamefont {Forker}}, \bibinfo {author} {\bibfnamefont
  {J.-L.}\ \bibnamefont {Bredas}}, \bibinfo {author} {\bibfnamefont
  {T.}~\bibnamefont {Fritz}}, \ and\ \bibinfo {author} {\bibfnamefont
  {O.}~\bibnamefont {Monti}},\ }\href@noop {} {\bibfield  {journal} {\bibinfo
  {journal} {J.~Phys.~Chem.~C}\ }\textbf {\bibinfo {volume} {119}},\ \bibinfo
  {pages} {4865} (\bibinfo {year} {2015})}\BibitemShut {NoStop}%
\bibitem [{\citenamefont {Wei}\ \emph {et~al.}(2019)\citenamefont {Wei},
  \citenamefont {Jin}, \citenamefont {Chen}, \citenamefont {Ma}, \citenamefont
  {Liu},\ and\ \citenamefont {Ma}}]{wei+19jpcc}%
  \BibitemOpen
  \bibfield  {author} {\bibinfo {author} {\bibfnamefont {M.}~\bibnamefont
  {Wei}}, \bibinfo {author} {\bibfnamefont {F.}~\bibnamefont {Jin}}, \bibinfo
  {author} {\bibfnamefont {T.}~\bibnamefont {Chen}}, \bibinfo {author}
  {\bibfnamefont {H.}~\bibnamefont {Ma}}, \bibinfo {author} {\bibfnamefont
  {C.}~\bibnamefont {Liu}}, \ and\ \bibinfo {author} {\bibfnamefont
  {Y.}~\bibnamefont {Ma}},\ }\href@noop {} {\bibfield  {journal} {\bibinfo
  {journal} {J.~Phys.~Chem.~C}\ }\textbf {\bibinfo {volume} {123}},\ \bibinfo
  {pages} {3541} (\bibinfo {year} {2019})}\BibitemShut {NoStop}%
\bibitem [{\citenamefont {Jono}\ \emph {et~al.}(2020)\citenamefont {Jono},
  \citenamefont {Awai}, \citenamefont {Kondo}, \citenamefont {Kawaraya},
  \citenamefont {Nakazaki}, \citenamefont {Bessho},\ and\ \citenamefont
  {Segawa}}]{jono+20jpcc}%
  \BibitemOpen
  \bibfield  {author} {\bibinfo {author} {\bibfnamefont {R.}~\bibnamefont
  {Jono}}, \bibinfo {author} {\bibfnamefont {F.}~\bibnamefont {Awai}}, \bibinfo
  {author} {\bibfnamefont {T.}~\bibnamefont {Kondo}}, \bibinfo {author}
  {\bibfnamefont {M.}~\bibnamefont {Kawaraya}}, \bibinfo {author}
  {\bibfnamefont {J.}~\bibnamefont {Nakazaki}}, \bibinfo {author}
  {\bibfnamefont {T.}~\bibnamefont {Bessho}}, \ and\ \bibinfo {author}
  {\bibfnamefont {H.}~\bibnamefont {Segawa}},\ }\href@noop {} {\bibfield
  {journal} {\bibinfo  {journal} {J.~Phys.~Chem.~C}\ }\textbf {\bibinfo
  {volume} {124}},\ \bibinfo {pages} {13535} (\bibinfo {year}
  {2020})}\BibitemShut {NoStop}%
\bibitem [{\citenamefont {Hofmann}\ \emph {et~al.}(2013)\citenamefont
  {Hofmann}, \citenamefont {Deinert}, \citenamefont {Xu}, \citenamefont
  {Rinke}, \citenamefont {St{\"a}hler}, \citenamefont {Wolf},\ and\
  \citenamefont {Scheffler}}]{hofm+13jcp}%
  \BibitemOpen
  \bibfield  {author} {\bibinfo {author} {\bibfnamefont {O.~T.}\ \bibnamefont
  {Hofmann}}, \bibinfo {author} {\bibfnamefont {J.-C.}\ \bibnamefont
  {Deinert}}, \bibinfo {author} {\bibfnamefont {Y.}~\bibnamefont {Xu}},
  \bibinfo {author} {\bibfnamefont {P.}~\bibnamefont {Rinke}}, \bibinfo
  {author} {\bibfnamefont {J.}~\bibnamefont {St{\"a}hler}}, \bibinfo {author}
  {\bibfnamefont {M.}~\bibnamefont {Wolf}}, \ and\ \bibinfo {author}
  {\bibfnamefont {M.}~\bibnamefont {Scheffler}},\ }\href@noop {} {\bibfield
  {journal} {\bibinfo  {journal} {J.~Chem.~Phys.~}\ }\textbf {\bibinfo {volume}
  {139}},\ \bibinfo {pages} {174701} (\bibinfo {year} {2013})}\BibitemShut
  {NoStop}%
\bibitem [{\citenamefont {Tremel}\ and\ \citenamefont
  {Hoffmann}(1987)}]{trem-hoff87jacs}%
  \BibitemOpen
  \bibfield  {author} {\bibinfo {author} {\bibfnamefont {W.}~\bibnamefont
  {Tremel}}\ and\ \bibinfo {author} {\bibfnamefont {R.}~\bibnamefont
  {Hoffmann}},\ }\href {\doibase 10.1021/ja00235a021} {\bibfield  {journal}
  {\bibinfo  {journal} {J.~Am.~Chem.~Soc.~}\ }\textbf {\bibinfo {volume}
  {109}},\ \bibinfo {pages} {124} (\bibinfo {year} {1987})},\ \Eprint
  {http://arxiv.org/abs/https://doi.org/10.1021/ja00235a021}
  {https://doi.org/10.1021/ja00235a021} \BibitemShut {NoStop}%
\bibitem [{\citenamefont {Yang}\ \emph {et~al.}(2018)\citenamefont {Yang},
  \citenamefont {Yang}, \citenamefont {Derunova}, \citenamefont {Parkin},
  \citenamefont {Yan},\ and\ \citenamefont {Ali}}]{yang+18apx}%
  \BibitemOpen
  \bibfield  {author} {\bibinfo {author} {\bibfnamefont {S.-Y.}\ \bibnamefont
  {Yang}}, \bibinfo {author} {\bibfnamefont {H.}~\bibnamefont {Yang}}, \bibinfo
  {author} {\bibfnamefont {E.}~\bibnamefont {Derunova}}, \bibinfo {author}
  {\bibfnamefont {S.~S.~P.}\ \bibnamefont {Parkin}}, \bibinfo {author}
  {\bibfnamefont {B.}~\bibnamefont {Yan}}, \ and\ \bibinfo {author}
  {\bibfnamefont {M.~N.}\ \bibnamefont {Ali}},\ }\href {\doibase
  10.1080/23746149.2017.1414631} {\bibfield  {journal} {\bibinfo  {journal}
  {Adv.~Phys.~X}\ }\textbf {\bibinfo {volume} {3}},\ \bibinfo {pages} {1414631}
  (\bibinfo {year} {2018})},\ \Eprint
  {http://arxiv.org/abs/https://doi.org/10.1080/23746149.2017.1414631}
  {https://doi.org/10.1080/23746149.2017.1414631} \BibitemShut {NoStop}%
\bibitem [{\citenamefont {Dargam}\ \emph {et~al.}(1997)\citenamefont {Dargam},
  \citenamefont {Capaz},\ and\ \citenamefont {Koiller}}]{darg+97prb}%
  \BibitemOpen
  \bibfield  {author} {\bibinfo {author} {\bibfnamefont {T.~G.}\ \bibnamefont
  {Dargam}}, \bibinfo {author} {\bibfnamefont {R.~B.}\ \bibnamefont {Capaz}}, \
  and\ \bibinfo {author} {\bibfnamefont {B.}~\bibnamefont {Koiller}},\ }\href
  {\doibase 10.1103/PhysRevB.56.9625} {\bibfield  {journal} {\bibinfo
  {journal} {Phys.~Rev.~B}\ }\textbf {\bibinfo {volume} {56}},\ \bibinfo
  {pages} {9625} (\bibinfo {year} {1997})}\BibitemShut {NoStop}%
\bibitem [{\citenamefont {Wang}\ \emph {et~al.}(1998)\citenamefont {Wang},
  \citenamefont {Bellaiche}, \citenamefont {Wei},\ and\ \citenamefont
  {Zunger}}]{wang+98prl}%
  \BibitemOpen
  \bibfield  {author} {\bibinfo {author} {\bibfnamefont {L.-W.}\ \bibnamefont
  {Wang}}, \bibinfo {author} {\bibfnamefont {L.}~\bibnamefont {Bellaiche}},
  \bibinfo {author} {\bibfnamefont {S.-H.}\ \bibnamefont {Wei}}, \ and\
  \bibinfo {author} {\bibfnamefont {A.}~\bibnamefont {Zunger}},\ }\href
  {\doibase 10.1103/PhysRevLett.80.4725} {\bibfield  {journal} {\bibinfo
  {journal} {Phys.~Rev.~Lett.~}\ }\textbf {\bibinfo {volume} {80}},\ \bibinfo
  {pages} {4725} (\bibinfo {year} {1998})}\BibitemShut {NoStop}%
\bibitem [{\citenamefont {Boykin}\ and\ \citenamefont
  {Klimeck}(2005)}]{boyk-gerh05prb}%
  \BibitemOpen
  \bibfield  {author} {\bibinfo {author} {\bibfnamefont {T.~B.}\ \bibnamefont
  {Boykin}}\ and\ \bibinfo {author} {\bibfnamefont {G.}~\bibnamefont
  {Klimeck}},\ }\href@noop {} {\bibfield  {journal} {\bibinfo  {journal}
  {Phys.~Rev.~B}\ }\textbf {\bibinfo {volume} {71}},\ \bibinfo {pages} {115215}
  (\bibinfo {year} {2005})}\BibitemShut {NoStop}%
\bibitem [{\citenamefont {Boykin}\ \emph {et~al.}(2007)\citenamefont {Boykin},
  \citenamefont {Kharche}, \citenamefont {Klimeck},\ and\ \citenamefont
  {Korkusinski}}]{boyk+07jpcm}%
  \BibitemOpen
  \bibfield  {author} {\bibinfo {author} {\bibfnamefont {T.~B.}\ \bibnamefont
  {Boykin}}, \bibinfo {author} {\bibfnamefont {N.}~\bibnamefont {Kharche}},
  \bibinfo {author} {\bibfnamefont {G.}~\bibnamefont {Klimeck}}, \ and\
  \bibinfo {author} {\bibfnamefont {M.}~\bibnamefont {Korkusinski}},\
  }\href@noop {} {\bibfield  {journal} {\bibinfo  {journal}
  {J.~Phys.~Condens.~Matter.~}\ }\textbf {\bibinfo {volume} {19}},\ \bibinfo
  {pages} {036203} (\bibinfo {year} {2007})}\BibitemShut {NoStop}%
\bibitem [{\citenamefont {Ku}\ \emph {et~al.}(2010)\citenamefont {Ku},
  \citenamefont {Berlijn},\ and\ \citenamefont {Lee}}]{ku+10prl}%
  \BibitemOpen
  \bibfield  {author} {\bibinfo {author} {\bibfnamefont {W.}~\bibnamefont
  {Ku}}, \bibinfo {author} {\bibfnamefont {T.}~\bibnamefont {Berlijn}}, \ and\
  \bibinfo {author} {\bibfnamefont {C.-C.}\ \bibnamefont {Lee}},\ }\href
  {\doibase 10.1103/PhysRevLett.104.216401} {\bibfield  {journal} {\bibinfo
  {journal} {Phys. Rev. Lett.}\ }\textbf {\bibinfo {volume} {104}},\ \bibinfo
  {pages} {216401} (\bibinfo {year} {2010})}\BibitemShut {NoStop}%
\bibitem [{\citenamefont {Mayo}\ \emph {et~al.}(2020)\citenamefont {Mayo},
  \citenamefont {Yndurain},\ and\ \citenamefont {Soler}}]{mayo+20jpcm}%
  \BibitemOpen
  \bibfield  {author} {\bibinfo {author} {\bibfnamefont {S.~G.}\ \bibnamefont
  {Mayo}}, \bibinfo {author} {\bibfnamefont {F.}~\bibnamefont {Yndurain}}, \
  and\ \bibinfo {author} {\bibfnamefont {J.~M.}\ \bibnamefont {Soler}},\
  }\href@noop {} {\bibfield  {journal} {\bibinfo  {journal}
  {J.~Phys.~Condens.~Matter.~}\ }\textbf {\bibinfo {volume} {32}},\ \bibinfo
  {pages} {205902} (\bibinfo {year} {2020})}\BibitemShut {NoStop}%
\bibitem [{\citenamefont {Popescu}\ and\ \citenamefont
  {Zunger}(2012)}]{popescuZunger2012prb}%
  \BibitemOpen
  \bibfield  {author} {\bibinfo {author} {\bibfnamefont {V.}~\bibnamefont
  {Popescu}}\ and\ \bibinfo {author} {\bibfnamefont {A.}~\bibnamefont
  {Zunger}},\ }\href {\doibase 10.1103/PhysRevB.85.085201} {\bibfield
  {journal} {\bibinfo  {journal} {Phys.~Rev.~B}\ }\textbf {\bibinfo {volume}
  {85}},\ \bibinfo {pages} {085201} (\bibinfo {year} {2012})}\BibitemShut
  {NoStop}%
\bibitem [{\citenamefont {Medeiros}\ \emph {et~al.}(2014)\citenamefont
  {Medeiros}, \citenamefont {Stafstr\"om},\ and\ \citenamefont
  {Bj\"ork}}]{made+14prb}%
  \BibitemOpen
  \bibfield  {author} {\bibinfo {author} {\bibfnamefont {P.~V.~C.}\
  \bibnamefont {Medeiros}}, \bibinfo {author} {\bibfnamefont {S.}~\bibnamefont
  {Stafstr\"om}}, \ and\ \bibinfo {author} {\bibfnamefont {J.}~\bibnamefont
  {Bj\"ork}},\ }\href {\doibase 10.1103/PhysRevB.89.041407} {\bibfield
  {journal} {\bibinfo  {journal} {Phys.~Rev.~B}\ }\textbf {\bibinfo {volume}
  {89}},\ \bibinfo {pages} {041407} (\bibinfo {year} {2014})}\BibitemShut
  {NoStop}%
\bibitem [{\citenamefont {Liu}\ \emph {et~al.}(2016)\citenamefont {Liu},
  \citenamefont {Reticcioli}, \citenamefont {Kim}, \citenamefont {Continenza},
  \citenamefont {Kresse}, \citenamefont {Sarma}, \citenamefont {Chen},\ and\
  \citenamefont {Franchini}}]{liu+16prb}%
  \BibitemOpen
  \bibfield  {author} {\bibinfo {author} {\bibfnamefont {P.}~\bibnamefont
  {Liu}}, \bibinfo {author} {\bibfnamefont {M.}~\bibnamefont {Reticcioli}},
  \bibinfo {author} {\bibfnamefont {B.}~\bibnamefont {Kim}}, \bibinfo {author}
  {\bibfnamefont {A.}~\bibnamefont {Continenza}}, \bibinfo {author}
  {\bibfnamefont {G.}~\bibnamefont {Kresse}}, \bibinfo {author} {\bibfnamefont
  {D.}~\bibnamefont {Sarma}}, \bibinfo {author} {\bibfnamefont {X.-Q.}\
  \bibnamefont {Chen}}, \ and\ \bibinfo {author} {\bibfnamefont
  {C.}~\bibnamefont {Franchini}},\ }\href@noop {} {\bibfield  {journal}
  {\bibinfo  {journal} {Phys.~Rev.~B}\ }\textbf {\bibinfo {volume} {94}},\
  \bibinfo {pages} {195145} (\bibinfo {year} {2016})}\BibitemShut {NoStop}%
\bibitem [{\citenamefont {Tan}\ \emph {et~al.}(2016)\citenamefont {Tan},
  \citenamefont {Chen},\ and\ \citenamefont {Ghosh}}]{tan+16apl}%
  \BibitemOpen
  \bibfield  {author} {\bibinfo {author} {\bibfnamefont {Y.}~\bibnamefont
  {Tan}}, \bibinfo {author} {\bibfnamefont {F.~W.}\ \bibnamefont {Chen}}, \
  and\ \bibinfo {author} {\bibfnamefont {A.~W.}\ \bibnamefont {Ghosh}},\
  }\href@noop {} {\bibfield  {journal} {\bibinfo  {journal}
  {Appl.~Phys.~Lett.~}\ }\textbf {\bibinfo {volume} {109}},\ \bibinfo {pages}
  {101601} (\bibinfo {year} {2016})}\BibitemShut {NoStop}%
\bibitem [{\citenamefont {Chen}\ \emph {et~al.}(2017)\citenamefont {Chen},
  \citenamefont {Chen}, \citenamefont {Zhang},\ and\ \citenamefont
  {Weinert}}]{chen+17prb}%
  \BibitemOpen
  \bibfield  {author} {\bibinfo {author} {\bibfnamefont {M.}~\bibnamefont
  {Chen}}, \bibinfo {author} {\bibfnamefont {W.}~\bibnamefont {Chen}}, \bibinfo
  {author} {\bibfnamefont {Z.}~\bibnamefont {Zhang}}, \ and\ \bibinfo {author}
  {\bibfnamefont {M.}~\bibnamefont {Weinert}},\ }\href@noop {} {\bibfield
  {journal} {\bibinfo  {journal} {Phys.~Rev.~B}\ }\textbf {\bibinfo {volume}
  {96}},\ \bibinfo {pages} {245111} (\bibinfo {year} {2017})}\BibitemShut
  {NoStop}%
\bibitem [{\citenamefont {Iwata}\ \emph {et~al.}(2017)\citenamefont {Iwata},
  \citenamefont {Matsushita}, \citenamefont {Nishi}, \citenamefont {Guo},\ and\
  \citenamefont {Oshiyama}}]{iwat+17prb}%
  \BibitemOpen
  \bibfield  {author} {\bibinfo {author} {\bibfnamefont {J.-I.}\ \bibnamefont
  {Iwata}}, \bibinfo {author} {\bibfnamefont {Y.-i.}\ \bibnamefont
  {Matsushita}}, \bibinfo {author} {\bibfnamefont {H.}~\bibnamefont {Nishi}},
  \bibinfo {author} {\bibfnamefont {Z.-X.}\ \bibnamefont {Guo}}, \ and\
  \bibinfo {author} {\bibfnamefont {A.}~\bibnamefont {Oshiyama}},\ }\href@noop
  {} {\bibfield  {journal} {\bibinfo  {journal} {Phys.~Rev.~B}\ }\textbf
  {\bibinfo {volume} {96}},\ \bibinfo {pages} {235442} (\bibinfo {year}
  {2017})}\BibitemShut {NoStop}%
\bibitem [{\citenamefont {Bhowmick}\ \emph {et~al.}(2016)\citenamefont
  {Bhowmick}, \citenamefont {Stegemann}, \citenamefont {Bartsch}, \citenamefont
  {Strassert},\ and\ \citenamefont {Zacharias}}]{bhow+16jpcc}%
  \BibitemOpen
  \bibfield  {author} {\bibinfo {author} {\bibfnamefont {D.~K.}\ \bibnamefont
  {Bhowmick}}, \bibinfo {author} {\bibfnamefont {L.}~\bibnamefont {Stegemann}},
  \bibinfo {author} {\bibfnamefont {M.}~\bibnamefont {Bartsch}}, \bibinfo
  {author} {\bibfnamefont {C.~A.}\ \bibnamefont {Strassert}}, \ and\ \bibinfo
  {author} {\bibfnamefont {H.}~\bibnamefont {Zacharias}},\ }\href@noop {}
  {\bibfield  {journal} {\bibinfo  {journal} {J.~Phys.~Chem.~C}\ }\textbf
  {\bibinfo {volume} {120}},\ \bibinfo {pages} {3275} (\bibinfo {year}
  {2016})}\BibitemShut {NoStop}%
\bibitem [{\citenamefont {Ritter}\ \emph {et~al.}(2020)\citenamefont {Ritter},
  \citenamefont {Caldas}, \citenamefont {da~Silva}, \citenamefont {Calzolari},\
  and\ \citenamefont {McCluskey}}]{ritt+20acsaem}%
  \BibitemOpen
  \bibfield  {author} {\bibinfo {author} {\bibfnamefont {J.~R.}\ \bibnamefont
  {Ritter}}, \bibinfo {author} {\bibfnamefont {M.~J.}\ \bibnamefont {Caldas}},
  \bibinfo {author} {\bibfnamefont {T.~J.}\ \bibnamefont {da~Silva}}, \bibinfo
  {author} {\bibfnamefont {A.}~\bibnamefont {Calzolari}}, \ and\ \bibinfo
  {author} {\bibfnamefont {M.~D.}\ \bibnamefont {McCluskey}},\ }\href {\doibase
  10.1021/acsaelm.0c00482} {\bibfield  {journal} {\bibinfo  {journal}
  {ACS~Appl.~Energy~Mater.}\ }\textbf {\bibinfo {volume} {2}},\ \bibinfo
  {pages} {2806} (\bibinfo {year} {2020})}\BibitemShut {NoStop}%
\bibitem [{\citenamefont {Azuma}\ \emph {et~al.}(2002)\citenamefont {Azuma},
  \citenamefont {Iwasawa}, \citenamefont {Kurihara}, \citenamefont {Okudaira},
  \citenamefont {Harada},\ and\ \citenamefont {Ueno}}]{azum+02jap}%
  \BibitemOpen
  \bibfield  {author} {\bibinfo {author} {\bibfnamefont {Y.}~\bibnamefont
  {Azuma}}, \bibinfo {author} {\bibfnamefont {K.}~\bibnamefont {Iwasawa}},
  \bibinfo {author} {\bibfnamefont {T.}~\bibnamefont {Kurihara}}, \bibinfo
  {author} {\bibfnamefont {K.~K.}\ \bibnamefont {Okudaira}}, \bibinfo {author}
  {\bibfnamefont {Y.}~\bibnamefont {Harada}}, \ and\ \bibinfo {author}
  {\bibfnamefont {N.}~\bibnamefont {Ueno}},\ }\href@noop {} {\bibfield
  {journal} {\bibinfo  {journal} {J.~Appl.~Phys.~}\ }\textbf {\bibinfo {volume}
  {91}},\ \bibinfo {pages} {5024} (\bibinfo {year} {2002})}\BibitemShut
  {NoStop}%
\bibitem [{\citenamefont {Neubauer}\ \emph {et~al.}(2011)\citenamefont
  {Neubauer}, \citenamefont {Szarko}, \citenamefont {Bartelt}, \citenamefont
  {Eichberger},\ and\ \citenamefont {Hannappel}}]{neub+11jpcc}%
  \BibitemOpen
  \bibfield  {author} {\bibinfo {author} {\bibfnamefont {A.}~\bibnamefont
  {Neubauer}}, \bibinfo {author} {\bibfnamefont {J.~M.}\ \bibnamefont
  {Szarko}}, \bibinfo {author} {\bibfnamefont {A.~F.}\ \bibnamefont {Bartelt}},
  \bibinfo {author} {\bibfnamefont {R.}~\bibnamefont {Eichberger}}, \ and\
  \bibinfo {author} {\bibfnamefont {T.}~\bibnamefont {Hannappel}},\ }\href@noop
  {} {\bibfield  {journal} {\bibinfo  {journal} {J.~Phys.~Chem.~C}\ }\textbf
  {\bibinfo {volume} {115}},\ \bibinfo {pages} {5683} (\bibinfo {year}
  {2011})}\BibitemShut {NoStop}%
\bibitem [{\citenamefont {Abd-Ellah}\ \emph {et~al.}(2019)\citenamefont
  {Abd-Ellah}, \citenamefont {Cann}, \citenamefont {Dayneko}, \citenamefont
  {Laventure}, \citenamefont {Cieplechowicz},\ and\ \citenamefont
  {Welch}}]{abde+19acsaem}%
  \BibitemOpen
  \bibfield  {author} {\bibinfo {author} {\bibfnamefont {M.}~\bibnamefont
  {Abd-Ellah}}, \bibinfo {author} {\bibfnamefont {J.}~\bibnamefont {Cann}},
  \bibinfo {author} {\bibfnamefont {S.~V.}\ \bibnamefont {Dayneko}}, \bibinfo
  {author} {\bibfnamefont {A.}~\bibnamefont {Laventure}}, \bibinfo {author}
  {\bibfnamefont {E.}~\bibnamefont {Cieplechowicz}}, \ and\ \bibinfo {author}
  {\bibfnamefont {G.~C.}\ \bibnamefont {Welch}},\ }\href@noop {} {\bibfield
  {journal} {\bibinfo  {journal} {ACS~Appl.~Energy~Mater.}\ }\textbf {\bibinfo
  {volume} {1}},\ \bibinfo {pages} {1590} (\bibinfo {year} {2019})}\BibitemShut
  {NoStop}%
\bibitem [{\citenamefont {Hohenberg}\ and\ \citenamefont
  {Kohn}(1964)}]{hohenbergKohn1964pr}%
  \BibitemOpen
  \bibfield  {author} {\bibinfo {author} {\bibfnamefont {P.}~\bibnamefont
  {Hohenberg}}\ and\ \bibinfo {author} {\bibfnamefont {W.}~\bibnamefont
  {Kohn}},\ }\href {\doibase 10.1103/PhysRev.136.B864} {\bibfield  {journal}
  {\bibinfo  {journal} {Phys.~Rev.~}\ }\textbf {\bibinfo {volume} {136}},\
  \bibinfo {pages} {B864} (\bibinfo {year} {1964})}\BibitemShut {NoStop}%
\bibitem [{\citenamefont {Kohn}\ and\ \citenamefont
  {Sham}(1965)}]{kohnSham1965pr}%
  \BibitemOpen
  \bibfield  {author} {\bibinfo {author} {\bibfnamefont {W.}~\bibnamefont
  {Kohn}}\ and\ \bibinfo {author} {\bibfnamefont {L.~J.}\ \bibnamefont
  {Sham}},\ }\href {\doibase 10.1103/PhysRev.140.A1133} {\bibfield  {journal}
  {\bibinfo  {journal} {Phys.~Rev.~}\ }\textbf {\bibinfo {volume} {140}},\
  \bibinfo {pages} {A1133} (\bibinfo {year} {1965})}\BibitemShut {NoStop}%
\bibitem [{\citenamefont {Cocchi}\ \emph {et~al.}(2018)\citenamefont {Cocchi},
  \citenamefont {Breuer}, \citenamefont {Witte},\ and\ \citenamefont
  {Draxl}}]{cocc+18pccp}%
  \BibitemOpen
  \bibfield  {author} {\bibinfo {author} {\bibfnamefont {C.}~\bibnamefont
  {Cocchi}}, \bibinfo {author} {\bibfnamefont {T.}~\bibnamefont {Breuer}},
  \bibinfo {author} {\bibfnamefont {G.}~\bibnamefont {Witte}}, \ and\ \bibinfo
  {author} {\bibfnamefont {C.}~\bibnamefont {Draxl}},\ }\href@noop {}
  {\bibfield  {journal} {\bibinfo  {journal} {Physical Chemistry Chemical
  Physics}\ }\textbf {\bibinfo {volume} {20}},\ \bibinfo {pages} {29724}
  (\bibinfo {year} {2018})}\BibitemShut {NoStop}%
\bibitem [{\citenamefont {Giannozzi}\ \emph {et~al.}(2009)\citenamefont
  {Giannozzi}, \citenamefont {Baroni}, \citenamefont {Bonini}, \citenamefont
  {Calandra}, \citenamefont {Car}, \citenamefont {Cavazzoni}, \citenamefont
  {Ceresoli}, \citenamefont {Chiarotti}, \citenamefont {Cococcioni},
  \citenamefont {Dabo}, \citenamefont {Corso}, \citenamefont {de~Gironcoli},
  \citenamefont {Fabris}, \citenamefont {Fratesi}, \citenamefont {Gebauer},
  \citenamefont {Gerstmann}, \citenamefont {Gougoussis}, \citenamefont
  {Kokalj}, \citenamefont {Lazzeri}, \citenamefont {Martin-Samos},
  \citenamefont {Marzari}, \citenamefont {Mauri}, \citenamefont {Mazzarello},
  \citenamefont {Paolini}, \citenamefont {Pasquarello}, \citenamefont
  {Paulatto}, \citenamefont {Sbraccia}, \citenamefont {Scandolo}, \citenamefont
  {Sclauzero}, \citenamefont {Seitsonen}, \citenamefont {Smogunov},
  \citenamefont {Umari},\ and\ \citenamefont {Wentzcovitch}}]{qe2009}%
  \BibitemOpen
  \bibfield  {author} {\bibinfo {author} {\bibfnamefont {P.}~\bibnamefont
  {Giannozzi}}, \bibinfo {author} {\bibfnamefont {S.}~\bibnamefont {Baroni}},
  \bibinfo {author} {\bibfnamefont {N.}~\bibnamefont {Bonini}}, \bibinfo
  {author} {\bibfnamefont {M.}~\bibnamefont {Calandra}}, \bibinfo {author}
  {\bibfnamefont {R.}~\bibnamefont {Car}}, \bibinfo {author} {\bibfnamefont
  {C.}~\bibnamefont {Cavazzoni}}, \bibinfo {author} {\bibfnamefont
  {D.}~\bibnamefont {Ceresoli}}, \bibinfo {author} {\bibfnamefont {G.~L.}\
  \bibnamefont {Chiarotti}}, \bibinfo {author} {\bibfnamefont {M.}~\bibnamefont
  {Cococcioni}}, \bibinfo {author} {\bibfnamefont {I.}~\bibnamefont {Dabo}},
  \bibinfo {author} {\bibfnamefont {A.~D.}\ \bibnamefont {Corso}}, \bibinfo
  {author} {\bibfnamefont {S.}~\bibnamefont {de~Gironcoli}}, \bibinfo {author}
  {\bibfnamefont {S.}~\bibnamefont {Fabris}}, \bibinfo {author} {\bibfnamefont
  {G.}~\bibnamefont {Fratesi}}, \bibinfo {author} {\bibfnamefont
  {R.}~\bibnamefont {Gebauer}}, \bibinfo {author} {\bibfnamefont
  {U.}~\bibnamefont {Gerstmann}}, \bibinfo {author} {\bibfnamefont
  {C.}~\bibnamefont {Gougoussis}}, \bibinfo {author} {\bibfnamefont
  {A.}~\bibnamefont {Kokalj}}, \bibinfo {author} {\bibfnamefont
  {M.}~\bibnamefont {Lazzeri}}, \bibinfo {author} {\bibfnamefont
  {L.}~\bibnamefont {Martin-Samos}}, \bibinfo {author} {\bibfnamefont
  {N.}~\bibnamefont {Marzari}}, \bibinfo {author} {\bibfnamefont
  {F.}~\bibnamefont {Mauri}}, \bibinfo {author} {\bibfnamefont
  {R.}~\bibnamefont {Mazzarello}}, \bibinfo {author} {\bibfnamefont
  {S.}~\bibnamefont {Paolini}}, \bibinfo {author} {\bibfnamefont
  {A.}~\bibnamefont {Pasquarello}}, \bibinfo {author} {\bibfnamefont
  {L.}~\bibnamefont {Paulatto}}, \bibinfo {author} {\bibfnamefont
  {C.}~\bibnamefont {Sbraccia}}, \bibinfo {author} {\bibfnamefont
  {S.}~\bibnamefont {Scandolo}}, \bibinfo {author} {\bibfnamefont
  {G.}~\bibnamefont {Sclauzero}}, \bibinfo {author} {\bibfnamefont {A.~P.}\
  \bibnamefont {Seitsonen}}, \bibinfo {author} {\bibfnamefont {A.}~\bibnamefont
  {Smogunov}}, \bibinfo {author} {\bibfnamefont {P.}~\bibnamefont {Umari}}, \
  and\ \bibinfo {author} {\bibfnamefont {R.~M.}\ \bibnamefont {Wentzcovitch}},\
  }\href {\doibase 10.1088/0953-8984/21/39/395502} {\bibfield  {journal}
  {\bibinfo  {journal} {J.~Phys.~Condens.~Matter.~}\ }\textbf {\bibinfo
  {volume} {21}},\ \bibinfo {pages} {395502} (\bibinfo {year}
  {2009})}\BibitemShut {NoStop}%
\bibitem [{\citenamefont {Giannozzi}\ \emph {et~al.}(2020)\citenamefont
  {Giannozzi}, \citenamefont {Baseggio}, \citenamefont {Bonfà}, \citenamefont
  {Brunato}, \citenamefont {Car}, \citenamefont {Carnimeo}, \citenamefont
  {Cavazzoni}, \citenamefont {de~Gironcoli}, \citenamefont {Delugas},
  \citenamefont {Ferrari~Ruffino}, \citenamefont {Ferretti}, \citenamefont
  {Marzari}, \citenamefont {Timrov}, \citenamefont {Urru},\ and\ \citenamefont
  {Baroni}}]{qe2020}%
  \BibitemOpen
  \bibfield  {author} {\bibinfo {author} {\bibfnamefont {P.}~\bibnamefont
  {Giannozzi}}, \bibinfo {author} {\bibfnamefont {O.}~\bibnamefont {Baseggio}},
  \bibinfo {author} {\bibfnamefont {P.}~\bibnamefont {Bonfà}}, \bibinfo
  {author} {\bibfnamefont {D.}~\bibnamefont {Brunato}}, \bibinfo {author}
  {\bibfnamefont {R.}~\bibnamefont {Car}}, \bibinfo {author} {\bibfnamefont
  {I.}~\bibnamefont {Carnimeo}}, \bibinfo {author} {\bibfnamefont
  {C.}~\bibnamefont {Cavazzoni}}, \bibinfo {author} {\bibfnamefont
  {S.}~\bibnamefont {de~Gironcoli}}, \bibinfo {author} {\bibfnamefont
  {P.}~\bibnamefont {Delugas}}, \bibinfo {author} {\bibfnamefont
  {F.}~\bibnamefont {Ferrari~Ruffino}}, \bibinfo {author} {\bibfnamefont
  {A.}~\bibnamefont {Ferretti}}, \bibinfo {author} {\bibfnamefont
  {N.}~\bibnamefont {Marzari}}, \bibinfo {author} {\bibfnamefont
  {I.}~\bibnamefont {Timrov}}, \bibinfo {author} {\bibfnamefont
  {A.}~\bibnamefont {Urru}}, \ and\ \bibinfo {author} {\bibfnamefont
  {S.}~\bibnamefont {Baroni}},\ }\href {\doibase 10.1063/5.0005082} {\bibfield
  {journal} {\bibinfo  {journal} {J.~Chem.~Phys.~}\ }\textbf {\bibinfo {volume}
  {152}},\ \bibinfo {pages} {154105} (\bibinfo {year} {2020})},\ \Eprint
  {http://arxiv.org/abs/https://doi.org/10.1063/5.0005082}
  {https://doi.org/10.1063/5.0005082} \BibitemShut {NoStop}%
\bibitem [{\citenamefont {Perdew}\ \emph {et~al.}(1996)\citenamefont {Perdew},
  \citenamefont {Burke},\ and\ \citenamefont {Ernzerhof}}]{pbe}%
  \BibitemOpen
  \bibfield  {author} {\bibinfo {author} {\bibfnamefont {J.~P.}\ \bibnamefont
  {Perdew}}, \bibinfo {author} {\bibfnamefont {K.}~\bibnamefont {Burke}}, \
  and\ \bibinfo {author} {\bibfnamefont {M.}~\bibnamefont {Ernzerhof}},\ }\href
  {\doibase 10.1103/PhysRevLett.77.3865} {\bibfield  {journal} {\bibinfo
  {journal} {Phys.~Rev.~Lett.~}\ }\textbf {\bibinfo {volume} {77}},\ \bibinfo
  {pages} {3865} (\bibinfo {year} {1996})}\BibitemShut {NoStop}%
\bibitem [{\citenamefont {Schlipf}\ and\ \citenamefont
  {Gygi}(2015)}]{sg15_2015}%
  \BibitemOpen
  \bibfield  {author} {\bibinfo {author} {\bibfnamefont {M.}~\bibnamefont
  {Schlipf}}\ and\ \bibinfo {author} {\bibfnamefont {F.}~\bibnamefont {Gygi}},\
  }\href {\doibase https://doi.org/10.1016/j.cpc.2015.05.011} {\bibfield
  {journal} {\bibinfo  {journal} {Comput.~Phys.~Commun.~}\ }\textbf {\bibinfo
  {volume} {196}},\ \bibinfo {pages} {36} (\bibinfo {year} {2015})}\BibitemShut
  {NoStop}%
\bibitem [{\citenamefont {Tkatchenko}\ and\ \citenamefont
  {Scheffler}(2009)}]{ts}%
  \BibitemOpen
  \bibfield  {author} {\bibinfo {author} {\bibfnamefont {A.}~\bibnamefont
  {Tkatchenko}}\ and\ \bibinfo {author} {\bibfnamefont {M.}~\bibnamefont
  {Scheffler}},\ }\href {\doibase 10.1103/PhysRevLett.102.073005} {\bibfield
  {journal} {\bibinfo  {journal} {Phys.~Rev.~Lett.~}\ }\textbf {\bibinfo
  {volume} {102}},\ \bibinfo {pages} {073005} (\bibinfo {year}
  {2009})}\BibitemShut {NoStop}%
\bibitem [{\citenamefont {Heyd}\ \emph {et~al.}(2003)\citenamefont {Heyd},
  \citenamefont {Scuseria},\ and\ \citenamefont {Ernzerhof}}]{hse06}%
  \BibitemOpen
  \bibfield  {author} {\bibinfo {author} {\bibfnamefont {J.}~\bibnamefont
  {Heyd}}, \bibinfo {author} {\bibfnamefont {G.~E.}\ \bibnamefont {Scuseria}},
  \ and\ \bibinfo {author} {\bibfnamefont {M.}~\bibnamefont {Ernzerhof}},\
  }\href {\doibase 10.1063/1.1564060} {\bibfield  {journal} {\bibinfo
  {journal} {J.~Chem.~Phys.~}\ }\textbf {\bibinfo {volume} {118}},\ \bibinfo
  {pages} {8207} (\bibinfo {year} {2003})},\ \Eprint
  {http://arxiv.org/abs/https://doi.org/10.1063/1.1564060}
  {https://doi.org/10.1063/1.1564060} \BibitemShut {NoStop}%
\bibitem [{\citenamefont {Scherpelz}\ \emph {et~al.}(2016)\citenamefont
  {Scherpelz}, \citenamefont {Govoni}, \citenamefont {Hamada},\ and\
  \citenamefont {Galli}}]{sg15_2016}%
  \BibitemOpen
  \bibfield  {author} {\bibinfo {author} {\bibfnamefont {P.}~\bibnamefont
  {Scherpelz}}, \bibinfo {author} {\bibfnamefont {M.}~\bibnamefont {Govoni}},
  \bibinfo {author} {\bibfnamefont {I.}~\bibnamefont {Hamada}}, \ and\ \bibinfo
  {author} {\bibfnamefont {G.}~\bibnamefont {Galli}},\ }\href {\doibase
  10.1021/acs.jctc.6b00114} {\bibfield  {journal} {\bibinfo  {journal}
  {J.~Chem.~Theory.~Comput.~}\ }\textbf {\bibinfo {volume} {12}},\ \bibinfo
  {pages} {3523} (\bibinfo {year} {2016})},\ \bibinfo {note} {pMID: 27331614},\
  \Eprint {http://arxiv.org/abs/https://doi.org/10.1021/acs.jctc.6b00114}
  {https://doi.org/10.1021/acs.jctc.6b00114} \BibitemShut {NoStop}%
\bibitem [{\citenamefont {Mostofi}\ \emph {et~al.}(2014)\citenamefont
  {Mostofi}, \citenamefont {Yates}, \citenamefont {Pizzi}, \citenamefont {Lee},
  \citenamefont {Souza}, \citenamefont {Vanderbilt},\ and\ \citenamefont
  {Marzari}}]{wannier90}%
  \BibitemOpen
  \bibfield  {author} {\bibinfo {author} {\bibfnamefont {A.~A.}\ \bibnamefont
  {Mostofi}}, \bibinfo {author} {\bibfnamefont {J.~R.}\ \bibnamefont {Yates}},
  \bibinfo {author} {\bibfnamefont {G.}~\bibnamefont {Pizzi}}, \bibinfo
  {author} {\bibfnamefont {Y.-S.}\ \bibnamefont {Lee}}, \bibinfo {author}
  {\bibfnamefont {I.}~\bibnamefont {Souza}}, \bibinfo {author} {\bibfnamefont
  {D.}~\bibnamefont {Vanderbilt}}, \ and\ \bibinfo {author} {\bibfnamefont
  {N.}~\bibnamefont {Marzari}},\ }\href {\doibase
  https://doi.org/10.1016/j.cpc.2014.05.003} {\bibfield  {journal} {\bibinfo
  {journal} {Comput.~Phys.~Commun.~}\ }\textbf {\bibinfo {volume} {185}},\
  \bibinfo {pages} {2309} (\bibinfo {year} {2014})}\BibitemShut {NoStop}%
\bibitem [{\citenamefont {Marzari}\ \emph {et~al.}(2012)\citenamefont
  {Marzari}, \citenamefont {Mostofi}, \citenamefont {Yates}, \citenamefont
  {Souza},\ and\ \citenamefont {Vanderbilt}}]{marzari+2012rmp}%
  \BibitemOpen
  \bibfield  {author} {\bibinfo {author} {\bibfnamefont {N.}~\bibnamefont
  {Marzari}}, \bibinfo {author} {\bibfnamefont {A.~A.}\ \bibnamefont
  {Mostofi}}, \bibinfo {author} {\bibfnamefont {J.~R.}\ \bibnamefont {Yates}},
  \bibinfo {author} {\bibfnamefont {I.}~\bibnamefont {Souza}}, \ and\ \bibinfo
  {author} {\bibfnamefont {D.}~\bibnamefont {Vanderbilt}},\ }\href {\doibase
  10.1103/RevModPhys.84.1419} {\bibfield  {journal} {\bibinfo  {journal}
  {Rev.~Mod.~Phys.~}\ }\textbf {\bibinfo {volume} {84}},\ \bibinfo {pages}
  {1419} (\bibinfo {year} {2012})}\BibitemShut {NoStop}%
\bibitem [{\citenamefont {Dresselhaus}\ \emph {et~al.}(2007)\citenamefont
  {Dresselhaus}, \citenamefont {Dresselhaus},\ and\ \citenamefont
  {Jorio}}]{dresselhaus2007group}%
  \BibitemOpen
  \bibfield  {author} {\bibinfo {author} {\bibfnamefont {M.}~\bibnamefont
  {Dresselhaus}}, \bibinfo {author} {\bibfnamefont {G.}~\bibnamefont
  {Dresselhaus}}, \ and\ \bibinfo {author} {\bibfnamefont {A.}~\bibnamefont
  {Jorio}},\ }\href {https://books.google.de/books?id=sKaH8vrfmnQC} {\emph
  {\bibinfo {title} {Group Theory: Application to the Physics of Condensed
  Matter}}}\ (\bibinfo  {publisher} {Springer Berlin Heidelberg},\ \bibinfo
  {year} {2007})\BibitemShut {NoStop}%
\bibitem [{\citenamefont {Xiao}\ \emph {et~al.}(2012)\citenamefont {Xiao},
  \citenamefont {Liu}, \citenamefont {Feng}, \citenamefont {Xu},\ and\
  \citenamefont {Yao}}]{xiao2012prl}%
  \BibitemOpen
  \bibfield  {author} {\bibinfo {author} {\bibfnamefont {D.}~\bibnamefont
  {Xiao}}, \bibinfo {author} {\bibfnamefont {G.-B.}\ \bibnamefont {Liu}},
  \bibinfo {author} {\bibfnamefont {W.}~\bibnamefont {Feng}}, \bibinfo {author}
  {\bibfnamefont {X.}~\bibnamefont {Xu}}, \ and\ \bibinfo {author}
  {\bibfnamefont {W.}~\bibnamefont {Yao}},\ }\href {\doibase
  10.1103/PhysRevLett.108.196802} {\bibfield  {journal} {\bibinfo  {journal}
  {Phys.~Rev.~Lett.~}\ }\textbf {\bibinfo {volume} {108}},\ \bibinfo {pages}
  {196802} (\bibinfo {year} {2012})}\BibitemShut {NoStop}%
\bibitem [{\citenamefont {B\"oker}\ \emph {et~al.}(2001)\citenamefont
  {B\"oker}, \citenamefont {Severin}, \citenamefont {M\"uller}, \citenamefont
  {Janowitz}, \citenamefont {Manzke}, \citenamefont {Vo\ss{}}, \citenamefont
  {Kr\"uger}, \citenamefont {Mazur},\ and\ \citenamefont
  {Pollmann}}]{boek+01prb}%
  \BibitemOpen
  \bibfield  {author} {\bibinfo {author} {\bibfnamefont {T.}~\bibnamefont
  {B\"oker}}, \bibinfo {author} {\bibfnamefont {R.}~\bibnamefont {Severin}},
  \bibinfo {author} {\bibfnamefont {A.}~\bibnamefont {M\"uller}}, \bibinfo
  {author} {\bibfnamefont {C.}~\bibnamefont {Janowitz}}, \bibinfo {author}
  {\bibfnamefont {R.}~\bibnamefont {Manzke}}, \bibinfo {author} {\bibfnamefont
  {D.}~\bibnamefont {Vo\ss{}}}, \bibinfo {author} {\bibfnamefont
  {P.}~\bibnamefont {Kr\"uger}}, \bibinfo {author} {\bibfnamefont
  {A.}~\bibnamefont {Mazur}}, \ and\ \bibinfo {author} {\bibfnamefont
  {J.}~\bibnamefont {Pollmann}},\ }\href {\doibase 10.1103/PhysRevB.64.235305}
  {\bibfield  {journal} {\bibinfo  {journal} {Phys. Rev. B}\ }\textbf {\bibinfo
  {volume} {64}},\ \bibinfo {pages} {235305} (\bibinfo {year}
  {2001})}\BibitemShut {NoStop}%
\bibitem [{\citenamefont {Pisarra}\ \emph {et~al.}(2021)\citenamefont
  {Pisarra}, \citenamefont {D\'{\i}az},\ and\ \citenamefont
  {Mart\'{\i}n}}]{pisa+21prb}%
  \BibitemOpen
  \bibfield  {author} {\bibinfo {author} {\bibfnamefont {M.}~\bibnamefont
  {Pisarra}}, \bibinfo {author} {\bibfnamefont {C.}~\bibnamefont {D\'{\i}az}},
  \ and\ \bibinfo {author} {\bibfnamefont {F.}~\bibnamefont {Mart\'{\i}n}},\
  }\href {\doibase 10.1103/PhysRevB.103.195416} {\bibfield  {journal} {\bibinfo
   {journal} {Phys. Rev. B}\ }\textbf {\bibinfo {volume} {103}},\ \bibinfo
  {pages} {195416} (\bibinfo {year} {2021})}\BibitemShut {NoStop}%
\bibitem [{\citenamefont {Ma}\ \emph {et~al.}(2011)\citenamefont {Ma},
  \citenamefont {Dai}, \citenamefont {Guo}, \citenamefont {Niu},\ and\
  \citenamefont {Huang}}]{ma+2011ns}%
  \BibitemOpen
  \bibfield  {author} {\bibinfo {author} {\bibfnamefont {Y.}~\bibnamefont
  {Ma}}, \bibinfo {author} {\bibfnamefont {Y.}~\bibnamefont {Dai}}, \bibinfo
  {author} {\bibfnamefont {M.}~\bibnamefont {Guo}}, \bibinfo {author}
  {\bibfnamefont {C.}~\bibnamefont {Niu}}, \ and\ \bibinfo {author}
  {\bibfnamefont {B.}~\bibnamefont {Huang}},\ }\href {\doibase
  10.1039/C1NR10577A} {\bibfield  {journal} {\bibinfo  {journal} {Nanoscale}\
  }\textbf {\bibinfo {volume} {3}},\ \bibinfo {pages} {3883} (\bibinfo {year}
  {2011})}\BibitemShut {NoStop}%
\bibitem [{\citenamefont {Sun}\ \emph {et~al.}(2019)\citenamefont {Sun},
  \citenamefont {Chu}, \citenamefont {Li}, \citenamefont {Zhao}, \citenamefont
  {Li},\ and\ \citenamefont {Li}}]{sun+2019md}%
  \BibitemOpen
  \bibfield  {author} {\bibinfo {author} {\bibfnamefont {Z.}~\bibnamefont
  {Sun}}, \bibinfo {author} {\bibfnamefont {H.}~\bibnamefont {Chu}}, \bibinfo
  {author} {\bibfnamefont {Y.}~\bibnamefont {Li}}, \bibinfo {author}
  {\bibfnamefont {S.}~\bibnamefont {Zhao}}, \bibinfo {author} {\bibfnamefont
  {G.}~\bibnamefont {Li}}, \ and\ \bibinfo {author} {\bibfnamefont
  {D.}~\bibnamefont {Li}},\ }\href {\doibase
  https://doi.org/10.1016/j.matdes.2019.108129} {\bibfield  {journal} {\bibinfo
   {journal} {Materials \& Design}\ }\textbf {\bibinfo {volume} {183}},\
  \bibinfo {pages} {108129} (\bibinfo {year} {2019})}\BibitemShut {NoStop}%
\bibitem [{\citenamefont {Kang}\ \emph {et~al.}(2013)\citenamefont {Kang},
  \citenamefont {Tongay}, \citenamefont {Zhou}, \citenamefont {Li},\ and\
  \citenamefont {Wu}}]{kang+2013apl}%
  \BibitemOpen
  \bibfield  {author} {\bibinfo {author} {\bibfnamefont {J.}~\bibnamefont
  {Kang}}, \bibinfo {author} {\bibfnamefont {S.}~\bibnamefont {Tongay}},
  \bibinfo {author} {\bibfnamefont {J.}~\bibnamefont {Zhou}}, \bibinfo {author}
  {\bibfnamefont {J.}~\bibnamefont {Li}}, \ and\ \bibinfo {author}
  {\bibfnamefont {J.}~\bibnamefont {Wu}},\ }\href {\doibase 10.1063/1.4774090}
  {\bibfield  {journal} {\bibinfo  {journal} {Appl.~Phys.~Lett.~}\ }\textbf
  {\bibinfo {volume} {102}},\ \bibinfo {pages} {012111} (\bibinfo {year}
  {2013})},\ \Eprint {http://arxiv.org/abs/https://doi.org/10.1063/1.4774090}
  {https://doi.org/10.1063/1.4774090} \BibitemShut {NoStop}%
\bibitem [{\citenamefont {Gusakova}\ \emph {et~al.}(2017)\citenamefont
  {Gusakova}, \citenamefont {Wang}, \citenamefont {Shiau}, \citenamefont
  {Krivosheeva}, \citenamefont {Shaposhnikov}, \citenamefont {Borisenko},
  \citenamefont {Gusakov},\ and\ \citenamefont {Tay}}]{gusakova+2017pa}%
  \BibitemOpen
  \bibfield  {author} {\bibinfo {author} {\bibfnamefont {J.}~\bibnamefont
  {Gusakova}}, \bibinfo {author} {\bibfnamefont {X.}~\bibnamefont {Wang}},
  \bibinfo {author} {\bibfnamefont {L.~L.}\ \bibnamefont {Shiau}}, \bibinfo
  {author} {\bibfnamefont {A.}~\bibnamefont {Krivosheeva}}, \bibinfo {author}
  {\bibfnamefont {V.}~\bibnamefont {Shaposhnikov}}, \bibinfo {author}
  {\bibfnamefont {V.}~\bibnamefont {Borisenko}}, \bibinfo {author}
  {\bibfnamefont {V.}~\bibnamefont {Gusakov}}, \ and\ \bibinfo {author}
  {\bibfnamefont {B.~K.}\ \bibnamefont {Tay}},\ }\href {\doibase
  https://doi.org/10.1002/pssa.201700218} {\bibfield  {journal} {\bibinfo
  {journal} {Physica A}\ }\textbf {\bibinfo {volume} {214}},\ \bibinfo {pages}
  {1700218} (\bibinfo {year} {2017})},\ \Eprint
  {http://arxiv.org/abs/https://onlinelibrary.wiley.com/doi/pdf/10.1002/pssa.201700218}
  {https://onlinelibrary.wiley.com/doi/pdf/10.1002/pssa.201700218} \BibitemShut
  {NoStop}%
\bibitem [{\citenamefont {Ramasubramaniam}(2012)}]{ramasubramaniam2012prb}%
  \BibitemOpen
  \bibfield  {author} {\bibinfo {author} {\bibfnamefont {A.}~\bibnamefont
  {Ramasubramaniam}},\ }\href {\doibase 10.1103/PhysRevB.86.115409} {\bibfield
  {journal} {\bibinfo  {journal} {Phys.~Rev.~B}\ }\textbf {\bibinfo {volume}
  {86}},\ \bibinfo {pages} {115409} (\bibinfo {year} {2012})}\BibitemShut
  {NoStop}%
\bibitem [{\citenamefont {Zhang}\ \emph {et~al.}(2015)\citenamefont {Zhang},
  \citenamefont {Chen}, \citenamefont {Johnson}, \citenamefont {Li},
  \citenamefont {Li}, \citenamefont {Mende}, \citenamefont {Feenstra},\ and\
  \citenamefont {Shih}}]{zhang+2015nl}%
  \BibitemOpen
  \bibfield  {author} {\bibinfo {author} {\bibfnamefont {C.}~\bibnamefont
  {Zhang}}, \bibinfo {author} {\bibfnamefont {Y.}~\bibnamefont {Chen}},
  \bibinfo {author} {\bibfnamefont {A.}~\bibnamefont {Johnson}}, \bibinfo
  {author} {\bibfnamefont {M.-Y.}\ \bibnamefont {Li}}, \bibinfo {author}
  {\bibfnamefont {L.-J.}\ \bibnamefont {Li}}, \bibinfo {author} {\bibfnamefont
  {P.~C.}\ \bibnamefont {Mende}}, \bibinfo {author} {\bibfnamefont {R.~M.}\
  \bibnamefont {Feenstra}}, \ and\ \bibinfo {author} {\bibfnamefont {C.-K.}\
  \bibnamefont {Shih}},\ }\href {\doibase 10.1021/acs.nanolett.5b01968}
  {\bibfield  {journal} {\bibinfo  {journal} {Nano~Lett.~}\ }\textbf {\bibinfo
  {volume} {15}},\ \bibinfo {pages} {6494} (\bibinfo {year} {2015})},\ \bibinfo
  {note} {pMID: 26389585},\ \Eprint
  {http://arxiv.org/abs/https://doi.org/10.1021/acs.nanolett.5b01968}
  {https://doi.org/10.1021/acs.nanolett.5b01968} \BibitemShut {NoStop}%
\bibitem [{\citenamefont {Hsu}\ \emph {et~al.}(2017)\citenamefont {Hsu},
  \citenamefont {Lu}, \citenamefont {Wang}, \citenamefont {Huang},
  \citenamefont {Li}, \citenamefont {Chang}, \citenamefont {Chou},
  \citenamefont {Juang}, \citenamefont {Jeng}, \citenamefont {Li},\ and\
  \citenamefont {Chang}}]{hsu+2017natcomm}%
  \BibitemOpen
  \bibfield  {author} {\bibinfo {author} {\bibfnamefont {W.-T.}\ \bibnamefont
  {Hsu}}, \bibinfo {author} {\bibfnamefont {L.-S.}\ \bibnamefont {Lu}},
  \bibinfo {author} {\bibfnamefont {D.}~\bibnamefont {Wang}}, \bibinfo {author}
  {\bibfnamefont {J.-K.}\ \bibnamefont {Huang}}, \bibinfo {author}
  {\bibfnamefont {M.-Y.}\ \bibnamefont {Li}}, \bibinfo {author} {\bibfnamefont
  {T.-R.}\ \bibnamefont {Chang}}, \bibinfo {author} {\bibfnamefont {Y.-C.}\
  \bibnamefont {Chou}}, \bibinfo {author} {\bibfnamefont {Z.-Y.}\ \bibnamefont
  {Juang}}, \bibinfo {author} {\bibfnamefont {H.-T.}\ \bibnamefont {Jeng}},
  \bibinfo {author} {\bibfnamefont {L.-J.}\ \bibnamefont {Li}}, \ and\ \bibinfo
  {author} {\bibfnamefont {W.-H.}\ \bibnamefont {Chang}},\ }\href {\doibase
  10.1038/s41467-017-01012-6} {\bibfield  {journal} {\bibinfo  {journal}
  {Nature~Comm.}\ }\textbf {\bibinfo {volume} {8}},\ \bibinfo {pages} {929}
  (\bibinfo {year} {2017})}\BibitemShut {NoStop}%
\bibitem [{\citenamefont {Tiwari}\ \emph {et~al.}(2008)\citenamefont {Tiwari},
  \citenamefont {Yang}, \citenamefont {Saeys},\ and\ \citenamefont
  {Joachim}}]{tiwa+08ss}%
  \BibitemOpen
  \bibfield  {author} {\bibinfo {author} {\bibfnamefont {R.~K.}\ \bibnamefont
  {Tiwari}}, \bibinfo {author} {\bibfnamefont {J.}~\bibnamefont {Yang}},
  \bibinfo {author} {\bibfnamefont {M.}~\bibnamefont {Saeys}}, \ and\ \bibinfo
  {author} {\bibfnamefont {C.}~\bibnamefont {Joachim}},\ }\href@noop {}
  {\bibfield  {journal} {\bibinfo  {journal} {Surf.~Sci.~}\ }\textbf {\bibinfo
  {volume} {602}},\ \bibinfo {pages} {2628} (\bibinfo {year}
  {2008})}\BibitemShut {NoStop}%
\bibitem [{\citenamefont {Szabo}\ and\ \citenamefont
  {Ostlund}(1996)}]{szabo1082}%
  \BibitemOpen
  \bibfield  {author} {\bibinfo {author} {\bibfnamefont {A.}~\bibnamefont
  {Szabo}}\ and\ \bibinfo {author} {\bibfnamefont {N.~S.}\ \bibnamefont
  {Ostlund}},\ }\href@noop {} {\emph {\bibinfo {title} {Modern Quantum
  Chemistry: Introduction to Advanced Electronic Structure Theory}}},\ \bibinfo
  {edition} {1st}\ ed.\ (\bibinfo  {publisher} {Dover Publications, Inc.},\
  \bibinfo {address} {Mineola},\ \bibinfo {year} {1996})\BibitemShut {NoStop}%
\bibitem [{\citenamefont {Breuer}\ \emph {et~al.}(2016)\citenamefont {Breuer},
  \citenamefont {Ma{\ss}meyer}, \citenamefont {M{\"a}nz}, \citenamefont
  {Zoerb}, \citenamefont {Harbrecht},\ and\ \citenamefont
  {Witte}}]{breu+16pssrrl}%
  \BibitemOpen
  \bibfield  {author} {\bibinfo {author} {\bibfnamefont {T.}~\bibnamefont
  {Breuer}}, \bibinfo {author} {\bibfnamefont {T.}~\bibnamefont
  {Ma{\ss}meyer}}, \bibinfo {author} {\bibfnamefont {A.}~\bibnamefont
  {M{\"a}nz}}, \bibinfo {author} {\bibfnamefont {S.}~\bibnamefont {Zoerb}},
  \bibinfo {author} {\bibfnamefont {B.}~\bibnamefont {Harbrecht}}, \ and\
  \bibinfo {author} {\bibfnamefont {G.}~\bibnamefont {Witte}},\ }\href@noop {}
  {\bibfield  {journal} {\bibinfo  {journal} {Phys.~Status~Solidi~--RRL}\
  }\textbf {\bibinfo {volume} {10}},\ \bibinfo {pages} {905} (\bibinfo {year}
  {2016})}\BibitemShut {NoStop}%
\bibitem [{\citenamefont {Mrkyvkova}\ \emph {et~al.}(2019)\citenamefont
  {Mrkyvkova}, \citenamefont {Hodas}, \citenamefont {Hagara}, \citenamefont
  {Nadazdy}, \citenamefont {Halahovets}, \citenamefont {Bodik}, \citenamefont
  {Tokar}, \citenamefont {Chai}, \citenamefont {Wang}, \citenamefont {Chi},
  \citenamefont {Chumakov}, \citenamefont {Konovalov}, \citenamefont
  {Hinderhofer}, \citenamefont {Jergel}, \citenamefont {Majkova}, \citenamefont
  {Siffalovic},\ and\ \citenamefont {Schreiber}}]{mrky+19apl}%
  \BibitemOpen
  \bibfield  {author} {\bibinfo {author} {\bibfnamefont {N.}~\bibnamefont
  {Mrkyvkova}}, \bibinfo {author} {\bibfnamefont {M.}~\bibnamefont {Hodas}},
  \bibinfo {author} {\bibfnamefont {J.}~\bibnamefont {Hagara}}, \bibinfo
  {author} {\bibfnamefont {P.}~\bibnamefont {Nadazdy}}, \bibinfo {author}
  {\bibfnamefont {Y.}~\bibnamefont {Halahovets}}, \bibinfo {author}
  {\bibfnamefont {M.}~\bibnamefont {Bodik}}, \bibinfo {author} {\bibfnamefont
  {K.}~\bibnamefont {Tokar}}, \bibinfo {author} {\bibfnamefont {J.~W.}\
  \bibnamefont {Chai}}, \bibinfo {author} {\bibfnamefont {S.~J.}\ \bibnamefont
  {Wang}}, \bibinfo {author} {\bibfnamefont {D.~Z.}\ \bibnamefont {Chi}},
  \bibinfo {author} {\bibfnamefont {A.}~\bibnamefont {Chumakov}}, \bibinfo
  {author} {\bibfnamefont {O.}~\bibnamefont {Konovalov}}, \bibinfo {author}
  {\bibfnamefont {A.}~\bibnamefont {Hinderhofer}}, \bibinfo {author}
  {\bibfnamefont {M.}~\bibnamefont {Jergel}}, \bibinfo {author} {\bibfnamefont
  {E.}~\bibnamefont {Majkova}}, \bibinfo {author} {\bibfnamefont
  {P.}~\bibnamefont {Siffalovic}}, \ and\ \bibinfo {author} {\bibfnamefont
  {F.}~\bibnamefont {Schreiber}},\ }\href {\doibase 10.1063/1.5100282}
  {\bibfield  {journal} {\bibinfo  {journal} {Appl.~Phys.~Lett.~}\ }\textbf
  {\bibinfo {volume} {114}},\ \bibinfo {pages} {251906} (\bibinfo {year}
  {2019})},\ \Eprint {http://arxiv.org/abs/https://doi.org/10.1063/1.5100282}
  {https://doi.org/10.1063/1.5100282} \BibitemShut {NoStop}%
\bibitem [{\citenamefont {Padgaonkar}\ \emph {et~al.}(2019)\citenamefont
  {Padgaonkar}, \citenamefont {Amsterdam}, \citenamefont {Bergeron},
  \citenamefont {Su}, \citenamefont {Marks}, \citenamefont {Hersam},\ and\
  \citenamefont {Weiss}}]{padg+19jpcc}%
  \BibitemOpen
  \bibfield  {author} {\bibinfo {author} {\bibfnamefont {S.}~\bibnamefont
  {Padgaonkar}}, \bibinfo {author} {\bibfnamefont {S.~H.}\ \bibnamefont
  {Amsterdam}}, \bibinfo {author} {\bibfnamefont {H.}~\bibnamefont {Bergeron}},
  \bibinfo {author} {\bibfnamefont {K.}~\bibnamefont {Su}}, \bibinfo {author}
  {\bibfnamefont {T.~J.}\ \bibnamefont {Marks}}, \bibinfo {author}
  {\bibfnamefont {M.~C.}\ \bibnamefont {Hersam}}, \ and\ \bibinfo {author}
  {\bibfnamefont {E.~A.}\ \bibnamefont {Weiss}},\ }\href@noop {} {\bibfield
  {journal} {\bibinfo  {journal} {J.~Phys.~Chem.~C}\ }\textbf {\bibinfo
  {volume} {123}},\ \bibinfo {pages} {13337} (\bibinfo {year}
  {2019})}\BibitemShut {NoStop}%
\bibitem [{\citenamefont {Kachel}\ \emph {et~al.}(2021)\citenamefont {Kachel},
  \citenamefont {Dombrowski}, \citenamefont {Breuer}, \citenamefont
  {Gottfried},\ and\ \citenamefont {Witte}}]{kach+21cs}%
  \BibitemOpen
  \bibfield  {author} {\bibinfo {author} {\bibfnamefont {S.~R.}\ \bibnamefont
  {Kachel}}, \bibinfo {author} {\bibfnamefont {P.-M.}\ \bibnamefont
  {Dombrowski}}, \bibinfo {author} {\bibfnamefont {T.}~\bibnamefont {Breuer}},
  \bibinfo {author} {\bibfnamefont {J.~M.}\ \bibnamefont {Gottfried}}, \ and\
  \bibinfo {author} {\bibfnamefont {G.}~\bibnamefont {Witte}},\ }\href@noop {}
  {\bibfield  {journal} {\bibinfo  {journal} {Chem.~Sci.~}\ }\textbf {\bibinfo
  {volume} {12}},\ \bibinfo {pages} {2575} (\bibinfo {year}
  {2021})}\BibitemShut {NoStop}%
\bibitem [{\citenamefont {Amsterdam}\ \emph {et~al.}(2021)\citenamefont
  {Amsterdam}, \citenamefont {Marks},\ and\ \citenamefont
  {Hersam}}]{amst+21jpcl}%
  \BibitemOpen
  \bibfield  {author} {\bibinfo {author} {\bibfnamefont {S.~H.}\ \bibnamefont
  {Amsterdam}}, \bibinfo {author} {\bibfnamefont {T.~J.}\ \bibnamefont
  {Marks}}, \ and\ \bibinfo {author} {\bibfnamefont {M.~C.}\ \bibnamefont
  {Hersam}},\ }\href@noop {} {\bibfield  {journal} {\bibinfo  {journal}
  {J.~Phys.~Chem.~Lett.}\ }\textbf {\bibinfo {volume} {12}},\ \bibinfo {pages}
  {4543} (\bibinfo {year} {2021})}\BibitemShut {NoStop}%
\bibitem [{\citenamefont {Puschnig}\ \emph {et~al.}(2009)\citenamefont
  {Puschnig}, \citenamefont {Berkebile}, \citenamefont {Fleming}, \citenamefont
  {Koller}, \citenamefont {Emtsev}, \citenamefont {Seyller}, \citenamefont
  {Riley}, \citenamefont {Ambrosch-Draxl}, \citenamefont {Netzer},\ and\
  \citenamefont {Ramsey}}]{puschnig+2009sci}%
  \BibitemOpen
  \bibfield  {author} {\bibinfo {author} {\bibfnamefont {P.}~\bibnamefont
  {Puschnig}}, \bibinfo {author} {\bibfnamefont {S.}~\bibnamefont {Berkebile}},
  \bibinfo {author} {\bibfnamefont {A.~J.}\ \bibnamefont {Fleming}}, \bibinfo
  {author} {\bibfnamefont {G.}~\bibnamefont {Koller}}, \bibinfo {author}
  {\bibfnamefont {K.}~\bibnamefont {Emtsev}}, \bibinfo {author} {\bibfnamefont
  {T.}~\bibnamefont {Seyller}}, \bibinfo {author} {\bibfnamefont {J.~D.}\
  \bibnamefont {Riley}}, \bibinfo {author} {\bibfnamefont {C.}~\bibnamefont
  {Ambrosch-Draxl}}, \bibinfo {author} {\bibfnamefont {F.~P.}\ \bibnamefont
  {Netzer}}, \ and\ \bibinfo {author} {\bibfnamefont {M.~G.}\ \bibnamefont
  {Ramsey}},\ }\href {\doibase 10.1126/science.1176105} {\bibfield  {journal}
  {\bibinfo  {journal} {Science}\ }\textbf {\bibinfo {volume} {326}},\ \bibinfo
  {pages} {702} (\bibinfo {year} {2009})},\ \Eprint
  {http://arxiv.org/abs/https://science.sciencemag.org/content/326/5953/702.full.pdf}
  {https://science.sciencemag.org/content/326/5953/702.full.pdf} \BibitemShut
  {NoStop}%
\bibitem [{\citenamefont {S\"attele}\ \emph {et~al.}(2021)\citenamefont
  {S\"attele}, \citenamefont {Windischbacher}, \citenamefont {Egger},
  \citenamefont {Haags}, \citenamefont {Hurdax}, \citenamefont {Kirschner},
  \citenamefont {Gottwald}, \citenamefont {Richter}, \citenamefont {Bocquet},
  \citenamefont {Soubatch}, \citenamefont {Tautz}, \citenamefont {Bettinger},
  \citenamefont {Peisert}, \citenamefont {Chassé}, \citenamefont {Ramsey},
  \citenamefont {Puschnig},\ and\ \citenamefont {Koller}}]{saet+21jpcc}%
  \BibitemOpen
  \bibfield  {author} {\bibinfo {author} {\bibfnamefont {M.~S.}\ \bibnamefont
  {S\"attele}}, \bibinfo {author} {\bibfnamefont {A.}~\bibnamefont
  {Windischbacher}}, \bibinfo {author} {\bibfnamefont {L.}~\bibnamefont
  {Egger}}, \bibinfo {author} {\bibfnamefont {A.}~\bibnamefont {Haags}},
  \bibinfo {author} {\bibfnamefont {P.}~\bibnamefont {Hurdax}}, \bibinfo
  {author} {\bibfnamefont {H.}~\bibnamefont {Kirschner}}, \bibinfo {author}
  {\bibfnamefont {A.}~\bibnamefont {Gottwald}}, \bibinfo {author}
  {\bibfnamefont {M.}~\bibnamefont {Richter}}, \bibinfo {author} {\bibfnamefont
  {F.~C.}\ \bibnamefont {Bocquet}}, \bibinfo {author} {\bibfnamefont
  {S.}~\bibnamefont {Soubatch}}, \bibinfo {author} {\bibfnamefont {F.~S.}\
  \bibnamefont {Tautz}}, \bibinfo {author} {\bibfnamefont {H.~F.}\ \bibnamefont
  {Bettinger}}, \bibinfo {author} {\bibfnamefont {H.}~\bibnamefont {Peisert}},
  \bibinfo {author} {\bibfnamefont {T.}~\bibnamefont {Chassé}}, \bibinfo
  {author} {\bibfnamefont {M.~G.}\ \bibnamefont {Ramsey}}, \bibinfo {author}
  {\bibfnamefont {P.}~\bibnamefont {Puschnig}}, \ and\ \bibinfo {author}
  {\bibfnamefont {G.}~\bibnamefont {Koller}},\ }\href {\doibase
  10.1021/acs.jpcc.0c09062} {\bibfield  {journal} {\bibinfo  {journal}
  {J.~Phys.~Chem.~C}\ }\textbf {\bibinfo {volume} {125}},\ \bibinfo {pages}
  {2918} (\bibinfo {year} {2021})},\ \Eprint
  {http://arxiv.org/abs/https://doi.org/10.1021/acs.jpcc.0c09062}
  {https://doi.org/10.1021/acs.jpcc.0c09062} \BibitemShut {NoStop}%
\bibitem [{\citenamefont {Krumland}\ \emph {et~al.}(2021)\citenamefont
  {Krumland}, \citenamefont {Gil}, \citenamefont {Corni},\ and\ \citenamefont
  {Cocchi}}]{krum+21jcp}%
  \BibitemOpen
  \bibfield  {author} {\bibinfo {author} {\bibfnamefont {J.}~\bibnamefont
  {Krumland}}, \bibinfo {author} {\bibfnamefont {G.}~\bibnamefont {Gil}},
  \bibinfo {author} {\bibfnamefont {S.}~\bibnamefont {Corni}}, \ and\ \bibinfo
  {author} {\bibfnamefont {C.}~\bibnamefont {Cocchi}},\ }\href@noop {}
  {\bibfield  {journal} {\bibinfo  {journal} {J.~Chem.~Phys.~}\ }\textbf
  {\bibinfo {volume} {154}},\ \bibinfo {pages} {224114} (\bibinfo {year}
  {2021})}\BibitemShut {NoStop}%
\end{thebibliography}

%

\end{document}